\pdfoutput=1

\documentclass[11pt]{article}

\usepackage[final]{acl}

\usepackage{times}
\usepackage{latexsym}
\usepackage{amsmath,amsfonts}
\usepackage{algorithmic}
\usepackage{algorithm}
\usepackage{subcaption}
\usepackage{array}
\usepackage{tabularx}
\usepackage{booktabs}
\usepackage{textcomp}
\usepackage{stfloats}
\usepackage{url}
\usepackage{verbatim}
\usepackage{graphicx}
\usepackage{bm}
\usepackage{multirow}
\usepackage{amssymb}

\usepackage{xcolor}
\usepackage[T1]{fontenc}

\usepackage[utf8]{inputenc}

\usepackage{microtype}

\usepackage{inconsolata}

\title{PERCEIVE: A Benchmark for Personalized Emotion and Communication Behavior Understanding on Social Media}

\author{
	\textbf{Jian Liao\textsuperscript{\rm 1,3}},
	\textbf{Yujin Zheng\textsuperscript{\rm 1}},
	\textbf{Suge Wang\textsuperscript{\rm 2}$^*$},
	\textbf{Jianxing Zheng\textsuperscript{\rm 2}},
	\textbf{Deyu Li\textsuperscript{\rm 2}}
	\\
	\textsuperscript{\rm 1}School of Computer and Information Technology, Shanxi University, China\\
	\textsuperscript{\rm 2}Key Laboratory of Computational Intelligence and Chinese Information Processing of\\ Ministry of Education, Shanxi University, China\\	
	\textsuperscript{\rm 3}	Joint Laboratory of Tourism Big Data in Shanxi Province, China\\
	\small{
		\textbf{$^*$Correspondence:} \href{mailto:wsg@sxu.edu.cn}{wsg@sxu.edu.cn}
		}
}

\begin{document}
\maketitle
\begin{abstract}
Current emotion analysis in social media is predominantly author-centric, failing to capture the subjective nature of emotional responses across diverse readers. This paradigm overlooks the crucial link between individual perception, communication behavior, and the underlying social network. To bridge this gap, we introduce PERCEIVE, a novel bilingual (English and Chinese) large-scale benchmark that, to the best of our knowledge, is the first to integrate five critical dimensions for social perception: author-created content, genuine readers' emotional feedback (derived from their comments), communication behavior, user attributes, and the social graph. This benchmark enables a paradigm shift towards truly personalized, reader-centric analysis, where different readers' emotional responses to the same content are naturally captured through their real-world interactions. By annotating emotions from reader comments and synchronously capturing communication intent, PERCEIVE provides a unique resource to model the intrinsic coupling between emotion and behavior, grounded in social context. We establish a comprehensive evaluation protocol, testing state-of-the-art methods, including large language models (LLMs) with advanced reasoning enhancement. Our findings reveal significant shortcomings in existing approaches when handling this multifaceted, user-aware task. PERCEIVE offers a foundational resource and clear direction for future research in socially-intelligent NLP, pushing models towards a more unified understanding of emotion on social media.
\end{abstract}

\section{Introduction}
\begin{figure}[t]
	\centering
	\includegraphics[width=0.99\linewidth]{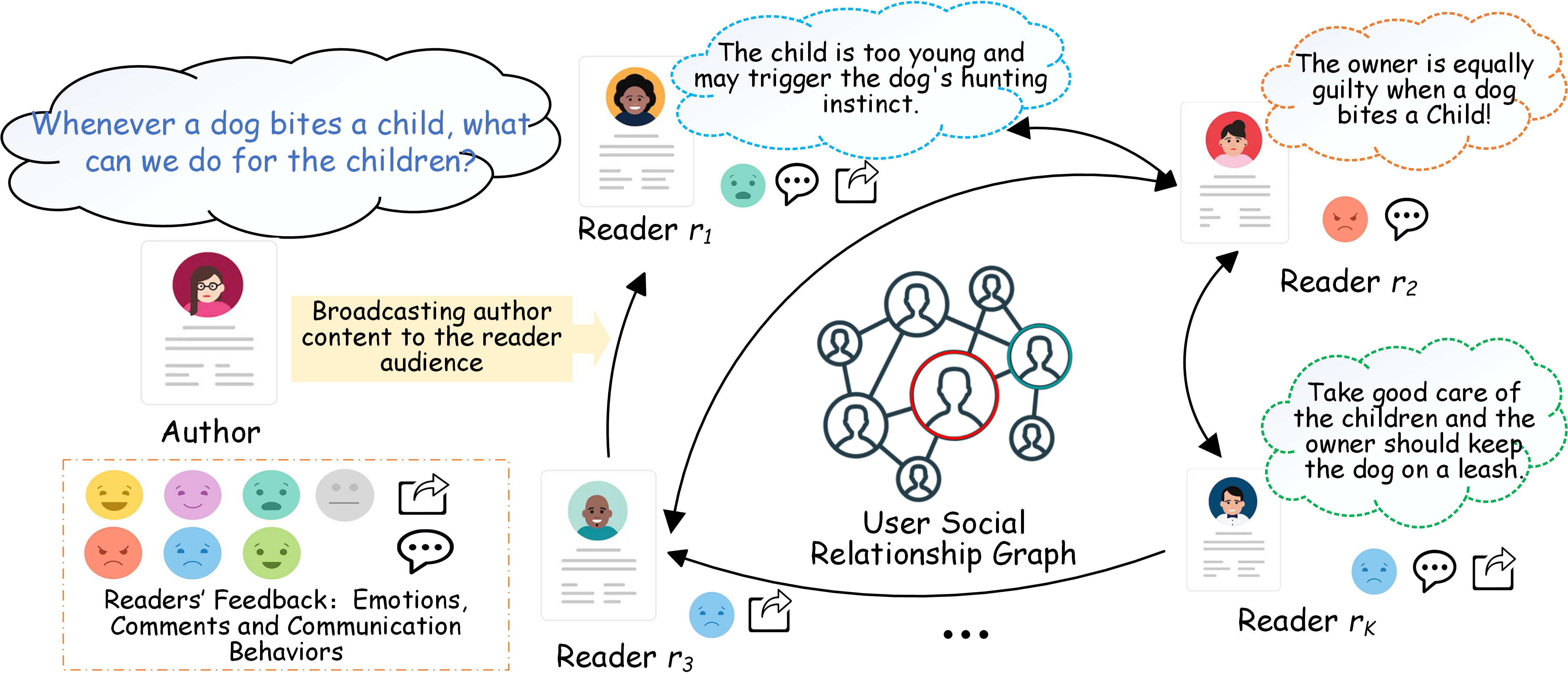}
	\caption{Illustration of personalized emotions and differential communication behaviors evoked by the same social media content from different reader perspectives.}
	\label{fig:challenges}
\end{figure}

Social media has become the primary arena for emotional expression and information dissemination in modern society. The analysis of emotions within the vast ocean of user-generated content is crucial for understanding public opinion, monitoring societal dynamics, and providing personalized services. However, a fundamental limitation persists in mainstream emotion analysis methodologies: they predominantly operate under an ``author-centric'' \cite{wei-etal-2025-mecot, hu-collier-2025-inews} or ``content-centric'' \cite{li2025exploring, jian2025simrp} paradigm, assigning a single, monolithic emotion label to an entire post. This paradigm overlooks the highly subjective and individualized nature of emotion perception, failing to capture that the same content can evoke disparate emotional responses and subsequent behaviors among different readers. For instance, a discussion about an emerging technology product might elicit ``excitement'' and a propensity to share among tech enthusiasts, while triggering ``anxiety'' and critical comments from skeptical users (as illustrated in Fig.~\ref{fig:challenges}). This exposes the core blind spot of existing research: a single label, detached from the ``reader's perspective,'' cannot capture the true dynamics of emotion, nor can it explain how internal states drive external social actions.

In contrast, annotating emotions on reader comments—each tied to a specific individual—preserves personalized responses without aggregation, shifting analysis from post-level to comment-level. Unlike previous datasets with isolated post-comment pairs, PERCEIVE adopts a post-comments structure: a single post linked to multiple genuine reader comments, each annotated with the reader's emotional reaction, capturing diverse responses to the same content.

Crucially, we do not rely on annotators to simulate reader emotions. Instead, PERCEIVE leverages natural social interactions: users who commented on or forwarded a post are treated as ``real readers,'' and emotion labels are annotated based on their genuine responses. This design grounds personalization in authentic user behavior, linking specific user profiles to their emotional reactions.

Although recent studies have begun to transcend conventional paradigms, their limitations remain significant. For instance, \citet{ye-etal-2025-generic} and \citet{jun-lee-2025-exploring} enhance personalized emotion understanding by incorporating the author's dynamic context or a user's historical content, yet they remain within an ``author-centric'' framework, lacking authentic emotional feedback data from the reader's standpoint. \citet{Liao2025} introduce an innovative approach that enhances personalized emotional modeling by integrating simulated feedback from reader agents powered by LLMs, but treat emotion prediction and behavior simulation as separate tasks, failing to capture their intrinsic coupling.
This decoupled approach lacks a firm theoretical underpinning. According to \textit{Social Exchange Theory} \cite{10.3389/fpsyg.2022.1015921}, an individual's social behavior is a rational decision-making process of weighing costs and benefits, where emotional state is a key internal variable influencing this decision. \textbf{Neglecting this mediating mechanism, models cannot explain why the same emotion might lead to different behaviors.} Furthermore, most existing models treat users as isolated nodes \cite{lyu2023, zhao2024rdgcn}, overlooking the essence of social media as a relational network. \textit{Emotion Contagion Theory} \cite{CHEN2024108304} posits that emotions spread like viruses within groups, with the rate and scope of communication profoundly influenced by network structure and node (user) influence. Consequently, without effective integration of the social graph, \textbf{models cannot answer the core scientific question: ``How do emotions propagate and evolve within a community?}''

These limitations lead to three interconnected core challenges:

\textbf{(1) How can we achieve personalized emotion annotation from a true reader's perspective?} Existing annotation paradigms inherently fail to capture inter-individual differences in emotional responses. A shift is required from labeling content itself to jointly annotating emotion and behavioral intent for ``user-content'' interaction pairs.

\textbf{(2) How can we effectively model the coupling between implicit emotions and overt communication behaviors?} A theory-driven framework is needed to establish a fine-grained mapping between abstract emotional states and concrete social choices (e.g., share, comment), thereby elucidating the mechanism by which emotion drives behavioral decisions.

\textbf{(3) How can we integrate user attributes with social structure to reveal the dynamics of group-level emotion communication?} The social graph must be treated as a core component of emotion analysis. By combining it with user profiles, we can investigate the communication pathways, sphere of influence, and evolutionary patterns of emotions in complex networks.

To address these challenges, we construct and release PERCEIVE, a large-scale, heterogeneous, bilingual (English and Chinese) benchmark from Weibo and Twitter. \textbf{To the best of our knowledge, it is the first benchmark integrating five dimensions of social perception: author content, genuine reader emotions (from comments), communication behaviors, user attributes, and social graph—capturing diverse reader reactions to the same content.} The main contributions of this paper are as follows:

(1) We construct and release the PERCEIVE dataset, a large, social media emotion understanding benchmark that integrates five core aspects: author, reader, behavior, profiles, and relationships, providing a solid foundation for studying emotion communication in social media.

(2) We introduce a novel annotation paradigm that annotates emotions from the perspective of readers' comments, achieving truly personalized emotion analysis by capturing real emotional responses to specific posts.

(3) We establish a fine-grained mapping between implicit emotions and overt behaviors by synchronously annotating comment emotions and communication intents, offering a valuable resource for emotion-driven behavior and intent understanding.

(4) We provide a comprehensive evaluation benchmark, systematically assessing the performance of various state-of-the-art (SOTA) methods, including LLMs, chain-of-reasoning, and debate-based reasoning. Our evaluation reveals the shortcomings of current techniques and provides clear directions for future research. The dataset will be released after the anonymous review period.

\section{Related Work}
\textbf{Sentiment/Emotion Datasets}.
Standard sentiment/emotion analysis benchmarks like SemEval-14 \cite{pontiki-etal-2014-semeval}, SemEval-17 \cite{rosenthal-etal-2017-semeval}, GoEmotions \cite{demszky-etal-2020-goemotions}, OATS-ABSA \cite{chebolu-etal-2024-oats}, M-ABSA \cite{wu-etal-2025-absa}, and BRIGHTER \cite{muhammad-etal-2025-brighter} provide large-scale annotations but are fundamentally author-centric, assigning a single emotion label to each post and ignoring the subjective variability among readers. While datasets like RAPPIE \cite{Liao2025} explore personalization. \textbf{They are often limited to ``content-centric'' settings, overlooking the joint annotation of implicit human emotions and explicit social behaviors.} Comparison of existing sentiment/emotion analysis datasets is illustrated in Table~\ref{tab:dataset_comparison}.

\begin{table*}[t]
	\centering
	\caption{Comparison of PERCEIVE with existing sentiment analysis datasets.}
	\label{tab:dataset_comparison}
	\footnotesize
	\tabcolsep 1.2pt
	\begin{tabularx}{\textwidth}{l|p{1.3cm}|p{2.1cm}|c|p{1.2cm}|p{1.6cm}|p{1.1cm}|p{0.7cm}|p{4cm}}
		\toprule
		\textbf{Dataset} & \textbf{Emotion Subject} & \textbf{Domain} & \textbf{Language} & \textbf{Behavior} & \textbf{User Profile} & \textbf{Social Graph} & \textbf{Scale} & \textbf{Applicable Tasks} \\
		\midrule
		SemEval-14 & author & product reviews & EN & $\times$ & $\times$ & $\times$ & 7.6k & ABSA \\\midrule
		ECC & author & conversation & EN & $\times$ & $\times$ & $\times$ & 2.4k & emotion cause extraction \\\midrule
		SemEval-17 & author & Twitter & EN, AR & $\times$ & $\times$ & $\times$ & 72k & ABSA \\
		GoEmotions & author & Reddit & EN & $\times$ & $\times$ & $\times$ & 58k & ABSA \\
		OATS-ABSA & author & social media & EN & $\times$ & $\times$ & $\times$ & 44k & open-target ABSA \\
		M-ABSA & author & social media & EN & $\times$ & $\times$ & $\times$ & 14k & multimodal ABSA \\
		BRIGHTER & author & social media & multiple & $\times$ & $\times$ & $\times$ & 100k & sentiment analysis \\
		RAPPIE & author & social media & EN, CN & $\times$ & \textbf{\checkmark} & CN only & 29k & personalized emotion analysis \\
		\midrule
		EmoBank & average reader & news & EN & $\times$ & $\times$ & $\times$ & 10k & dimensional emotion analysis \\
		MultiPICo & annotator & social media & multiple & $\times$ &  506 annotators' & $\times$ & 18.8k & irony detection\\
		INEWS & annotator & news & EN & $\times$ & 291 annotators' & $\times$ & 2.9k & emotional respond simulation \\
		\midrule
		\textbf{PERCEIVE} & individual reader & social media & EN, CN & \checkmark & 39349 real reader users & \checkmark & 113k & personalized emotion/behavior analysis, emotion propagation, group-level analysis \\
		\bottomrule
	\end{tabularx}
\end{table*}

\textbf{Reader Perspective and Subjectivity}.
Recent work has explored subjective annotation, as shown in Table~\ref{tab:dataset_comparison}. In contrast to EmoBank \cite{buechel-hahn-2017-emobank}, which assesses emotions evoked in an \textit{average reader} and thus captures only group-level feedback, our work distinguishes itself by achieving precise, individual-level reader emotion annotation.
Unlike MultiPICo \cite{casola-etal-2024-multipico} and INEWS \cite{hu-collier-2025-inews}, which ask annotators to \textit{imagine} or \textit{simulate} a reader's potential feelings, PERCEIVE leverages \textit{natural social interactions}. \textbf{We identify users who actively engaged with a post—by commenting or forwarding—as ``real readers,'' integrating their content, emotions, behaviors, profiles, and social graphs.}

\textbf{Personalization and Human Behavior Simulation}. 
Effective personalization relies on user models, which have evolved from traditional features like attributes and historical embeddings \cite{zhou-etal-2023-learning} to persona-based LLMs such as LaMP \cite{salemi-etal-2024-lamp} and PPlug \cite{liu-etal-2025-llms}. A primary challenge, however, is their dependence on detailed user profiles, which raises significant scalability and privacy concerns \cite{hellwig-etal-2025-still}. Recently, LLMs have demonstrated remarkable capabilities in generating user profiles, enabling new approaches to human behavior simulation. These methods, including prompt-based techniques \cite{wang-etal-2025-besimulator}, behavior-realism alignment \cite{wang-etal-2025-implicit}, and cognitive-behavioral fixation \cite{wang-etal-2025-evaluating-cognitive}, have proven valuable for downstream tasks such as recommendation systems \cite{bougie-watanabe-2025-simuser} and social simulation \cite{mou-etal-2024-unveiling}.
\textbf{However, the evaluation of these sophisticated models has been constrained by datasets that lack the personalization and behavioral context necessary for a comprehensive assessment of social intelligence.}

\section{Dataset Construction: PERCEIVE}

To address the three core challenges outlined in the Introduction, we construct and release PERCEIVE, a large-scale, heterogeneous, bilingual (English and Chinese) benchmark designed to study emotion communication from a truly personalized and networked perspective. The construction of PERCEIVE is meticulously designed to operationalize our theoretical commitments, specifically integrating the reader's viewpoint, modeling emotion-behavior coupling, and embedding social structures.

\subsection{Data Collection and Curation}

Our data collection is anchored by a \textbf{topic-driven} curation strategy, ensuring that every instance is grounded in authentic public discourse.

\textbf{Sources and Scope.} To capture culturally diverse perspectives, we collected data sourced from two high-impact platforms: Weibo (Chinese) and Twitter (English). This dataset's temporal scope (2009–2023) offers a dual advantage: it anchors the data firmly in a pre-LLM era, ensuring it is free from AI-generated content contamination, and its long-term span enables a novel longitudinal analysis of how these emotions originate, evolve, and propagate across socio-cultural developments.

\textbf{Topic Filtering.} To move beyond keyword-based search and ensure social relevance, we employed a multi-step programmatic analysis: (1) extracting hashtagged posts; (2) categorizing hashtags into domains (e.g., politics, sports) via keyword matching and LDA; (3) refining topics within domains based on post volume and engagement diversity; (4) manual expert review for relevance. Activity thresholds were then applied, retaining 116 Twitter topics and 139 Weibo topics. This ensures PERCEIVE is a curated archive of meaningful public conversations, not merely a text collection.

\textbf{Multi-Dimensional Heterogeneity.} To actualize the five-dimensional information integral to our research goals, our collection strategy systematically gathered:

\textbf{(1) Author-created Content:} The original posts that initiate or participate in discussions.

\textbf{(2) Readers' Emotional Feedback:} The textual comments from real users responding to the original posts. This is the core data unit for our reader-centric analysis. Collectively, they reflect the diverse emotional feedback of different users to the original post.

\textbf{(3) Communication Behaviors:} Overt user engagement, categorized as \textit{Repost \& Comment} (R\&C), \textit{Repost Only} (R), and \textit{Lurking} (L) (including likes and passive viewing).

\textbf{(4) User Attributes:} Desensitized profile information for all participating users (authors, commenters, reposters), including gender, region, personalized tag, and so on.

\textbf{(5) Social Graph:} The ``\textit{following}'' relationships between users, to model the underlying social network structure.

This comprehensive approach preserves the inherent heterogeneity and relational context of social media interactions from the ground up. Standard preprocessing, including URL removal, tag cleaning, and heuristic-based spam filtering, was applied to ensure data quality.

\subsection{Emotion-Behavior Annotation Paradigm}
The core innovation of PERCEIVE lies in its novel annotation paradigm, which directly tackles the challenge of modeling the coupling between internal emotions and overt behaviors. We move beyond annotating content in isolation; instead, we annotate \textbf{user-content interaction pairs}, establishing a fine-grained emotion-to-behavior mapping.
Our framework annotates each interaction on two concurrent fronts:

\textbf{(1) Personalized Emotion (Reader's Perspective):} We capture readers' emotional reactions in comments using seven categories: \textit{Happy, Anger, Sad, Fear, Surprise, Disgust, Neutral}. The six basic emotions follow Ekman's theory \cite{Ekman1992, Ekman1971}; \textit{Neutral} covers non-emotional responses, avoiding forced classification. This task applies only to \textit{R\&C} interactions, where the comment text serves as direct evidence of the reader's affective state. Crucially, our annotators are tasked with inferring the emotion conveyed in the \textit{comment}, not the original post. This operationalizes the ``reader's perspective'' by focusing on genuine participants rather than simulated readers.
	
\textbf{(2) Behavioral Intent:} We classify the user's communication intent into three categories: \textit{R\&C} (deep engagement), \textit{R} (information diffusion), and \textit{L} (passive acknowledgment). This provides a nuanced view of user engagement beyond simple binary actions.

By creating an \textbf{Emotion-Behavior Pair} for each \textit{R\&C} instance, PERCEIVE provides the empirical foundation needed to investigate the central tenets of Social Exchange Theory \cite{10.3389/fpsyg.2022.1015921}. It allows for a direct analysis of how specific emotional states translate into distinct behavioral choices, thereby bridging the gap between latent affect and manifest social actions.

\subsection{Human-LLM Collaborative Annotation}
To ensure both scalability and reliability, we employed a human-in-the-loop annotation pipeline leveraging the strengths of both LLMs and human experts. To prevent potential ``LLM hacking'' in subjective tasks \cite{baumann2025largelanguagemodelhacking}, we have designed our process to prioritize human judgment at every critical step. Additionally, we analyze the LLM-generated content in Task E (Section 5.5).

\textbf{(1) LLM-Based Pre-annotation.} To boost efficiency, we first generated initial emotion label suggestions for all \textit{R\&C} comments using GLM-4.5 \cite{glm2024chatglm}. Crucially, we engineered a stable, context-aware prompting strategy. The prompt included both the original post and the reader's comment, instructing the model to perform empathetic reasoning from the reader's standpoint. The model was tasked with providing a label along with a confidence score (0-1), encouraging fine-grained, interpretable outputs rather than black-box classifications. A detailed prompt is illustrated in Appendix~\ref{sec:labelprompt}.

\textbf{(2) Human Verification and Arbitration.} 
We conducted a 100\% human review of all LLM pre-annotations. Six trained postgraduate annotators independently revised each instance following a comprehensive Annotation Guideline, which included: (a) definitions and examples for each emotion category; (b) criteria emphasizing the perspective of a ``typical reader''; (c) instructions for handling edge cases (e.g., sarcasm, metaphor); and (d) guidance to prioritize their own judgment over LLM suggestions. All annotators underwent unified training prior to annotation. To quantify the reliability of this process, we measured inter-annotator agreement using Krippendorff's Alpha, which reached $\alpha = 0.814$ among the six annotators. Disagreements were resolved through discussion and arbitration by a senior researcher. This collaborative process synergizes the scalability of LLMs with the nuanced understanding of human cognition, ensuring high-quality, reliable annotations that capture the true complexity of emotional expression.

\subsection{Dataset Statistics and Analysis}
As shown in Table~\ref{tab:data_stat}, PERCEIVE comprises a substantial number of user-post interactions across the three behavior types (\textit{R\&C}, \textit{R}, and \textit{L}) and seven emotion categories (including \textit{Happy} (Ha), \textit{Anger} (An), \textit{Sad} (Sa), \textit{Disgust} (Di), \textit{Fear} (Fe), \textit{Surprise} (Su), and \textit{Neutral} (Ne)). To address the inherent class imbalance common in real-world social data, we employed LLM-based data augmentation (DA) on the training set, following the methodology of \citet{Liao2025}. Table~\ref{tab:user_stat} presents detailed statistics on user profiles and the social graph. Notably, the ``Follow'' graph $\mathcal{G}_f$, with its E$_f$ edges and degree per user (D/U), reflects the real-world sparsity of social networks, providing a realistic testbed for models incorporating relational information, as motivated by Emotion Contagion Theory.

\begin{table}[htb]
	\centering
	\footnotesize
	\tabcolsep 1.8pt 
	\caption{Distribution of behaviors and emtoions.}
	\label{tab:data_stat}
	\begin{tabular}{lcccccccc}
		\hline
		& \multicolumn{8}{c}{Communication Behavior}                         \\\cline{2-9}
		& \multicolumn{2}{c}{R\&C}  & \multicolumn{2}{c}{R}     & \multicolumn{2}{c}{L}    & \multicolumn{2}{c}{Total} \\
		\hline
		Twitter & \multicolumn{2}{c}{22580} & \multicolumn{2}{c}{22233} & \multicolumn{2}{c}{21975} & \multicolumn{2}{c}{66788} \\
		Weibo   & \multicolumn{2}{c}{15663} & \multicolumn{2}{c}{15358} & \multicolumn{2}{c}{15639} & \multicolumn{2}{c}{46660} \\
		\hline
		& \multicolumn{8}{c}{Emotion Distribution of Comments}               \\
		\cline{2-9}
		& Ha & An & Sa  & Fe & Su & Di & Ne & Total \\
		\hline
		Twitter      & 12338 & 1645  & 947  & 446  & 1862     & 614     & 4728    & 22580 \\
		Twitter$_{DA}$ & 12338 & 8838  & 9310 & 8817 & 8792     & 8864    & 8414    & 65373 \\
		Weibo        & 6769  & 1293  & 1170 & 462  & 912      & 386     & 4671    & 15663 \\
		Weibo$_{DA}$   & 6769  & 4976  & 4566 & 4687 & 4776     & 4370    & 4671    & 34815 \\
		\hline
	\end{tabular}%
\end{table}
\begin{table}[htb]
	\centering
	\footnotesize
	\tabcolsep 3pt 
	\caption{Statistics of user information}
	\label{tab:user_stat}
	\begin{tabular}{lccccccc}
		\hline
		& \multicolumn{2}{c}{Profiles} & \multicolumn{2}{c}{History} & \multirow{2}{*}{Total} & \multirow{2}{*}{E$_{f}$} & \multirow{2}{*}{D/U} \\\cline{2-5}
		& w/   & w/o & w/  & w/o  &       &    &     \\\hline
		Twitter & 18331 & 5142 & 8552 & 14921 & 23473 & 47127 & 2.01 \\
		Weibo   & 12139 & 3737 & 9946 & 5930  & 15876 & 540530 & 34.05\\
		\hline
	\end{tabular}%
\end{table}

\section{Tasks and Evaluation Protocol}
\subsection{Task Definitions for Social Perception}
To systematically evaluate a model's capacity for social perception, the PERCEIVE benchmark introduces a suite of five progressively defined tasks.

\textbf{Task A: Reader Emotion Prediction.} Establishes a baseline by classifying a reader's comment into one of seven discrete emotions. This task marks a paradigm shift from author-centric to reader-centric analysis (\textbf{Challenge 1}).

\textbf{Task B: Personalized Emotion Prediction.} Extends Task A by incorporating user profiles to account for inter-individual emotional variance. The performance gain over Task A directly measures the utility of personalization for nuanced emotion perception (\textbf{Challenge 1}).

\textbf{Task C: Reader Behavior Prediction.} Predicts a user's communication action (\textit{R\&C}, \textit{R}, \textit{L}) from a comment, leveraging profiles of both the author and the commenter. This task probes the crucial link between internal states and observable social actions (\textbf{Challenge 2}).

\textbf{Task D: Joint Emotion-Behavior Modeling.} Requires simultaneous inference of a target user's emotion and likely behavior. Evaluation prioritizes not only individual accuracy but also the \textit{consistency} of the predicted pair, grounding the model in unified theories of social cognition (\textbf{Challenge 3}).

\textbf{Task E: In-context Comment Generation.} The culminating generative task. Given a post and user profiles, the model must generate a persona-consistent comment, testing the ultimate goal of socially-intelligent, interactive generation (\textbf{Challenge 3}).

\subsection{Overview of Evaluation Protocol}
A comprehensive evaluation is conducted on the PERCEIVE benchmark, utilizing an 8:1:1 data set partitioning ratio. 

\textbf{Baselines} For a rigorous evaluation, we benchmark against a set of SOTA baselines chosen for their complementary reasoning strengths, encompassing advanced online LLMs like \textbf{GLM4.5-plus} (GLM) \cite{glm2024chatglm}, \textbf{Qwen3-plus} (Qwen) \cite{qwen3}, and \textbf{DeepSeek-V3.2} (DeepSeek) \cite{deepseekr1}, as well as established reasoning paradigms including \textbf{THOR} \cite{fei2023}, \textbf{TOC} \cite{weinzierl-harabagiu-2024-tree}, and \textbf{Debate} \cite{fang-etal-2025-counterfactual}. We evaluated all baselines on tasks A-C. For tasks D and E, to prevent error propagation and label leakage, we assessed the capabilities of three foundation LLMs. Detailed metrics, baseline model descriptions are provided in Appendix~\ref{appendix:evaluation}.

\textbf{Metrics} Our protocol is designed for fairness and multi-dimensional assessment. We evaluate Tasks A-C with Accuracy (Acc) and Macro-F1 (M-F1), use confusion matrices of emotion-emotion (EECM), behavior-behavior (BBCM), and emotion-behavior (EBCM) for Task D, and assess Task E with a hybrid of n-gram metrics (BLEU-4, ROUGE-L) and LLM-based judgments (User Intent Consistency (UIC), Propagation Function Consistency (PFC), and Emotional Expression Consistency (EEC)). The prompts for LLM-based evaluations are detailed in Appendix~\ref{appendix:prompts}.

\section{Benchmark Experiments Analysis}
Our comprehensive evaluation on the PERCEIVE benchmark yields several critical insights, answering the following core research questions:

\textbf{RQ1}: Does a reader-centric, emotion-behavior joint-modeling benchmark expose fundamental limitations in current models' ability for personalized social understanding? (supported by \textbf{Tasks A--D})
	
\textbf{RQ2}: How do SOTA modeling approaches, particularly those LLMs with complex reasoning, fare in capturing the intricate coupling between emotion and behavior? (supported by \textbf{Tasks C and D})
	
\textbf{RQ3}: What are the key takeaways and future directions for building socially intelligent NLP systems, as highlighted by our evaluation? (supported by \textbf{Tasks A--E})

\subsection{Results and Analysis of Task A}
The overall evaluation performance of Task A is shown in Table~\ref{tab:taska} (Detailed results for each emotion are presented in Tables~\ref{tab:drtaska_t} and \ref{tab:drtaska_w} in Appendix~\ref{appendix:drv}). Our evaluation yields three pivotal findings. First, we observe a ``striking platform dependency'', where models like Qwen significantly outperform others on Weibo but not on Twitter, validating our benchmark's cross-platform design for assessing model generalization. Second, contrary to common assumptions, chain-of-reasoning or debate-based reasoning does not consistently improve performance and often degrades it, challenging its universal applicability in socially nuanced tasks. Finally, a fine-grained analysis reveals that while models excel on unambiguous emotions like ``Happy'' (see Table~\ref{tab:drtaska_t} and \ref{tab:drtaska_w}), they struggle with high-context, nuanced emotions such as ``Disgust''. This exposes a fundamental limitation in capturing the subtle, context-dependent nature of emotional expression, highlighting the need for deeper, socially-aware modeling.
\begin{table}[hbt]
	\centering
	\tabcolsep 5pt
	\footnotesize
	\caption{Overall performance on Task A}
	\label{tab:taska}
		\begin{tabular}{lcccc}
			\hline
			\textbf{}        & \multicolumn{2}{c}{\textbf{Twitter}} & \multicolumn{2}{c}{\textbf{Weibo}} \\\cline{2-5}
			\textbf{models}   & \textbf{M-F1}     & \textbf{Acc}     & \textbf{M-F1}    & \textbf{Acc}    \\\hline
			GLM & 0.5346            & 0.7436           & 0.4684           & 0.6437          \\
			Qwen             & \textbf{0.5429}            & 0.7387           & \textbf{0.5205}           & \textbf{0.7056}          \\
			DeepSeek         & 0.5334            & \textbf{0.7520}            & 0.4916           & 0.6756          \\\hline
			GLM\_TOC         & 0.5096            & 0.7201           & 0.4430            & 0.6341          \\
			Qwen\_TOC        & 0.5186            & 0.7161           & 0.4856           & 0.6584          \\
			DeepSeek\_TOC    & 0.5238            & 0.7334           & 0.4807           & 0.6724          \\\hline
			GLM\_THOR        & 0.4671            & 0.6953           & 0.4164           & 0.6086          \\
			Qwen\_THOR       & 0.5092            & 0.7068           & 0.4872           & 0.6328          \\
			DeepSeek\_THOR   & 0.4940             & 0.7112           & 0.4425           & 0.6220           \\\hline
			GLM\_Debate      & 0.4833            & 0.7121           & 0.4557           & 0.6424          \\
			Qwen\_Debate     & 0.5369            & 0.7383           & 0.5046           & 0.6711          \\
			DeepSeek\_Debate & 0.5343            & 0.7378           & 0.4745           & 0.6379         \\\hline
		\end{tabular}%
\end{table}

\subsection{Results and Analysis of Task B}
\label{sec:rataskb}
The results for Task B, presented in Table~\ref{tab:taskb} (Detailed results are in Tables~\ref{tab:drtaskb_t} and \ref{tab:drtaskb_w} in Appendix~\ref{appendix:drv}), reveal two critical insights into personalized emotion analysis. First, the Debate-based reasoning models outperform their baselines on Twitter. We posit that this success stems from an emergent process of dynamic user profiling. By arguing different emotional interpretations, the models construct a more nuanced, context-aware mental model of the reader, leading to superior personalization.
Second, compared to Task A, we observe that naively concatenating user profile information often degrades performance. This finding exposes a fundamental weakness in current architectures, which fail to move beyond shallow feature fusion. Our analysis underscores that the path to effective personalization requires not just more data, but more sophisticated, theory-driven reasoning mechanisms that can deeply integrate user context.
\begin{table}[htb]
	\centering
	\tabcolsep 5pt
	\footnotesize
	\caption{Overall performance on Task B}
	\label{tab:taskb}
	\begin{tabular}{lcccc}
		\hline
		\textbf{}        & \multicolumn{2}{c}{\textbf{Twitter}} & \multicolumn{2}{c}{\textbf{Weibo}} \\\cline{2-5}
		\textbf{models}   & \textbf{M-F1}     & \textbf{Acc}     & \textbf{M-F1}    & \textbf{Acc}    \\\hline
		GLM              & 0.4907  & 0.7152 & 0.4471 & 0.6073 \\
		Qwen             & 0.5456  & 0.7436 & \textbf{0.4952} & \textbf{0.6628} \\
		DeepSeek         & 0.5226  & 0.7484 & 0.4525 & 0.6424 \\\hline
		GLM\_TOC         & 0.5126  & 0.7219 & 0.4318 & 0.6258 \\
		Qwen\_TOC        & 0.5085  & 0.7046 & 0.4841 & 0.6533 \\
		DeepSeek\_TOC    & 0.4909  & 0.7183 & 0.4743 & 0.6571 \\\hline
		GLM\_THOR        & 0.4669  & 0.6975 & 0.4566 & 0.6175 \\
		Qwen\_THOR       & 0.5591  & 0.7471 & 0.4866 & 0.6501 \\
		DeepSeek\_THOR   & 0.4896  & 0.7077 & 0.4045 & 0.6098 \\\hline
		GLM\_Debate      & 0.5326  & 0.7360  & 0.4823 & 0.6354 \\
		Qwen\_Debate     & \textbf{0.6221}  & \textbf{0.7484} & 0.4734 & 0.6411 \\
		DeepSeek\_Debate & 0.5872  & 0.7245 & 0.4282 & 0.6111    \\\hline
	\end{tabular}%
\end{table}

Moreover, our preliminary analysis of Task B reveals a subtle and counter-intuitive phenomenon regarding user activity and personalized emotion analysis. To investigate this, we stratified users into high-activity (top 20\%) and low-activity (bottom 20\%) groups based on their total historical interactions (posts and comments). As shown in Fig.~\ref{fig:taskB_sentiment_hl}, the model consistently achieves higher M-F1 scores on the low-activity user group across both datasets.

This finding challenges the common assumption that more user data invariably yields better personalization. We posit an explanation grounded in social psychology: high-activity users engage in more complex self-presentation \cite{10.1145/3686909}. They curate diverse online personas depending on social context and audience, leading to a historical record that is a collage of multifaceted emotional expressions. For a model, this results in a user profile rich in contextual ``noise,'' making it difficult to distill a single, stable emotional propensity from such a large yet varied corpus.
Conversely, the digital footprint of low-activity users is often simpler and more authentic. With fewer interactions, each data point carries a stronger signal, potentially reflecting their core emotional dispositions more faithfully.

\subsection{Results and Analysis of Task C}
As presented in Table~\ref{tab:taskc}, our evaluation of Task C reveals significant challenges for current SOTA models in predicting overt social actions (Detailed results for each behavior are presented in Tables~\ref{tab:drtaskc}).
Echoing findings from prior tasks, we confirm a significant platform dependency in behavior prediction. More importantly, our analysis of advanced reasoning methods uncovers a nuanced mechanism. While not universally beneficial, methods like TOC can dramatically improve the detection of specific high-engagement behaviors (e.g., nearly doubling the F1-score for R\&C on Twitter). However, this benefit comes at the expense of performance on other classes, revealing clear performance trade-offs and indicating that these frameworks are coarse proxies for genuine behavioral understanding.

Our user-activity analysis (Fig.~\ref{fig:taskC_behavior_hl} in Appendix~\ref{appendix:drv}) reveals a sharp contrast with Section~\ref{sec:rataskb}: prediction accuracy for highly active users significantly exceeds that for less active users. This discrepancy underscores a crucial hypothesis: behavioral patterns, as reflections of underlying personality, are more consistent than volatile emotional expressions. Consequently, the extensive history of highly active users is more conducive to modeling robust behavioral profiles. This accuracy gap exposes a fundamental limitation of current paradigms in data-sparse scenarios, where insufficient user history prevents the formation of reliable profiles. Collectively, Task C results demonstrate that even advanced models fail to effectively integrate the nuanced social signals within PERCEIVE, highlighting the urgent need for architectures that unify user profiles, content, and relational semantics.

\begin{table}[htb]
	\centering
	\footnotesize
	\tabcolsep 5pt
	\caption{Overall performance on Task C}
	\label{tab:taskc}
	\begin{tabular}{lcccc}
		\hline
		\textbf{}        & \multicolumn{2}{c}{\textbf{Twitter}} & \multicolumn{2}{c}{\textbf{Weibo}} \\\cline{2-5}
		\textbf{models}   & \textbf{M-F1}     & \textbf{Acc}     & \textbf{M-F1}    & \textbf{Acc}    \\\hline
		GLM & 0.3003  & 0.4047 & \textbf{0.3701} & \textbf{0.4896} \\
		Qwen             & \textbf{0.3738}  & \textbf{0.5320}  & 0.2678 & 0.3167 \\
		DeepSeek         & 0.3500    & 0.4745 & 0.2654 & 0.3384 \\\hline
		GLM\_TOC         & 0.3112  & 0.4304 & 0.3660  & 0.4879 \\
		Qwen\_TOC        & 0.3622  & 0.4874 & 0.2955 & 0.4227 \\
		DeepSeek\_TOC    & 0.3693  & 0.4925 & 0.3416 & 0.4666 \\\hline
		GLM\_THOR        & 0.2589  & 0.3764 & 0.3529 & 0.4542 \\
		Qwen\_THOR       & 0.3234  & 0.4576 & 0.3536 & 0.4587 \\
		DeepSeek\_THOR   & 0.3675  & 0.4819 & 0.3202 & 0.4417 \\\hline
		GLM\_Debate      & 0.2440   & 0.3511 & 0.1947 & 0.3135 \\
		Qwen\_Debate     & 0.3404  & 0.4497 & 0.3384 & 0.4689 \\
		DeepSeek\_Debate & 0.3513  & 0.4614 & 0.3365 & 0.4464  \\\hline
	\end{tabular}%
\end{table}

\subsection{Results and Analysis of Task D}
The results for Task D (shown as Figs.~\ref{fig:taskd-glm-t}-\ref{fig:taskd-ds-w} in Appendix~\ref{appendix:drv}), which requires the simultaneous and consistent prediction of emotion and behavior, reveal a fundamental limitation in current multi-task learning paradigms. While state-of-the-art models achieve moderate performance on individual sub-tasks, they exhibit a striking lack of inter-task consistency. We observe a systemic decoupling: emotion predictions often rely on superficial lexical cues (e.g., misclassifying high-arousal ``Anger'' as ``Happy'' due to charged language), while behavior predictions regress to a simplistic intensity ranking (e.g., collapsing ``Reply \& Comment'' into ``Reply'').
This failure underscores a core gap: contemporary models lack a unified, theory-driven reasoning framework to model the emotion-to-behavior causal chain. As posited by \textit{Social Exchange Theory} \cite{10.3389/fpsyg.2022.1015921}, an emotion is not a direct trigger but an internal variable that informs a cost-benefit analysis leading to a social action. Decoupled prediction heads cannot capture this mediating cognitive process.

A more profound challenge, revealed by our benchmark, pertains to modeling the ``silent majority''— the lurkers. These users, who consume content without overtly expressing emotions (e.g., liking or commenting), are a massive, unobserved variable in the emotion propagation system. Drawing from the \textit{Spiral of Silence Effect} \cite{Sohn2022Spiral}, their silence is not neutrality but a passive state influenced by perceived social pressure and dominant opinion. Neglecting this cohort leads to a significantly skewed understanding of group sentiment. Their latent, unexpressed affect is a critical component for accurately modeling the \textit{evolution of group-level emotion}, as its accumulation can trigger sudden, large-scale shifts in public opinion or social activism. Future work must therefore advance beyond modeling explicit signals to develop mechanisms for \textit{latent affect inference} and predict the activation threshold of this passive population, which is crucial for a holistic theory of emotion dynamics on social media.

\subsection{Results and Analysis of Task E}

The low scores for Task E on traditional n-gram metrics like BLEU-4 and ROUGE-L (Appendix~\ref{appendix:drv}, Fig.~\ref{fig:taske-bleurouge}) are not only unsurprising but validating. This gap confirms that surface-level similarity is a poor proxy for socially-aware generation. The core objective is not to replicate a single ground-truth comment, but to generate contextually appropriate, persona-consistent responses—where semantic and pragmatic fidelity are paramount. Thus, our proposed LLM-based evaluation metrics are essential for assessing true performance.
As shown in Fig.~\ref{fig:taske-upe}, while most baselines demonstrate a foundational capability (scoring between 3.5 and 4.5), a performance ceiling persists. This gap reveals that current LLMs struggle to internalize complex user profiles and social contexts. According to \textit{Uses and Gratifications Theory} \cite{XU2025103511}, users actively seek media to fulfill specific needs. Our results indicate that while models can simulate gratification (i.e., generate emotional responses in EEC), they are less adept at modeling strategic, goal-oriented use (i.e., achieving specific intent in UIC and social function in PFC). The relatively lower PFC scores compared to EEC suggest that generating text with a coherent social purpose (e.g., building solidarity, critiquing, seeking information) is more challenging than simply expressing a matching emotion. Furthermore, the varying performance across Twitter and Weibo datasets underscores that models have not yet mastered the platform-specific cultural scripts and interaction norms that govern online communication.

\section{Conclusion}
\label{sec:con}
This paper introduces PERCEIVE, a novel benchmark that shifts social media emotion analysis from an author-centric to a reader-centric perspective. By integrating five key dimensions—author content, reader emotion, behavior, user profiles, and the social graph—PERCEIVE provides a holistic platform for studying emotion as a social phenomenon. Our evaluation reveals a significant gap: even SOTA LLMs struggle to capture personalized emotional responses and the critical coupling between emotion and communication behavior. These findings highlight the urgent need for theory-driven models that fuse content, user, and graph-structural information to model emotion propagation. The PERCEIVE benchmark will serve as a vital catalyst for this new line of inquiry, paving the way for models that can not only understand what people feel, but also why they act, and how these emotions propagate through social media.

\section*{Limitations}
While our work introduces the novel PERCEIVE benchmark, several limitations open exciting avenues for future research.
First, our evaluation relies on LLMs, which, while scalable, may introduce inherent implicit biases and do not fully account for ethical considerations. Furthermore, our analysis lacks a deep dive into the distinct cultural and platform-specific dynamics of emotion communication on Weibo and Twitter. Future work should focus on developing bias-aware evaluation protocols and context-aware models that can reason about these cross-platform nuances.
Second, our dataset, like most based on observable interactions, is subject to the ``silent spiral effect''. It captures the emotions of active participants but not the vast population of ``lurkers,'' thus failing to model the critical transition from passive consumption to active engagement. A promising direction is to integrate passive signals (e.g., dwell time) or hybrid methodologies to infer the emotional states of the silent majority, enabling a more holistic view of emotion dynamics.

\section*{Ethical Considerations}
This research adheres to strict ethical guidelines concerning data, human annotators, and potential misuse. All data used in the PERCEIVE benchmark underwent rigorous anonymization. All personally identifiable information and traceable user identifiers were removed before public release to ensure user privacy. All annotators involved in this project were compensated fairly.
While this work aims to have a positive impact by advancing our understanding of social dynamics, we strongly condemn any malicious applications. We explicitly prohibit the use of our dataset and derivative models for emotion manipulation, behavioral intervention, non-consensual user tracking, or political manipulation. This work is intended strictly for academic research to prevent its misuse.

\bibliography{ref}

@inproceedings{ye-etal-2025-generic,
	title = "From Generic Empathy to Personalized Emotional Support: A Self-Evolution Framework for User Preference Alignment",
	author = "Ye, Jing  and
	Xiang, Lu  and
	Zhang, Yaping  and
	Zong, Chengqing",
	editor = "Christodoulopoulos, Christos  and
	Chakraborty, Tanmoy  and
	Rose, Carolyn  and
	Peng, Violet",
	booktitle = "Findings of the Association for Computational Linguistics: EMNLP 2025",
	month = nov,
	year = "2025",
	address = "Suzhou, China",
	publisher = "Association for Computational Linguistics",
	url = "https://aclanthology.org/2025.findings-emnlp.1024/",
	doi = "10.18653/v1/2025.findings-emnlp.1024",
	pages = "18826--18853",
	ISBN = "979-8-89176-335-7",
}

@inproceedings{jun-lee-2025-exploring,
	title = "Exploring Persona Sentiment Sensitivity in Personalized Dialogue Generation",
	author = "Jun, Yonghyun  and
	Lee, Hwanhee",
	editor = "Che, Wanxiang  and
	Nabende, Joyce  and
	Shutova, Ekaterina  and
	Pilehvar, Mohammad Taher",
	booktitle = "Proceedings of the 63rd Annual Meeting of the Association for Computational Linguistics (Volume 1: Long Papers)",
	month = jul,
	year = "2025",
	address = "Vienna, Austria",
	publisher = "Association for Computational Linguistics",
	url = "https://aclanthology.org/2025.acl-long.900/",
	doi = "10.18653/v1/2025.acl-long.900",
	pages = "18384--18402",
	ISBN = "979-8-89176-251-0",
}

@article{deepseekr1,
	title={DeepSeek-R1: Incentivizing Reasoning Capability in LLMs via Reinforcement Learning}, 
	author={DeepSeek-AI and Daya Guo and Dejian Yang and et al.},
	year={2025},
	journal={Nature},
	volume={645},
	year={2025},
	pages={633--638},
}

@inproceedings{lyu2023,
	title = "Exploiting Rich Textual User-Product Context for Improving Personalized Sentiment Analysis",
	author = "Lyu, Chenyang  and
	Yang, Linyi  and
	Zhang, Yue  and
	Graham, Yvette  and
	Foster, Jennifer",
	editor = "Rogers, Anna  and
	Boyd-Graber, Jordan  and
	Okazaki, Naoaki",
	booktitle = "Findings of the Association for Computational Linguistics: ACL 2023",
	month = jul,
	year = "2023",
	address = "Toronto, Canada",
	publisher = "Association for Computational Linguistics",
	doi = "10.18653/v1/2023.findings-acl.92",
	pages = "1419--1429",
}

@inproceedings{Liao2025,
	title = "My Words Imply Your Opinion: Reader Agent-based Propagation Enhancement for Personalized Implicit Emotion Analysis",
	author = "Jian Liao  and
	Yu Feng  and
	Yujin Zheng  and
	Jun Zhao  and
	Suge Wang and 
	Jianxing Zheng",
	booktitle = "Proceedings of the 63rd Annual Meeting of the Association for Computational Linguistics (Volume 1: Long Papers)",
	month = jul,
	year = "2025",
	address = "Vienna, Austria",
	publisher = "Association for Computational Linguistics",
}

@inproceedings{wu-etal-2025-absa,
	title = "{M}-{ABSA}: A Multilingual Dataset for Aspect-Based Sentiment Analysis",
	author = "Wu, ChengYan  and
	Ma, Bolei  and
	Liu, Yihong  and
	Zhang, Zheyu  and
	Deng, Ningyuan  and
	Li, Yanshu  and
	Chen, Baolan  and
	Zhang, Yi  and
	Xue, Yun  and
	Plank, Barbara",
	editor = "Christodoulopoulos, Christos  and
	Chakraborty, Tanmoy  and
	Rose, Carolyn  and
	Peng, Violet",
	booktitle = "Proceedings of the 2025 Conference on Empirical Methods in Natural Language Processing",
	month = nov,
	year = "2025",
	address = "Suzhou, China",
	publisher = "Association for Computational Linguistics",
	url = "https://aclanthology.org/2025.emnlp-main.128/",
	doi = "10.18653/v1/2025.emnlp-main.128",
	pages = "2530--2557",
	ISBN = "979-8-89176-332-6",
}

@inproceedings{chebolu-etal-2024-oats,
	title = "{OATS}: A Challenge Dataset for Opinion Aspect Target Sentiment Joint Detection for Aspect-Based Sentiment Analysis",
	author = "Chebolu, Siva Uday Sampreeth  and
	Dernoncourt, Franck  and
	Lipka, Nedim  and
	Solorio, Thamar",
	editor = "Calzolari, Nicoletta  and
	Kan, Min-Yen  and
	Hoste, Veronique  and
	Lenci, Alessandro  and
	Sakti, Sakriani  and
	Xue, Nianwen",
	booktitle = "Proceedings of the 2024 Joint International Conference on Computational Linguistics, Language Resources and Evaluation (LREC-COLING 2024)",
	month = may,
	year = "2024",
	address = "Torino, Italia",
	publisher = "ELRA and ICCL",
	url = "https://aclanthology.org/2024.lrec-main.1080/",
	pages = "12336--12347",
}

@inproceedings{pontiki-etal-2014-semeval,
	title = "{S}em{E}val-2014 Task 4: Aspect Based Sentiment Analysis",
	author = "Pontiki, Maria  and
	Galanis, Dimitris  and
	Pavlopoulos, John  and
	Papageorgiou, Harris  and
	Androutsopoulos, Ion  and
	Manandhar, Suresh",
	editor = "Nakov, Preslav  and
	Zesch, Torsten",
	booktitle = "Proceedings of the 8th International Workshop on Semantic Evaluation ({S}em{E}val 2014)",
	month = aug,
	year = "2014",
	address = "Dublin, Ireland",
	publisher = "Association for Computational Linguistics",
	url = "https://aclanthology.org/S14-2004/",
	doi = "10.3115/v1/S14-2004",
	pages = "27--35"
}

@inproceedings{bougie-watanabe-2025-simuser,
	title = "{S}im{USER}: Simulating User Behavior with Large Language Models for Recommender System Evaluation",
	author = "Bougie, Nicolas  and
	Watanabe, Narimawa",
	editor = "Rehm, Georg  and
	Li, Yunyao",
	booktitle = "Proceedings of the 63rd Annual Meeting of the Association for Computational Linguistics (Volume 6: Industry Track)",
	month = jul,
	year = "2025",
	address = "Vienna, Austria",
	publisher = "Association for Computational Linguistics",
	url = "https://aclanthology.org/2025.acl-industry.5/",
	doi = "10.18653/v1/2025.acl-industry.5",
	pages = "43--60",
	ISBN = "979-8-89176-288-6",
}

@inproceedings{wang-etal-2025-evaluating-cognitive,
	title = "Evaluating Cognitive-Behavioral Fixation via Multimodal User Viewing Patterns on Social Media",
	author = "Wang, Yujie  and
	Zhao, Yunwei  and
	Yang, Jing  and
	Han, Han  and
	Shan, Shiguang  and
	Zhang, Jie",
	editor = "Christodoulopoulos, Christos  and
	Chakraborty, Tanmoy  and
	Rose, Carolyn  and
	Peng, Violet",
	booktitle = "Proceedings of the 2025 Conference on Empirical Methods in Natural Language Processing",
	month = nov,
	year = "2025",
	address = "Suzhou, China",
	publisher = "Association for Computational Linguistics",
	url = "https://aclanthology.org/2025.emnlp-main.987/",
	doi = "10.18653/v1/2025.emnlp-main.987",
	pages = "19571--19583",
	ISBN = "979-8-89176-332-6",
}

@inproceedings{wang-etal-2025-implicit,
	title = "Implicit Behavioral Alignment of Language Agents in High-Stakes Crowd Simulations",
	author = "Wang, Yunzhe  and
	Lucas, Gale  and
	Becerik-Gerber, Burcin  and
	Ustun, Volkan",
	editor = "Christodoulopoulos, Christos  and
	Chakraborty, Tanmoy  and
	Rose, Carolyn  and
	Peng, Violet",
	booktitle = "Proceedings of the 2025 Conference on Empirical Methods in Natural Language Processing",
	month = nov,
	year = "2025",
	address = "Suzhou, China",
	publisher = "Association for Computational Linguistics",
	url = "https://aclanthology.org/2025.emnlp-main.1562/",
	doi = "10.18653/v1/2025.emnlp-main.1562",
	pages = "30669--30686",
	ISBN = "979-8-89176-332-6",
}

@inproceedings{wang-etal-2025-besimulator,
	title = "{B}e{S}imulator: A Large Language Model Powered Text-based Behavior Simulator",
	author = "Wang, Jianan  and
	Li, Bin  and
	Qi, Jingtao  and
	Wang, Xueying  and
	Li, Fu  and
	Lihanxun",
	editor = "Christodoulopoulos, Christos  and
	Chakraborty, Tanmoy  and
	Rose, Carolyn  and
	Peng, Violet",
	booktitle = "Proceedings of the 2025 Conference on Empirical Methods in Natural Language Processing",
	month = nov,
	year = "2025",
	address = "Suzhou, China",
	publisher = "Association for Computational Linguistics",
	url = "https://aclanthology.org/2025.emnlp-main.237/",
	doi = "10.18653/v1/2025.emnlp-main.237",
	pages = "4736--4754",
	ISBN = "979-8-89176-332-6",
}

@inproceedings{mou-etal-2024-unveiling,
	title = "Unveiling the Truth and Facilitating Change: Towards Agent-based Large-scale Social Movement Simulation",
	author = "Mou, Xinyi  and
	Wei, Zhongyu  and
	Huang, Xuanjing",
	editor = "Ku, Lun-Wei  and
	Martins, Andre  and
	Srikumar, Vivek",
	booktitle = "Findings of the Association for Computational Linguistics: ACL 2024",
	month = aug,
	year = "2024",
	address = "Bangkok, Thailand",
	publisher = "Association for Computational Linguistics",
	url = "https://aclanthology.org/2024.findings-acl.285/",
	doi = "10.18653/v1/2024.findings-acl.285",
	pages = "4789--4809",
}

@inproceedings{hellwig-etal-2025-still,
	title = "Do we still need Human Annotators? Prompting Large Language Models for Aspect Sentiment Quad Prediction",
	author = "Hellwig, Nils Constantin  and
	Fehle, Jakob  and
	Kruschwitz, Udo  and
	Wolff, Christian",
	editor = "Fei, Hao  and
	Tu, Kewei  and
	Zhang, Yuhui  and
	Hu, Xiang  and
	Han, Wenjuan  and
	Jia, Zixia  and
	Zheng, Zilong  and
	Cao, Yixin  and
	Zhang, Meishan  and
	Lu, Wei  and
	Siddharth, N.  and
	{\O}vrelid, Lilja  and
	Xue, Nianwen  and
	Zhang, Yue",
	booktitle = "Proceedings of the 1st Joint Workshop on Large Language Models and Structure Modeling (XLLM 2025)",
	month = aug,
	year = "2025",
	address = "Vienna, Austria",
	publisher = "Association for Computational Linguistics",
	url = "https://aclanthology.org/2025.xllm-1.15/",
	doi = "10.18653/v1/2025.xllm-1.15",
	pages = "153--172",
	ISBN = "979-8-89176-286-2",
}

@inproceedings{zhou-etal-2023-learning,
	title = "Learning to Predict Persona Information for Dialogue Personalization without Explicit Persona Description",
	author = "Zhou, Wangchunshu  and
	Li, Qifei  and
	Li, Chenle",
	editor = "Rogers, Anna  and
	Boyd-Graber, Jordan  and
	Okazaki, Naoaki",
	booktitle = "Findings of the Association for Computational Linguistics: ACL 2023",
	month = jul,
	year = "2023",
	address = "Toronto, Canada",
	publisher = "Association for Computational Linguistics",
	url = "https://aclanthology.org/2023.findings-acl.186/",
	doi = "10.18653/v1/2023.findings-acl.186",
	pages = "2979--2991",
}

@inproceedings{salemi-etal-2024-lamp,
	title = "{L}a{MP}: When Large Language Models Meet Personalization",
	author = "Salemi, Alireza  and
	Mysore, Sheshera  and
	Bendersky, Michael  and
	Zamani, Hamed",
	editor = "Ku, Lun-Wei  and
	Martins, Andre  and
	Srikumar, Vivek",
	booktitle = "Proceedings of the 62nd Annual Meeting of the Association for Computational Linguistics (Volume 1: Long Papers)",
	month = aug,
	year = "2024",
	address = "Bangkok, Thailand",
	publisher = "Association for Computational Linguistics",
	url = "https://aclanthology.org/2024.acl-long.399/",
	doi = "10.18653/v1/2024.acl-long.399",
	pages = "7370--7392",
}

@inproceedings{liu-etal-2025-llms,
	title = "{LLM}s + Persona-Plug = Personalized {LLM}s",
	author = "Liu, Jiongnan  and
	Zhu, Yutao  and
	Wang, Shuting  and
	Wei, Xiaochi  and
	Min, Erxue  and
	Lu, Yu  and
	Wang, Shuaiqiang  and
	Yin, Dawei  and
	Dou, Zhicheng",
	editor = "Che, Wanxiang  and
	Nabende, Joyce  and
	Shutova, Ekaterina  and
	Pilehvar, Mohammad Taher",
	booktitle = "Proceedings of the 63rd Annual Meeting of the Association for Computational Linguistics (Volume 1: Long Papers)",
	month = jul,
	year = "2025",
	address = "Vienna, Austria",
	publisher = "Association for Computational Linguistics",
	url = "https://aclanthology.org/2025.acl-long.461/",
	doi = "10.18653/v1/2025.acl-long.461",
	pages = "9373--9385",
	ISBN = "979-8-89176-251-0",
}

@article{CHEN2024108304,
	title = {Fostering YouTube followers’ stickiness through social contagion: The role of digital influencer' characteristics and followers’ compensation psychology},
	journal = {Computers in Human Behavior},
	volume = {158},
	pages = {108304},
	year = {2024},
	issn = {0747-5632},
	doi = {https://doi.org/10.1016/j.chb.2024.108304},
	url = {https://www.sciencedirect.com/science/article/pii/S0747563224001729},
	author = {Chien-Wen Chen and Duong Thuy Trang Nguyen and Mingchang Chih and Pei-Ying Chen},
}

@inproceedings{wei-etal-2025-mecot,
	title = "{MEC}o{T}: {M}arkov Emotional Chain-of-Thought for Personality-Consistent Role-Playing",
	author = "Wei, Yangbo  and
	Huang, Zhen  and
	Zhao, Fangzhou  and
	Feng, Qi  and
	Xing, Wei W.",
	editor = "Che, Wanxiang  and
	Nabende, Joyce  and
	Shutova, Ekaterina  and
	Pilehvar, Mohammad Taher",
	booktitle = "Findings of the Association for Computational Linguistics: ACL 2025",
	month = jul,
	year = "2025",
	address = "Vienna, Austria",
	publisher = "Association for Computational Linguistics",
	url = "https://aclanthology.org/2025.findings-acl.435/",
	doi = "10.18653/v1/2025.findings-acl.435",
	pages = "8297--8314",
	ISBN = "979-8-89176-256-5",
}

@ARTICLE{10.3389/fpsyg.2022.1015921,
	AUTHOR={Ahmad, Rehan  and Nawaz, Muhammad Rafay  and Ishaq, Muhammad Ishtiaq  and Khan, Mumtaz Muhammad  and Ashraf, Hafiz Ahmad },
	TITLE={Social exchange theory: Systematic review and future directions},
	JOURNAL={Frontiers in Psychology},
	VOLUME={Volume 13 - 2022},
	YEAR={2023},	
	DOI={10.3389/fpsyg.2022.1015921},	
	ISSN={1664-1078},
}

@inproceedings{hu-collier-2025-inews,
	title = "i{N}ews: A Multimodal Dataset for Modeling Personalized Affective Responses to News",
	author = "Hu, Tiancheng  and
	Collier, Nigel",
	editor = "Che, Wanxiang  and
	Nabende, Joyce  and
	Shutova, Ekaterina  and
	Pilehvar, Mohammad Taher",
	booktitle = "Proceedings of the 63rd Annual Meeting of the Association for Computational Linguistics (Volume 1: Long Papers)",
	month = jul,
	year = "2025",
	address = "Vienna, Austria",
	publisher = "Association for Computational Linguistics",
	url = "https://aclanthology.org/2025.acl-long.1217/",
	doi = "10.18653/v1/2025.acl-long.1217",
	pages = "25000--25040",
	ISBN = "979-8-89176-251-0",
}

@article{Sohn2022Spiral,
	title={Spiral of Silence in the Social Media Era: A Simulation Approach to the Interplay Between Social Networks and Mass Media},
	author={Dongyoung Sohn},
	journal={Communication Research},
	volume={49},
	number={1},
	year={2022},
	pages={139-166},
}

@article{glm2024chatglm,
	title={ChatGLM: A Family of Large Language Models from GLM-130B to GLM-4 All Tools}, 
	author={Team GLM and Aohan Zeng and Bin Xu and et al.},
	year={2024},
	journal={arXiv preprint arXiv:2406.12793},
	archivePrefix={arXiv},
}

@inproceedings{zhao2024rdgcn,
	title={Rdgcn: Reinforced dependency graph convolutional network for aspect-based sentiment analysis},
	author={Zhao, Xusheng and Peng, Hao and Dai, Qiong and Bai, Xu and Peng, Huailiang and Liu, Yanbing and Guo, Qinglang and Yu, Philip S},
	booktitle={Proceedings of the 17th ACM International Conference on Web Search and Data Mining},
	pages={976--984},
	year={2024}
}

@inproceedings{fei2023,
	title = "Reasoning Implicit Sentiment with Chain-of-Thought Prompting",
	author = "Fei, Hao  and
	Li, Bobo  and
	Liu, Qian  and
	Bing, Lidong  and
	Li, Fei  and
	Chua, Tat-Seng",
	editor = "Rogers, Anna  and
	Boyd-Graber, Jordan  and
	Okazaki, Naoaki",
	booktitle = "Proceedings of the 61st Annual Meeting of the Association for Computational Linguistics (Volume 2: Short Papers)",
	month = jul,
	year = "2023",
	address = "Toronto, Canada",
	publisher = "Association for Computational Linguistics",
	doi = "10.18653/v1/2023.acl-short.101",
	pages = "1171--1182",
}

@inproceedings{li2025exploring,
	title={Exploring Model Editing for LLM-based Aspect-Based Sentiment Classification},
	author={Li, Shichen and Wang, Zhongqing and Zhao, Zheyu and Zhang, Yue and Li, Peifeng},
	booktitle={Proceedings of the AAAI Conference on Artificial Intelligence},
	volume={39},
	number={23},
	pages={24467--24475},
	year={2025}
}

@inproceedings{jian2025simrp,
	title={SimRP: Syntactic and Semantic Similarity Retrieval Prompting Enhances Aspect Sentiment Quad Prediction},
	author={Jian, Zhongquan and Chen, Yanhao and Li, Jiajian and Wang, Shaopan and Zeng, Xiangjian and Yao, Junfeng and An, Xinying and Wu, Qingqiang},
	booktitle={Proceedings of the AAAI Conference on Artificial Intelligence},
	volume={39},
	number={23},
	pages={24248--24256},
	year={2025}
}

@inproceedings{demszky-etal-2020-goemotions,
	title = "{G}o{E}motions: A Dataset of Fine-Grained Emotions",
	author = "Demszky, Dorottya  and
	Movshovitz-Attias, Dana  and
	Ko, Jeongwoo  and
	Cowen, Alan  and
	Nemade, Gaurav  and
	Ravi, Sujith",
	editor = "Jurafsky, Dan  and
	Chai, Joyce  and
	Schluter, Natalie  and
	Tetreault, Joel",
	booktitle = "Proceedings of the 58th Annual Meeting of the Association for Computational Linguistics",
	month = jul,
	year = "2020",
	address = "Online",
	publisher = "Association for Computational Linguistics",
	url = "https://aclanthology.org/2020.acl-main.372/",
	doi = "10.18653/v1/2020.acl-main.372",
	pages = "4040--4054",
}

@article{XU2025103511,
	title = {Enhancing user information disclosure intention in dynamic conversations of intelligent recommendation systems based on large language models: A perspective of user gratification and privacy calculus},
	journal = {International Journal of Human-Computer Studies},
	volume = {200},
	pages = {103511},
	year = {2025},
	issn = {1071-5819},
	doi = {https://doi.org/10.1016/j.ijhcs.2025.103511},
	url = {https://www.sciencedirect.com/science/article/pii/S1071581925000680},
	author = {Chunze Xu and Fengqiang Gao and Lei Han},
}

@inproceedings{10.1145/3686909,
	author = {Kim, JaeWon and Wolfe, Robert and Chordia, Ishita and Davis, Katie and Hiniker, Alexis},
	title = {"Sharing, Not Showing Off": How BeReal Approaches Authentic Self-Presentation on Social Media Through Its Design},
	year = {2024},
	issue_date = {November 2024},
	publisher = {Association for Computing Machinery},
	address = {New York, NY, USA},
	volume = {8},
	number = {CSCW2},
	url = {https://doi.org/10.1145/3686909},
	doi = {10.1145/3686909},
	booktitle = {Proceedings of the ACM on Human-Computer Interaction},
	month = nov,
	pages = {1--32},
}

@inproceedings{rosenthal-etal-2017-semeval,
	title = "{S}em{E}val-2017 Task 4: Sentiment Analysis in {T}witter",
	author = "Rosenthal, Sara  and
	Farra, Noura  and
	Nakov, Preslav",
	editor = "Bethard, Steven  and
	Carpuat, Marine  and
	Apidianaki, Marianna  and
	Mohammad, Saif M.  and
	Cer, Daniel  and
	Jurgens, David",
	booktitle = "Proceedings of the 11th International Workshop on Semantic Evaluation ({S}em{E}val-2017)",
	month = aug,
	year = "2017",
	address = "Vancouver, Canada",
	publisher = "Association for Computational Linguistics",
	url = "https://aclanthology.org/S17-2088/",
	doi = "10.18653/v1/S17-2088",
	pages = "502--518",
}

@inproceedings{weinzierl-harabagiu-2024-tree,
	title = "Tree-of-Counterfactual Prompting for Zero-Shot Stance Detection",
	author = "Weinzierl, Maxwell  and
	Harabagiu, Sanda",
	editor = "Ku, Lun-Wei  and
	Martins, Andre  and
	Srikumar, Vivek",
	booktitle = "Proceedings of the 62nd Annual Meeting of the Association for Computational Linguistics (Volume 1: Long Papers)",
	month = aug,
	year = "2024",
	address = "Bangkok, Thailand",
	publisher = "Association for Computational Linguistics",
	doi = "10.18653/v1/2024.acl-long.49",
	pages = "861--880",
}

@article{qwen3,
	title={Qwen3 Technical Report}, 
	author={An Yang and Anfeng Li and Baosong Yang and Beichen Zhang and Binyuan Hui and Bo Zheng and Bowen Yu and Chang Gao and Chengen Huang and Chenxu Lv and Chujie Zheng and Dayiheng Liu and Fan Zhou and Fei Huang and Feng Hu and Hao Ge and Haoran Wei and Huan Lin and Jialong Tang and Jian Yang and Jianhong Tu and Jianwei Zhang and Jianxin Yang and Jiaxi Yang and Jing Zhou and Jingren Zhou and Junyang Lin and Kai Dang and Keqin Bao and Kexin Yang and Le Yu and Lianghao Deng and Mei Li and Mingfeng Xue and Mingze Li and Pei Zhang and Peng Wang and Qin Zhu and Rui Men and Ruize Gao and Shixuan Liu and Shuang Luo and Tianhao Li and Tianyi Tang and Wenbiao Yin and Xingzhang Ren and Xinyu Wang and Xinyu Zhang and Xuancheng Ren and Yang Fan and Yang Su and Yichang Zhang and Yinger Zhang and Yu Wan and Yuqiong Liu and Zekun Wang and Zeyu Cui and Zhenru Zhang and Zhipeng Zhou and Zihan Qiu},
	journal = {arXiv preprint arXiv:2505.09388},
	year={2025}
}

@inproceedings{fang-etal-2025-counterfactual,
	title = "Counterfactual Debating with Preset Stances for Hallucination Elimination of {LLM}s",
	author = "Fang, Yi  and
	Li, Moxin  and
	Wang, Wenjie  and
	Hui, Lin  and
	Feng, Fuli",
	editor = "Rambow, Owen  and
	Wanner, Leo  and
	Apidianaki, Marianna  and
	Al-Khalifa, Hend  and
	Eugenio, Barbara Di  and
	Schockaert, Steven",
	booktitle = "Proceedings of the 31st International Conference on Computational Linguistics",
	month = jan,
	year = "2025",
	address = "Abu Dhabi, UAE",
	publisher = "Association for Computational Linguistics",
	url = "https://aclanthology.org/2025.coling-main.703/",
	pages = "10554--10568",
}

@inproceedings{muhammad-etal-2025-brighter,
	title = "{BRIGHTER}: {BRI}dging the Gap in Human-Annotated Textual Emotion Recognition Datasets for 28 Languages",
	author = "Muhammad, Shamsuddeen Hassan  and
	Ousidhoum, Nedjma  and
	Abdulmumin, Idris  and
	Wahle, Jan Philip  and
	Ruas, Terry  and
	Beloucif, Meriem  and
	de Kock, Christine  and
	Surange, Nirmal  and
	Teodorescu, Daniela  and
	Ahmad, Ibrahim Said  and
	Adelani, David Ifeoluwa  and
	Aji, Alham Fikri  and
	Ali, Felermino D. M. A.  and
	Alimova, Ilseyar  and
	Araujo, Vladimir  and
	Babakov, Nikolay  and
	Baes, Naomi  and
	Bucur, Ana-Maria  and
	Bukula, Andiswa  and
	Cao, Guanqun  and
	Tufi{\~n}o, Rodrigo  and
	Chevi, Rendi  and
	Chukwuneke, Chiamaka Ijeoma  and
	Ciobotaru, Alexandra  and
	Dementieva, Daryna  and
	Gadanya, Murja Sani  and
	Geislinger, Robert  and
	Gipp, Bela  and
	Hourrane, Oumaima  and
	Ignat, Oana  and
	Lawan, Falalu Ibrahim  and
	Mabuya, Rooweither  and
	Mahendra, Rahmad  and
	Marivate, Vukosi  and
	Panchenko, Alexander  and
	Piper, Andrew  and
	Ferreira, Charles Henrique Porto  and
	Protasov, Vitaly  and
	Rutunda, Samuel  and
	Shrivastava, Manish  and
	Udrea, Aura Cristina  and
	Wanzare, Lilian Diana Awuor  and
	Wu, Sophie  and
	Wunderlich, Florian Valentin  and
	Zhafran, Hanif Muhammad  and
	Zhang, Tianhui  and
	Zhou, Yi  and
	Mohammad, Saif M.",
	editor = "Che, Wanxiang  and
	Nabende, Joyce  and
	Shutova, Ekaterina  and
	Pilehvar, Mohammad Taher",
	booktitle = "Proceedings of the 63rd Annual Meeting of the Association for Computational Linguistics (Volume 1: Long Papers)",
	month = jul,
	year = "2025",
	address = "Vienna, Austria",
	publisher = "Association for Computational Linguistics",
	url = "https://aclanthology.org/2025.acl-long.436/",
	doi = "10.18653/v1/2025.acl-long.436",
	pages = "8895--8916",
}

@inproceedings{casola-etal-2024-multipico,
	title = "{M}ulti{PIC}o: Multilingual Perspectivist Irony Corpus",
	author = "Casola, Silvia  and
	Frenda, Simona  and
	Lo, Soda Marem  and
	Sezerer, Erhan  and
	Uva, Antonio  and
	Basile, Valerio  and
	Bosco, Cristina  and
	Pedrani, Alessandro  and
	Rubagotti, Chiara  and
	Patti, Viviana  and
	Bernardi, Davide",
	editor = "Ku, Lun-Wei  and
	Martins, Andre  and
	Srikumar, Vivek",
	booktitle = "Proceedings of the 62nd Annual Meeting of the Association for Computational Linguistics (Volume 1: Long Papers)",
	month = aug,
	year = "2024",
	address = "Bangkok, Thailand",
	publisher = "Association for Computational Linguistics",
	url = "https://aclanthology.org/2024.acl-long.849/",
	doi = "10.18653/v1/2024.acl-long.849",
	pages = "16008--16021",
}

@inproceedings{buechel-hahn-2017-emobank,
	title = "{E}mo{B}ank: Studying the Impact of Annotation Perspective and Representation Format on Dimensional Emotion Analysis",
	author = "Buechel, Sven  and
	Hahn, Udo",
	editor = "Lapata, Mirella  and
	Blunsom, Phil  and
	Koller, Alexander",
	booktitle = "Proceedings of the 15th Conference of the {E}uropean Chapter of the Association for Computational Linguistics: Volume 2, Short Papers",
	month = apr,
	year = "2017",
	address = "Valencia, Spain",
	publisher = "Association for Computational Linguistics",
	url = "https://aclanthology.org/E17-2092/",
	pages = "578--585",
}

@article{Ekman1971,
	title     = {Constants across cultures in the face and emotion},
	author    = {Ekman, Paul and Friesen, Wallace V.},
	journal   = {Journal of Personality and Social Psychology},
	volume    = {17},
	number    = {2},
	pages     = {124--129},
	year      = {1971},
	doi       = {10.1037/h0030377}
}

@article{Ekman1992,
	title     = {An argument for basic emotions},
	author    = {Ekman, Paul},
	journal   = {Cognition \& Emotion},
	volume    = {6},
	number    = {3-4},
	pages     = {169--200},
	year      = {1992},
	doi       = {10.1080/02699939208411068}
}

@misc{baumann2025largelanguagemodelhacking,
	title={Large Language Model Hacking: Quantifying the Hidden Risks of Using LLMs for Text Annotation}, 
	author={Joachim Baumann and Paul Röttger and Aleksandra Urman and Albert Wendsjö and Flor Miriam Plaza-del-Arco and Johannes B. Gruber and Dirk Hovy},
	year={2025},
	eprint={2509.08825},
	archivePrefix={arXiv},
	primaryClass={cs.CL},
	url={https://arxiv.org/abs/2509.08825}, 
}

\appendix

\section{Prompt for LLM-based Pre-annotation}
\label{sec:labelprompt}
\begin{figure*}[htb]
	\centering
	\includegraphics[width=0.98\textwidth]{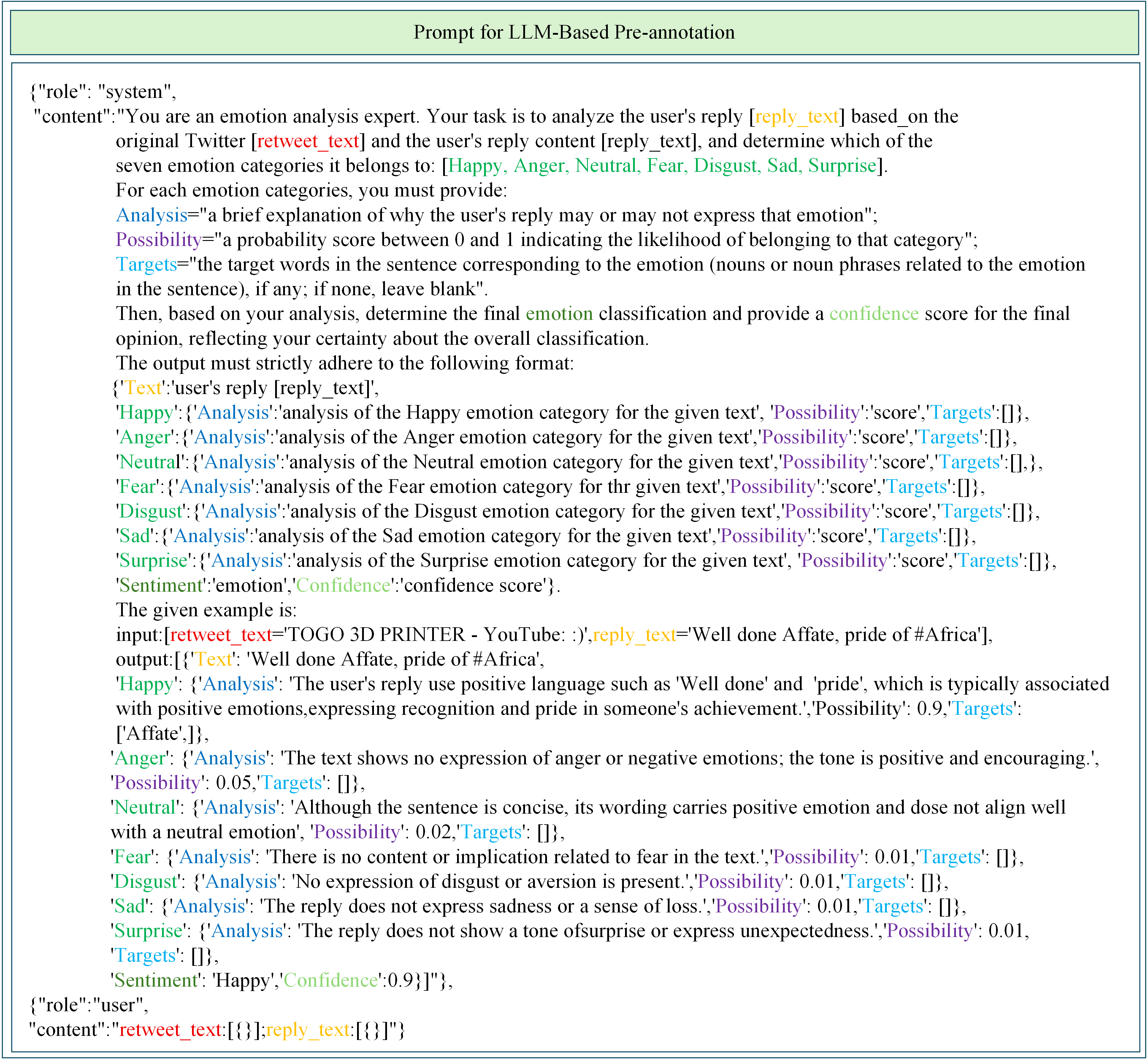}
	\caption{Prompt for LLM-based pre-annotation.}
	\label{fig:p_preann}
\end{figure*}

The prompt for LLM-based pre-annotation is illustrated in Fig.~\ref{fig:p_preann}.

\section{Detailed Evaluation Settings}
\label{appendix:evaluation}
\subsection{Task-Specific Evaluation Metrics}
The metrics are meticulously chosen to align with the objective of each task.
\begin{itemize}
	\item \textbf{Task A, B, \& C:} We report \textbf{Accuracy} and \textbf{Macro-F1}. Accuracy measures overall correctness, while Macro-F1 ensures balanced performance across all emotion/behavior categories, including minorities. We also report the F1 score for individual emotions/behaviors. The performance gain from Task A to B is a key indicator of a model's effectiveness in utilizing user profiles for personalization.
	\item \textbf{Task D:} We report \textbf{Confusion Matrix (CM)} for the emotion-emotion (EECM), behavior-behavior (BBCM), and emotion-behavior (EBCM) prediction independently. Recognizing that emotion annotations are exclusive to the \textit{R\&C} behavior, we compute the EBCM specifically for this category. This provides a controlled and meaningful assessment of a model's joint emotion-behavior reasoning capabilities, focusing on the interaction type with the richest semantic signal.
	\item \textbf{Task E:} We employ standard automatic metrics, including \textbf{BLEU-4} and \textbf{ROUGE-L}, to measure the n-gram overlap between the generated comments and the ground truth. Furthermore, we leverage GLM4.5-plus, Qwen3-plus, and DeepSeek-V3.2 to evaluate the generated content through the corresponding LLMs along three distinct dimensions: User Intent Consistency (\textbf{UIC}), Propagation Function Consistency (\textbf{PFC}), and Emotional Expression Consistency (\textbf{EEC}). Each dimension is rated on a 1-to-5 scale, with the mean score serving as the final evaluation metric.
\end{itemize}

\subsection{Baselines}
\label{appendix:baselines}
To ensure a rigorous evaluation, we benchmark our datasets against a carefully selected set of state-of-the-art models and reasoning paradigms, chosen for their complementary strengths.
\textbf{(1) Representative Online LLMs:} Our selection encompasses leading models recognized for their advanced reasoning capabilities.
\begin{itemize}
	\item \textbf{GLM4.5-plus} \cite{glm2024chatglm}: A model distinguished by its native integration of complex reasoning, coding, and agent capabilities, making it a robust baseline for multifaceted tasks.
	\item \textbf{Qwen3-plus} \cite{qwen3}: An open-weight ``thinking'' model that demonstrates state-of-the-art performance on tasks requiring expert-level, multi-step reasoning.
	\item \textbf{DeepSeek-V3.2} \cite{deepseekr1}: An architecture architected with a specific focus on deep, step-by-step logical deduction (``thinking'' ability), excelling in problems that demand extended reasoning chains.
\end{itemize}
\textbf{(2) Advanced Reasoning Paradigms:} We further investigate established frameworks designed to augment the intrinsic reasoning abilities of LLMs.
\begin{itemize}
	\item \textbf{THOR}~\cite{fei2023}: We employ a standard Chain-of-Thought prompting template, which guides the model to decompose the problem into a sequence of intermediate reasoning steps before yielding a final answer.
	\item \textbf{TOC}~\cite{weinzierl-harabagiu-2024-tree}: This method models the reasoning process as a tree structure, systematically exploring multiple solution paths and using self-evaluation to identify the most promising line of thought.
	\item \textbf{Debate-based Reasoning (Debate)}~\cite{fang-etal-2025-counterfactual}: A multi-agent framework that enhances reasoning by compelling models to defend counterfactual stances, thereby mitigating inherent biases and stimulating more thorough justification.
\end{itemize}

\section{Detailed Results and Visualization}
\label{appendix:drv}
\begin{figure*}[htb]
	\centering
	\begin{subfigure}[b]{0.49\textwidth}
		\centering
		\includegraphics[width=\textwidth]{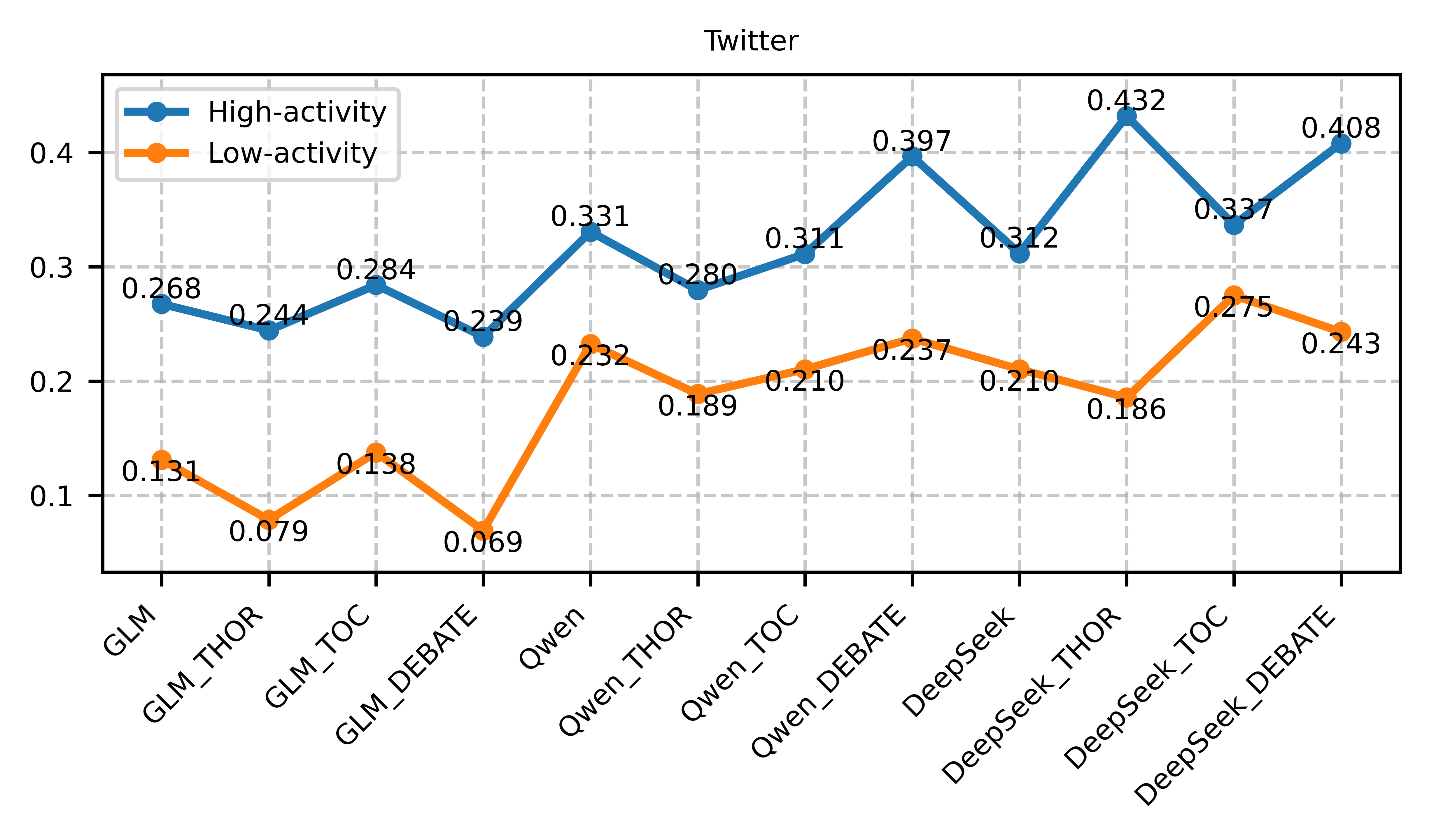}
	\end{subfigure}
	\hfill
	\begin{subfigure}[b]{0.49\textwidth}
		\centering
		\includegraphics[width=\textwidth]{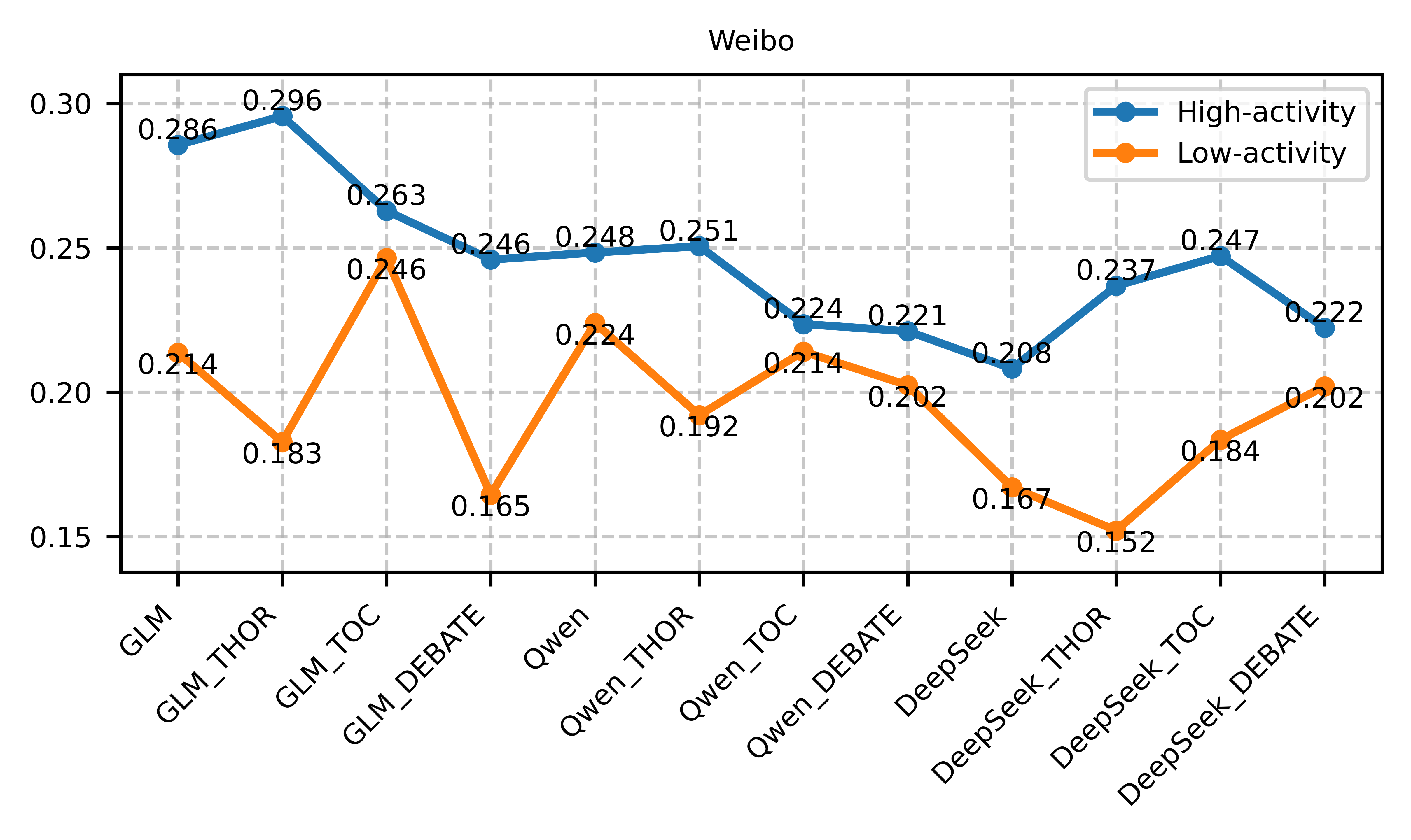}
	\end{subfigure}
	\caption{Comparison of high- and low- activity users for Task C.}
	\label{fig:taskC_behavior_hl}
\end{figure*}

\begin{figure*}[htb]
	\centering
	\begin{subfigure}[b]{0.33\textwidth}
		\centering
		\includegraphics[width=\textwidth]{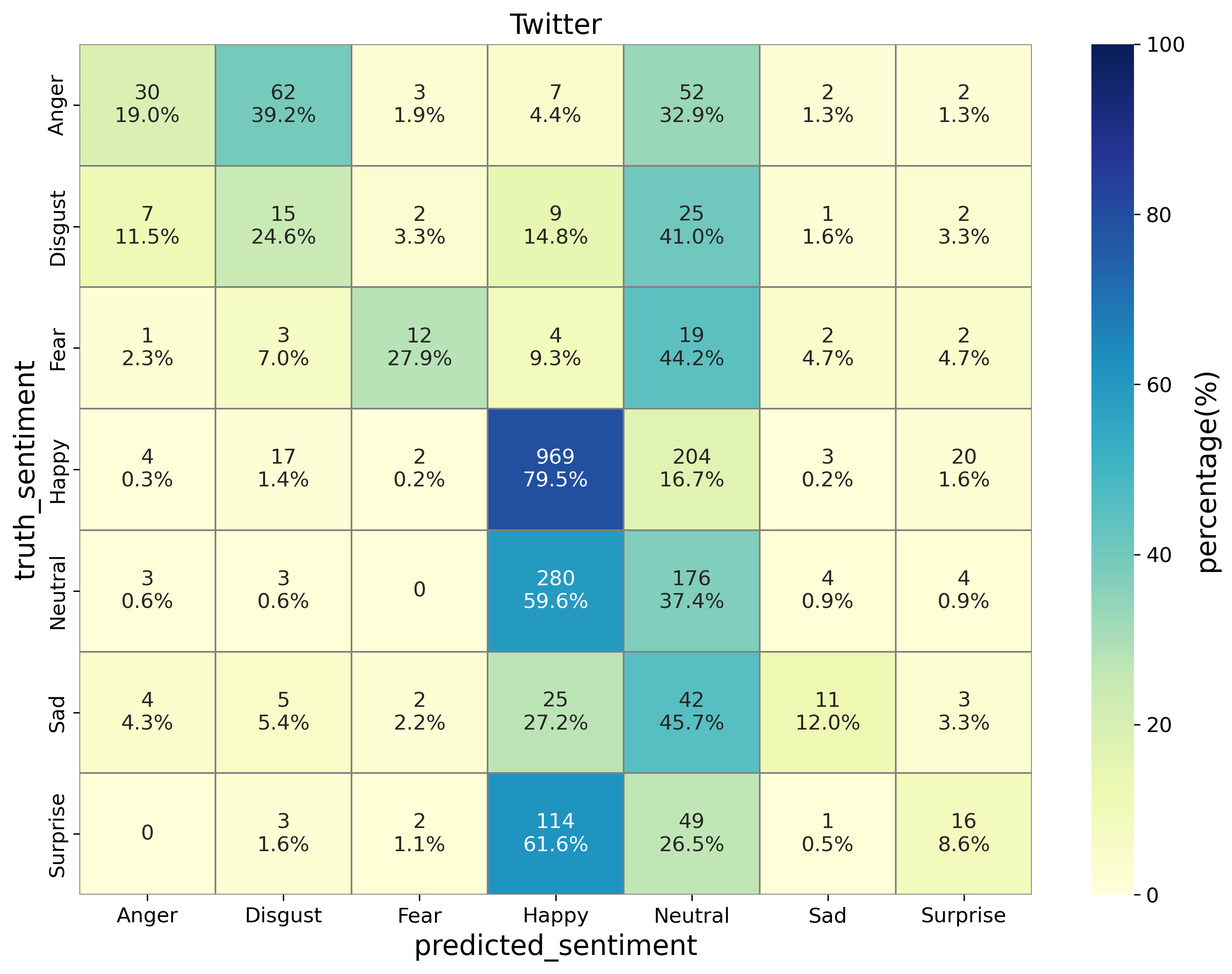}
		\caption{EECM}
	\end{subfigure}
	\hfill
	\begin{subfigure}[b]{0.32\textwidth}
		\centering
		\includegraphics[width=\textwidth]{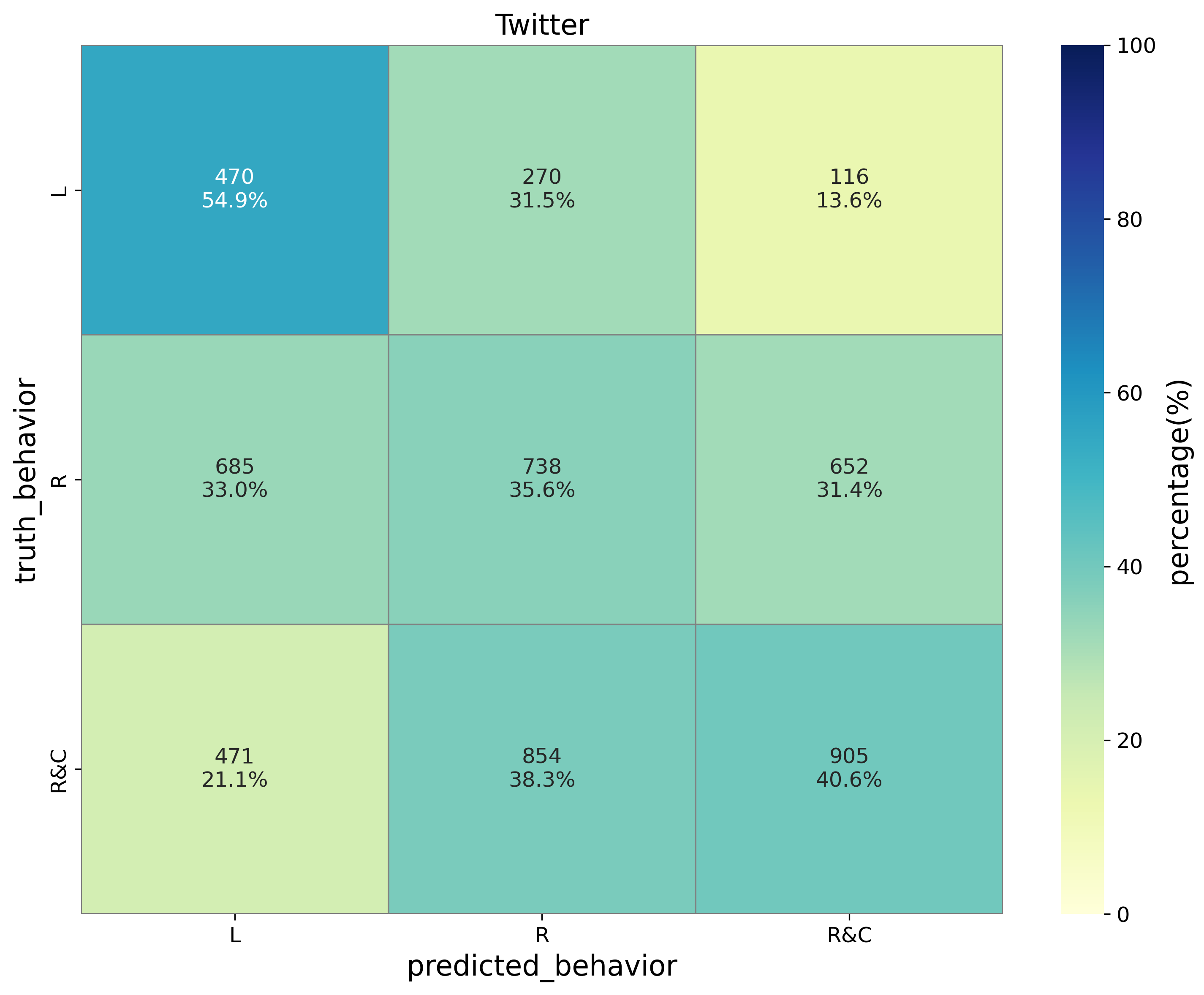}
		\caption{BBCM}
	\end{subfigure}
	\hfill
	\begin{subfigure}[b]{0.33\textwidth}
		\centering
		\includegraphics[width=\textwidth]{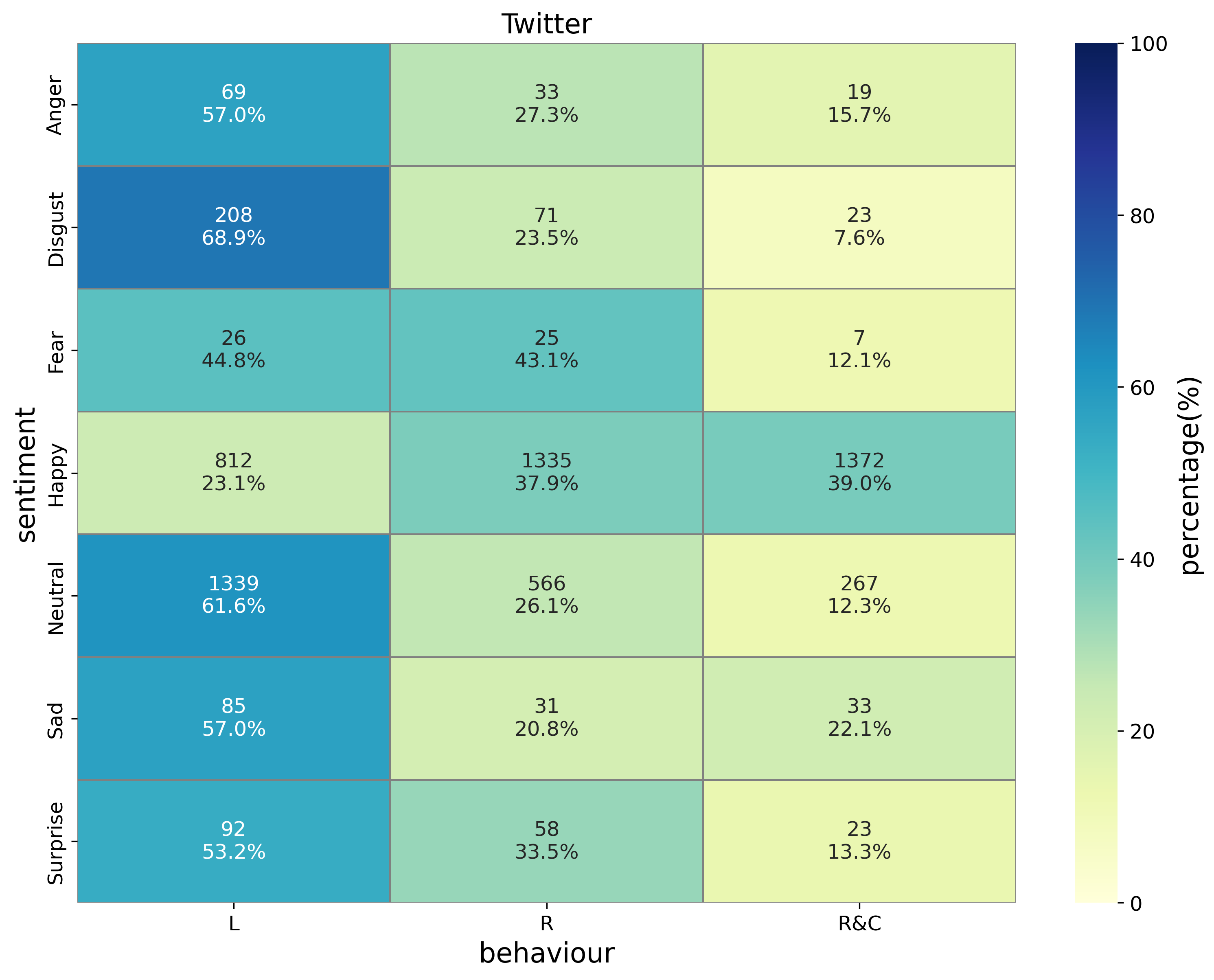}
		\caption{EBCM}
	\end{subfigure}
	\caption{Confusion matrices of emotion-behavior joint prediction using GLM in Twitter.}
	\label{fig:taskd-glm-t}
\end{figure*}

\begin{figure*}[htb]
	\centering
	\begin{subfigure}[b]{0.33\textwidth}
		\centering
		\includegraphics[width=\textwidth]{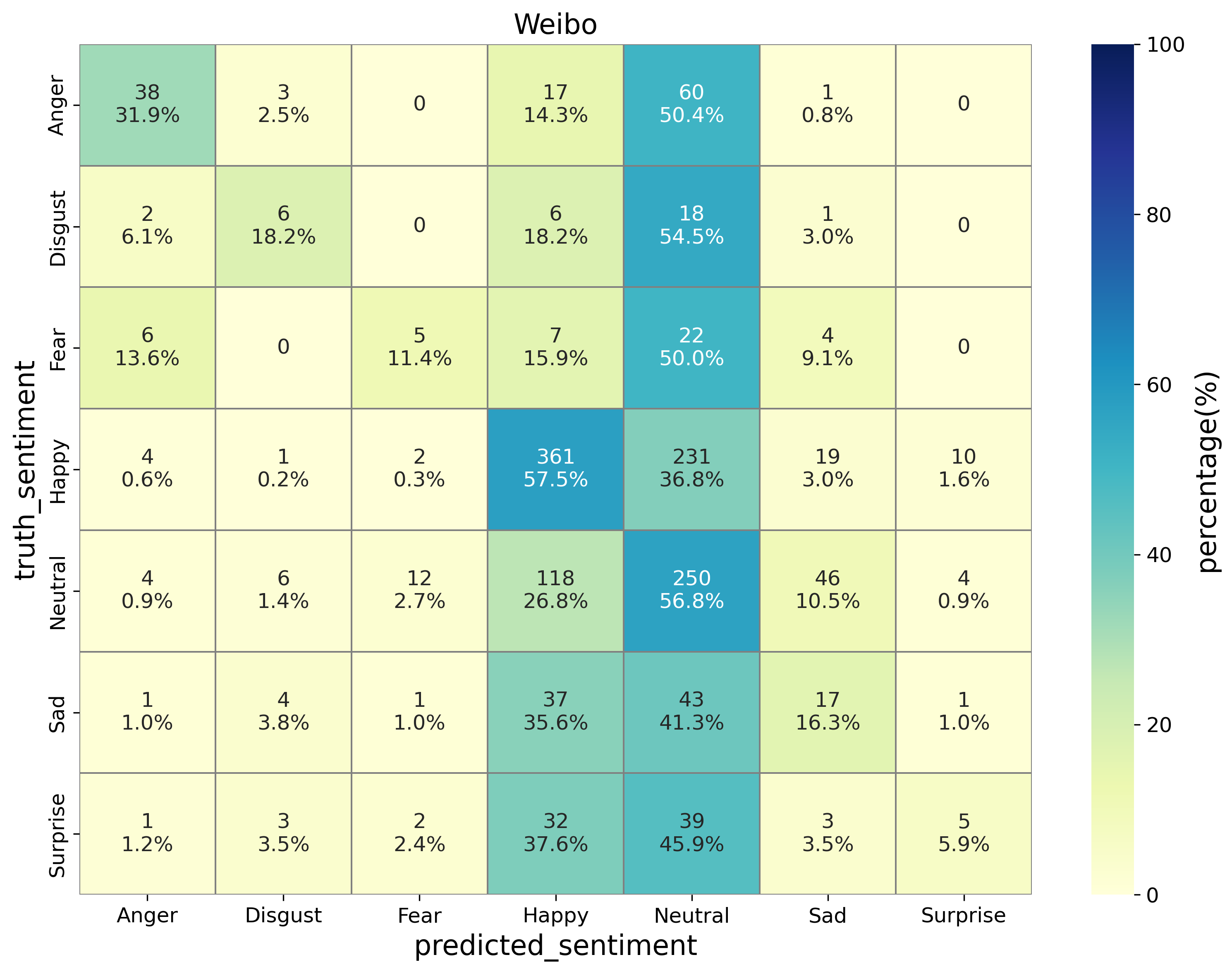}
		\caption{EECM}
	\end{subfigure}
	\hfill
	\begin{subfigure}[b]{0.32\textwidth}
		\centering
		\includegraphics[width=\textwidth]{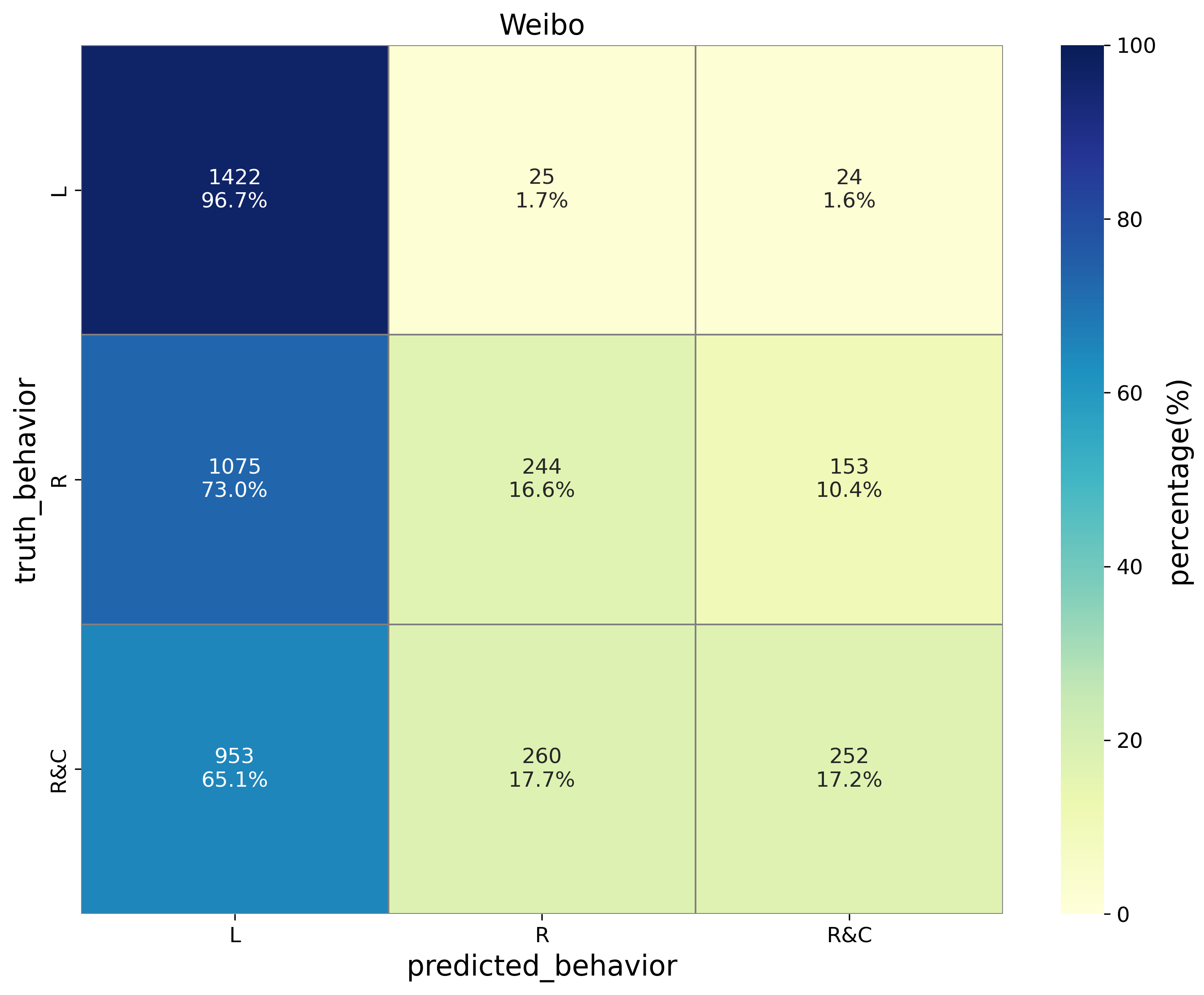}
		\caption{BBCM}
	\end{subfigure}
	\hfill
	\begin{subfigure}[b]{0.33\textwidth}
		\centering
		\includegraphics[width=\textwidth]{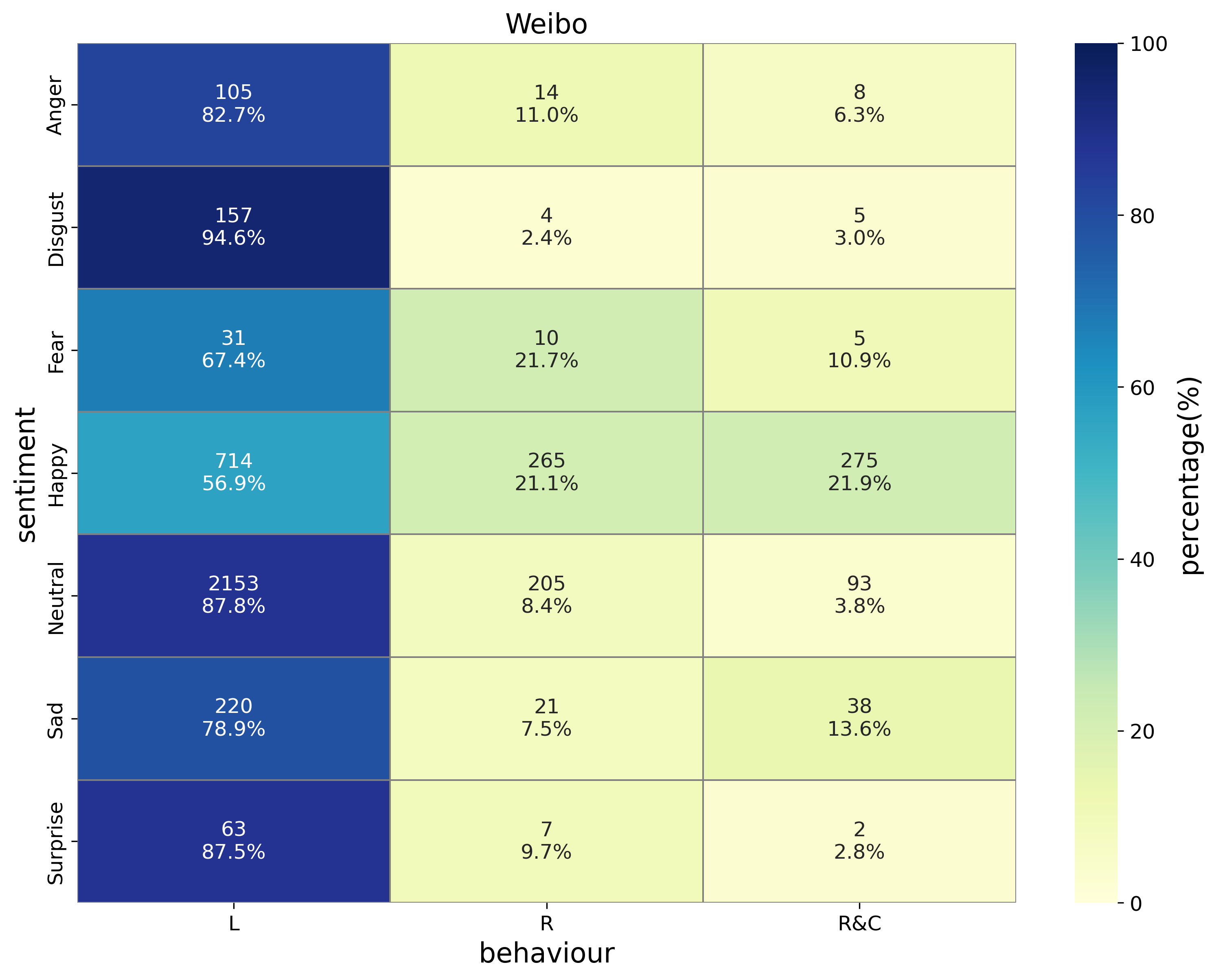}
		\caption{EBCM}
	\end{subfigure}
	\caption{Confusion matrices of emotion-behavior joint prediction using GLM in Weibo.}
	\label{fig:taskd-glm-w}
\end{figure*}

\begin{figure*}[htb]
	\centering
	\begin{subfigure}[b]{0.33\textwidth}
		\centering
		\includegraphics[width=\textwidth]{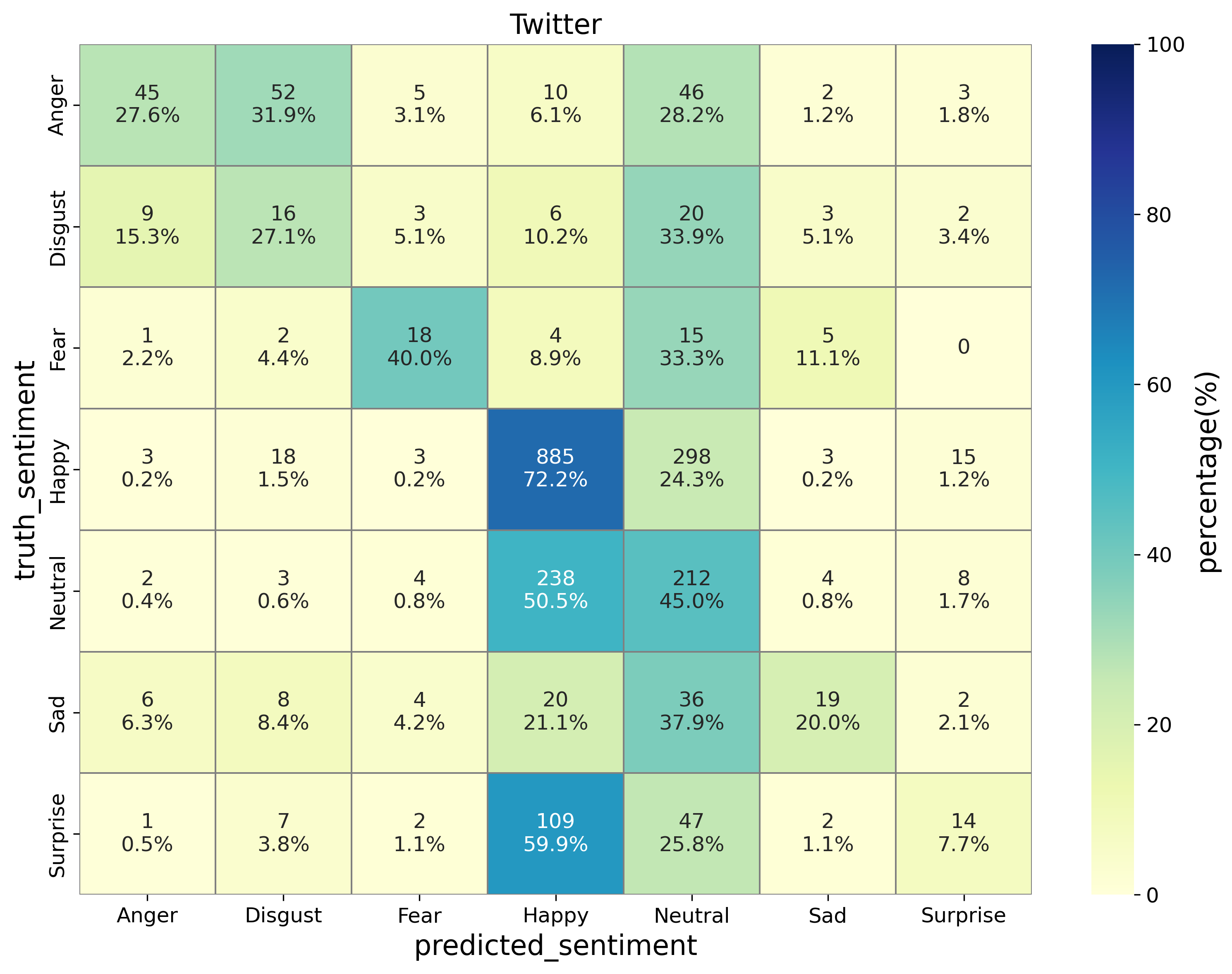}
		\caption{EECM}
	\end{subfigure}
	\hfill
	\begin{subfigure}[b]{0.32\textwidth}
		\centering
		\includegraphics[width=\textwidth]{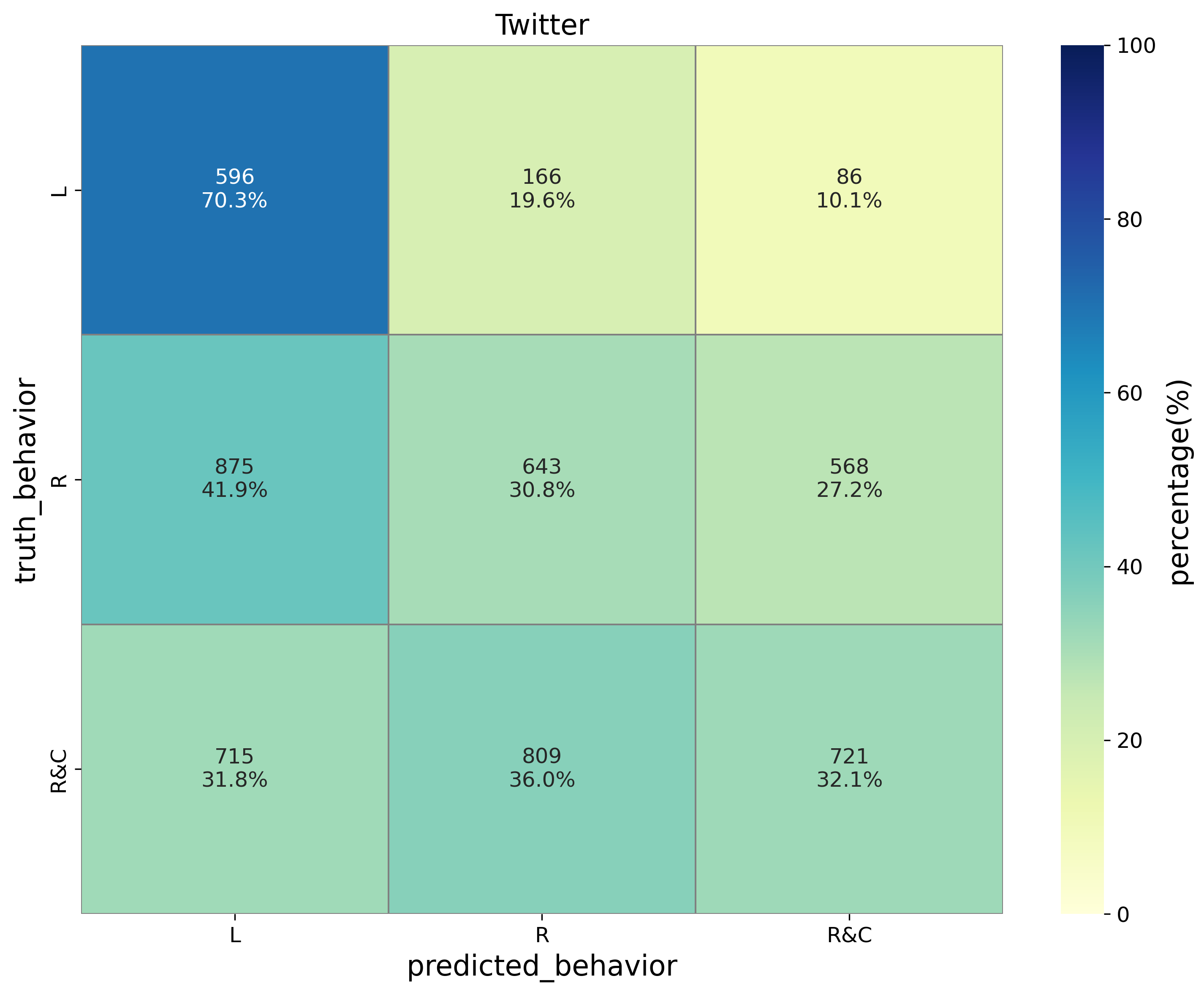}
		\caption{BBCM}
	\end{subfigure}
	\hfill
	\begin{subfigure}[b]{0.33\textwidth}
		\centering
		\includegraphics[width=\textwidth]{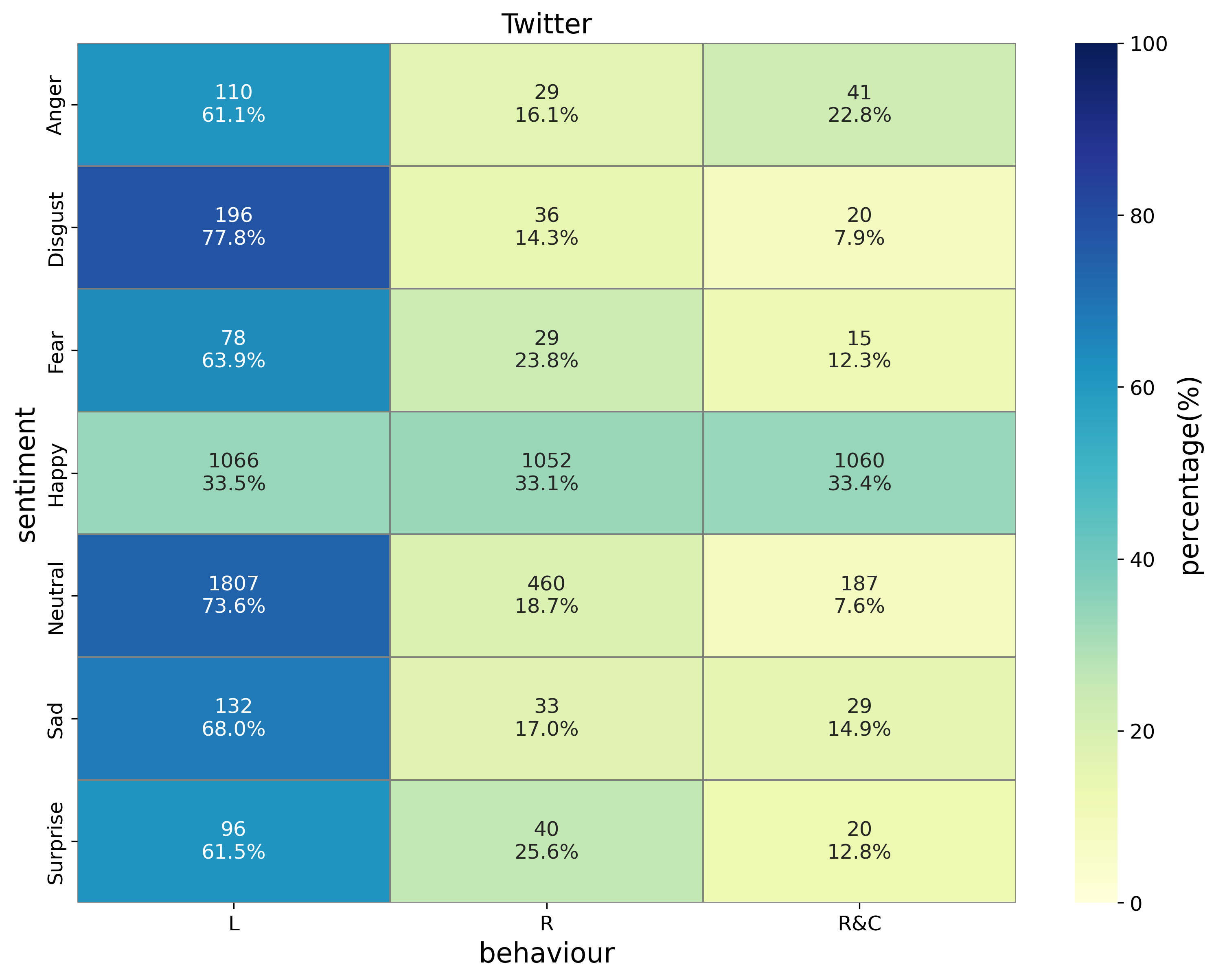}
		\caption{EBCM}
	\end{subfigure}
	\caption{Confusion matrices of emotion-behavior joint prediction using Qwen in Twitter.}
	\label{fig:taskd-qwen-t}
\end{figure*}

\begin{figure*}[htb]
	\centering
	\begin{subfigure}[b]{0.33\textwidth}
		\centering
		\includegraphics[width=\textwidth]{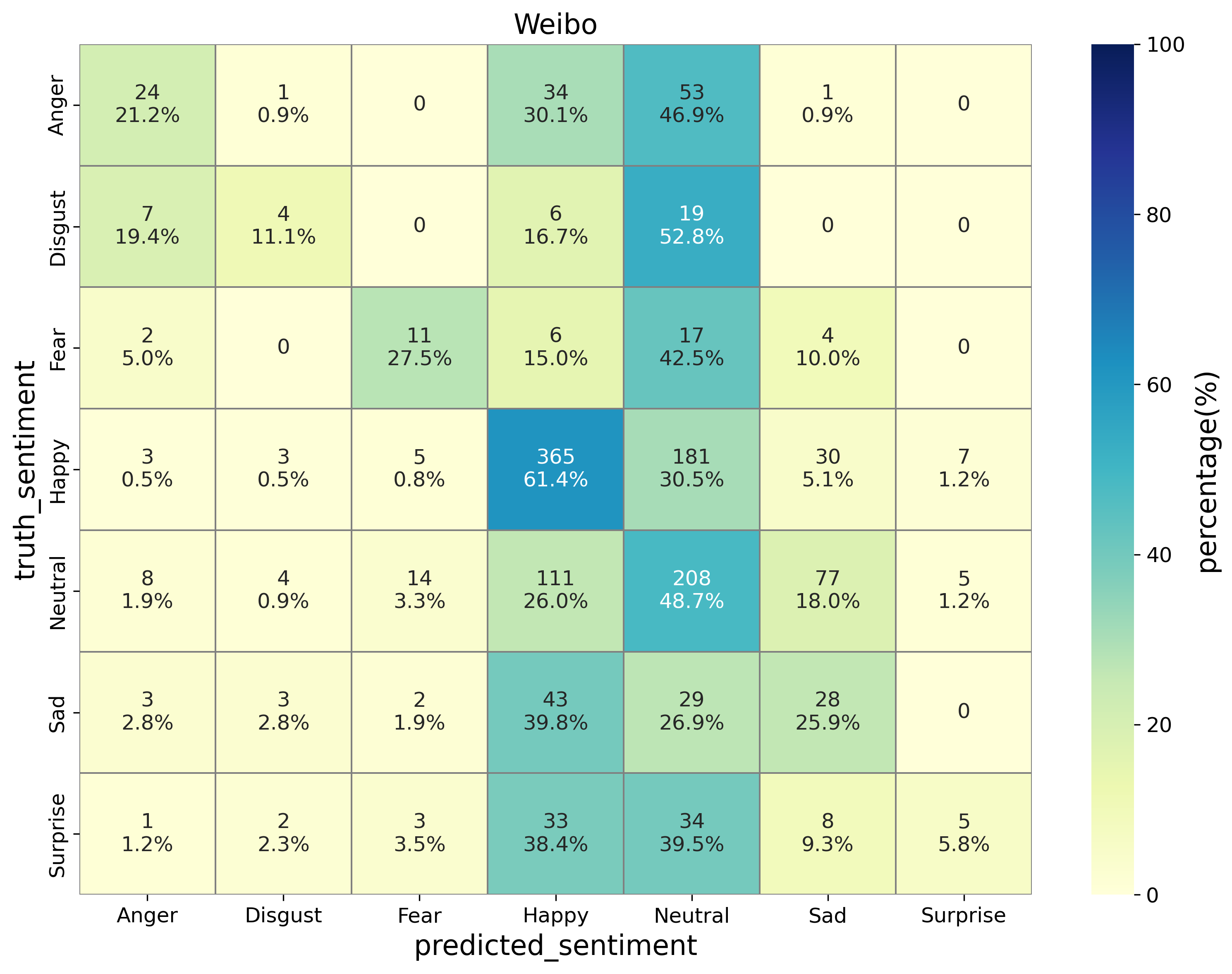}
		\caption{EECM}
	\end{subfigure}
	\hfill
	\begin{subfigure}[b]{0.32\textwidth}
		\centering
		\includegraphics[width=\textwidth]{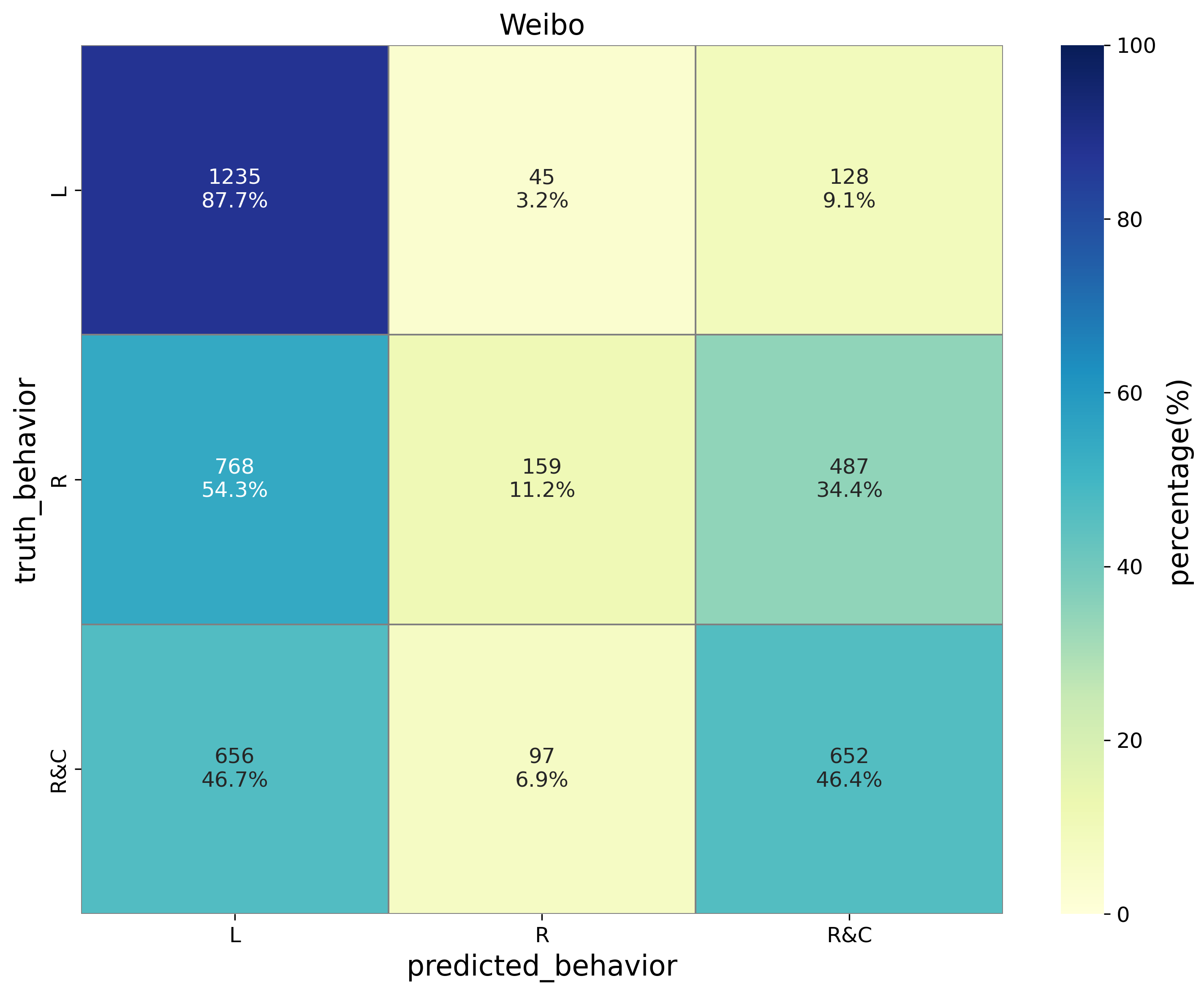}
		\caption{BBCM}
	\end{subfigure}
	\hfill
	\begin{subfigure}[b]{0.33\textwidth}
		\centering
		\includegraphics[width=\textwidth]{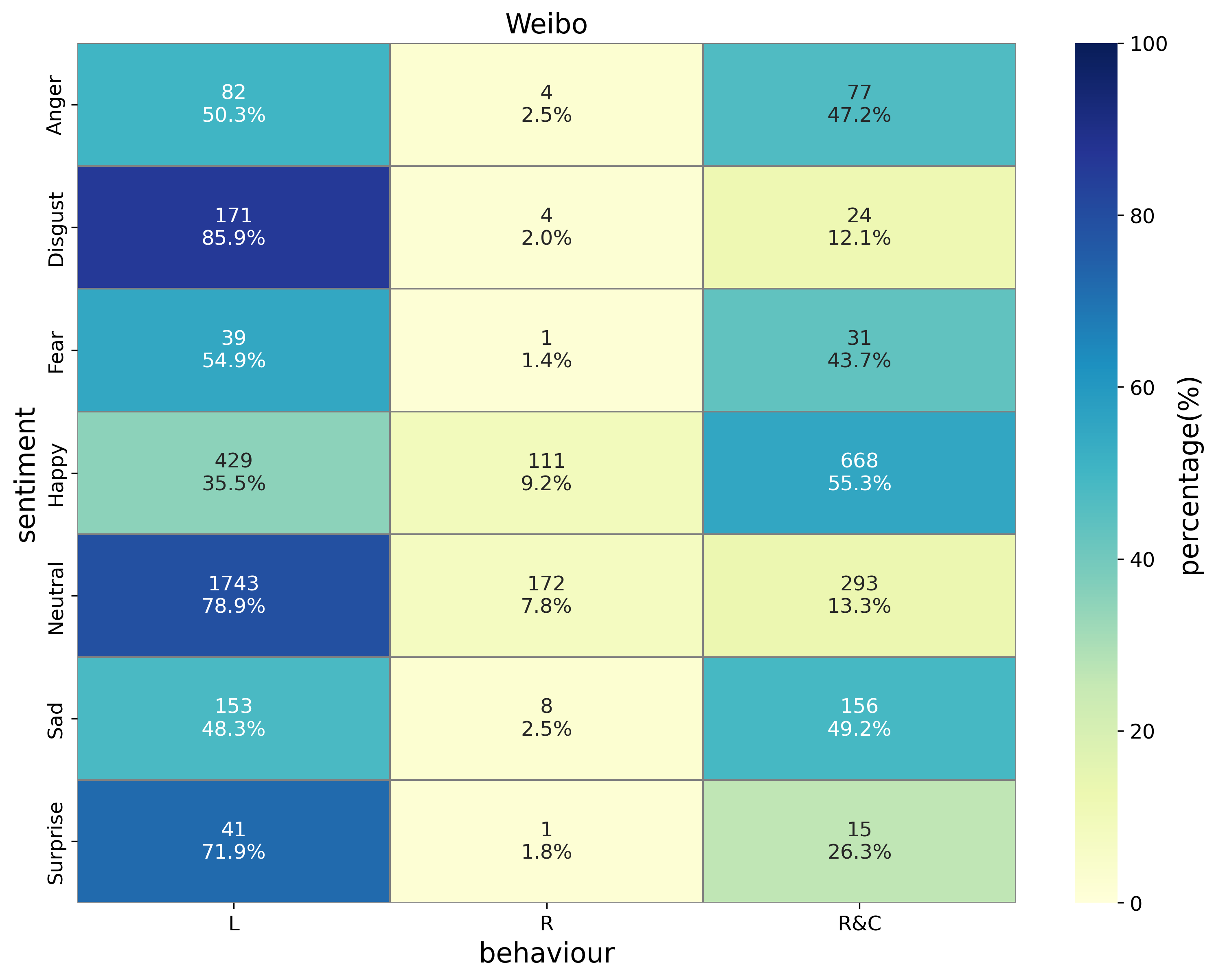}
		\caption{EBCM}
	\end{subfigure}
	\caption{Confusion matrices of emotion-behavior joint prediction using Qwen in Weibo.}
	\label{fig:taskd-qwen-w}
\end{figure*}

\begin{figure*}[htb]
	\centering
	\begin{subfigure}[b]{0.33\textwidth}
		\centering
		\includegraphics[width=\textwidth]{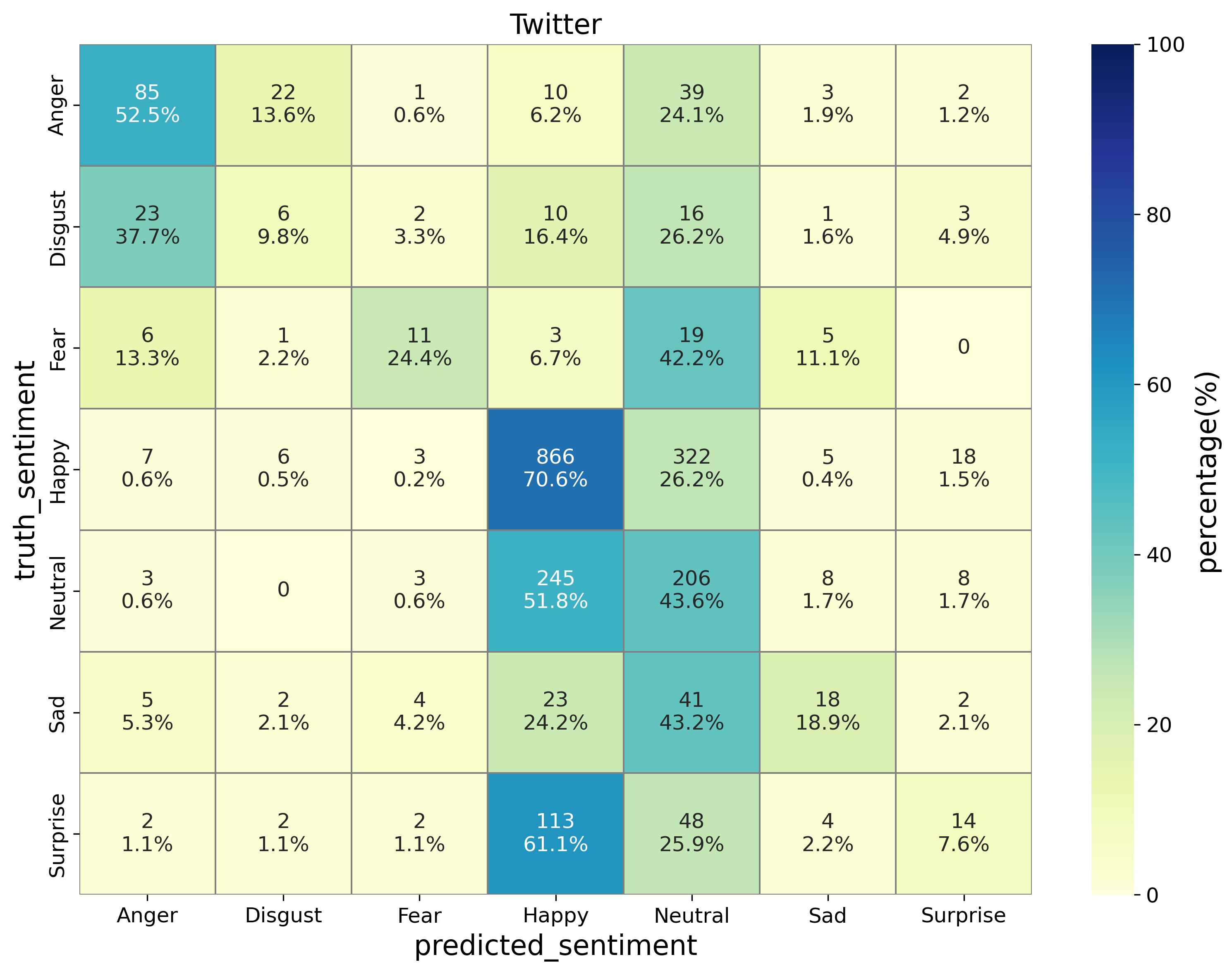}
		\caption{EECM}
	\end{subfigure}
	\hfill
	\begin{subfigure}[b]{0.32\textwidth}
		\centering
		\includegraphics[width=\textwidth]{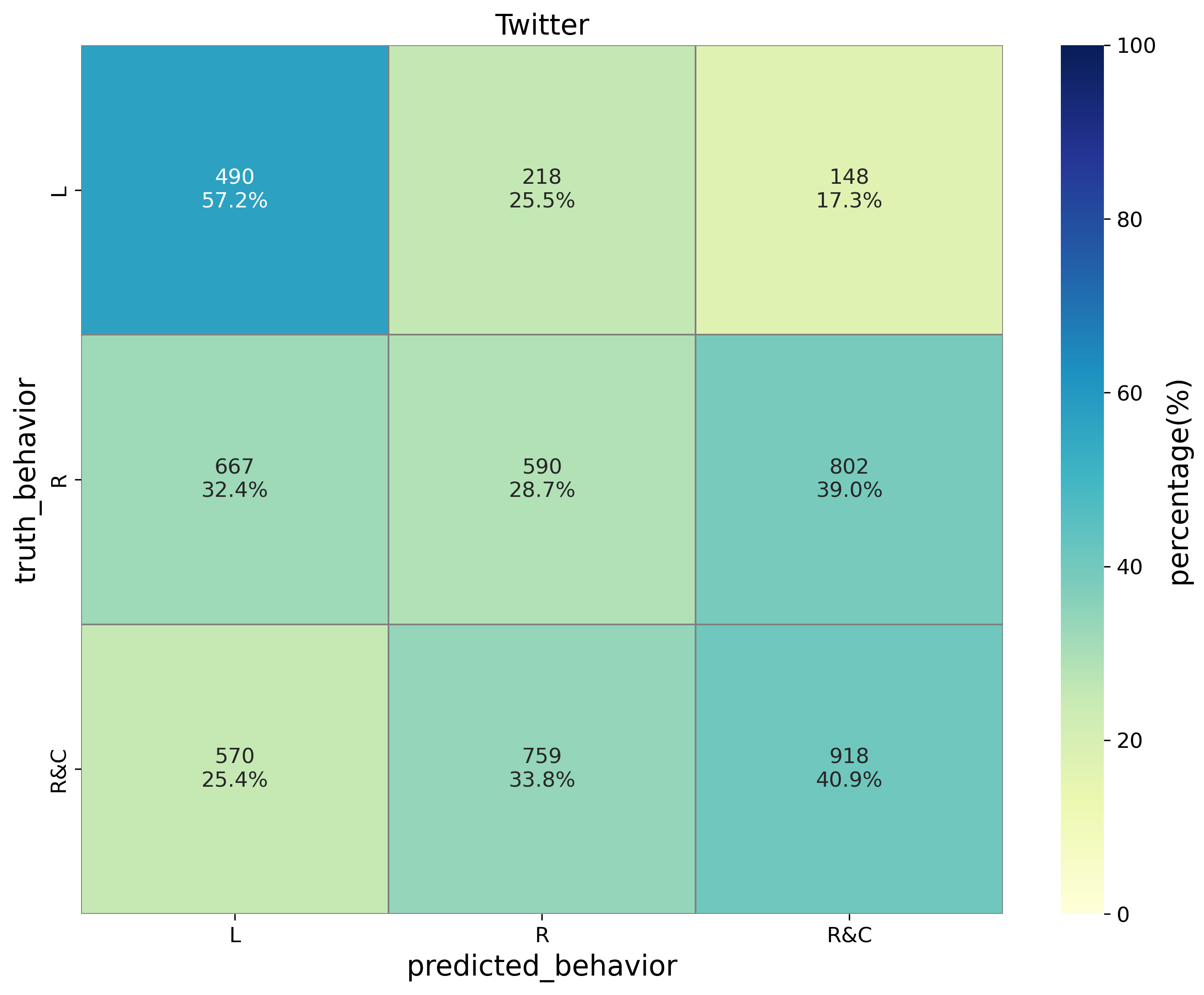}
		\caption{BBCM}
	\end{subfigure}
	\hfill
	\begin{subfigure}[b]{0.33\textwidth}
		\centering
		\includegraphics[width=\textwidth]{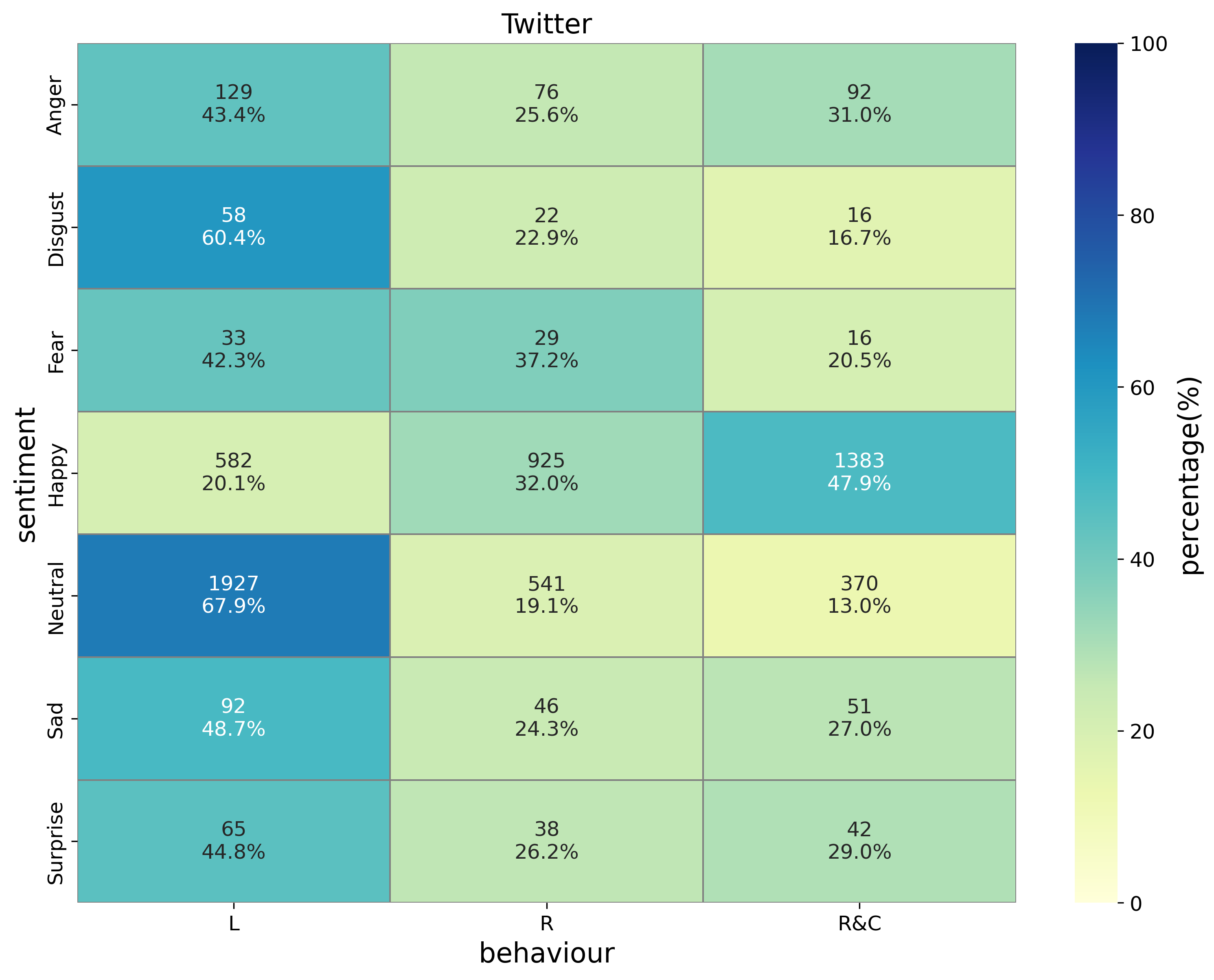}
		\caption{EBCM}
	\end{subfigure}
	\caption{Confusion matrices of emotion-behavior joint prediction using DeepSeek in Twitter.}
	\label{fig:taskd-ds-t}
\end{figure*}

\begin{figure*}[htb]
	\centering
	\begin{subfigure}[b]{0.33\textwidth}
		\centering
		\includegraphics[width=\textwidth]{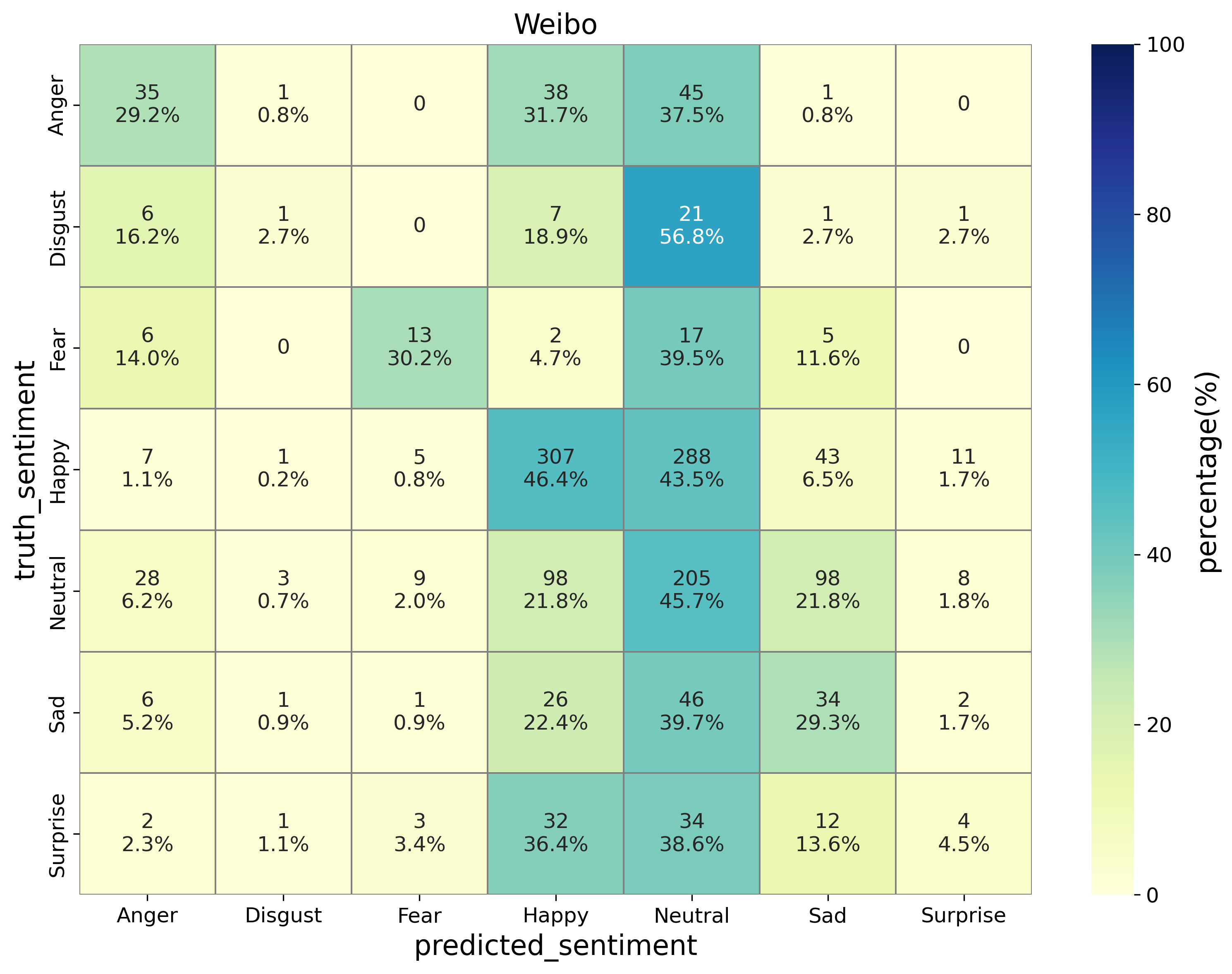}
		\caption{EECM}
	\end{subfigure}
	\hfill
	\begin{subfigure}[b]{0.32\textwidth}
		\centering
		\includegraphics[width=\textwidth]{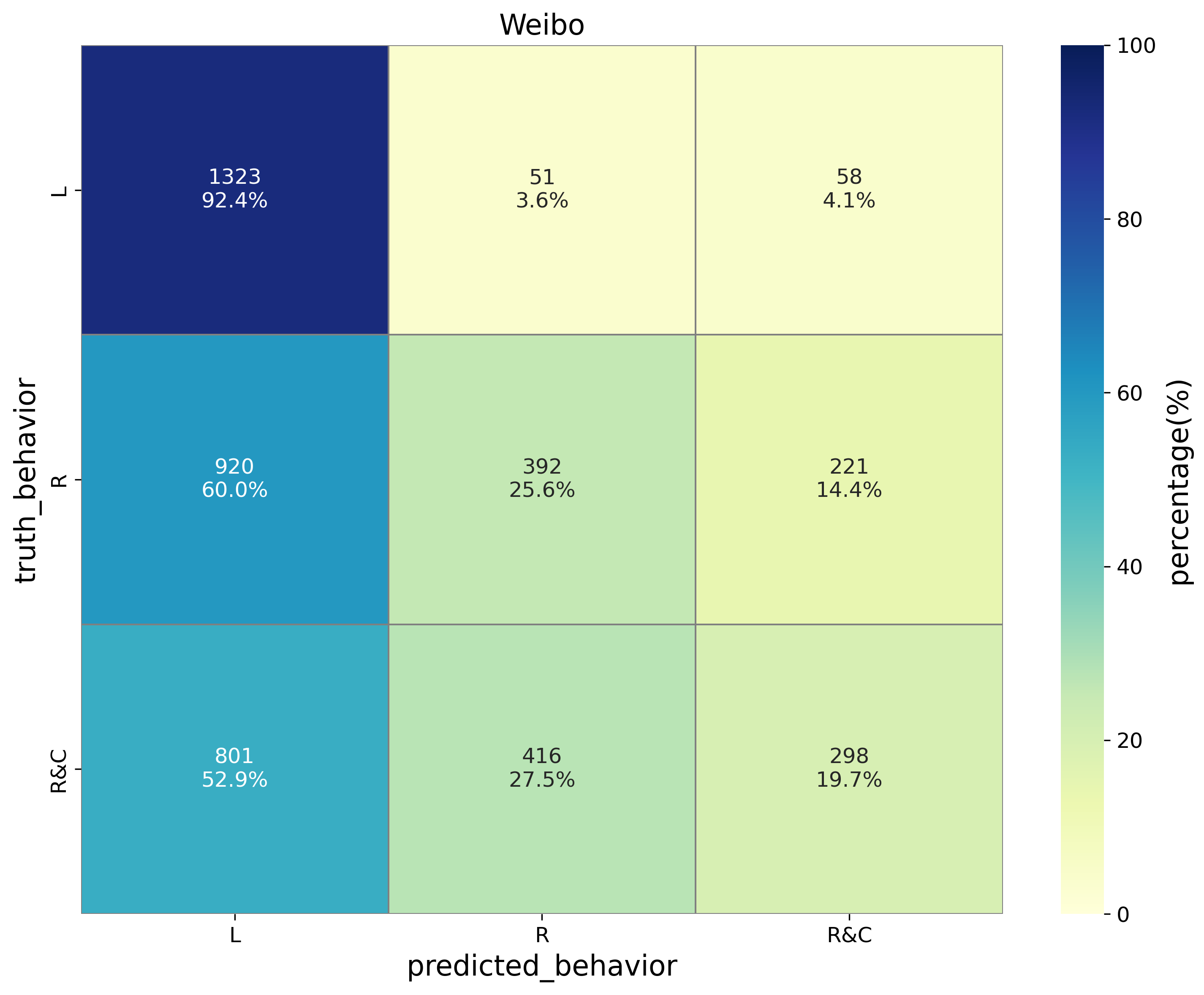}
		\caption{BBCM}
	\end{subfigure}
	\hfill
	\begin{subfigure}[b]{0.33\textwidth}
		\centering
		\includegraphics[width=\textwidth]{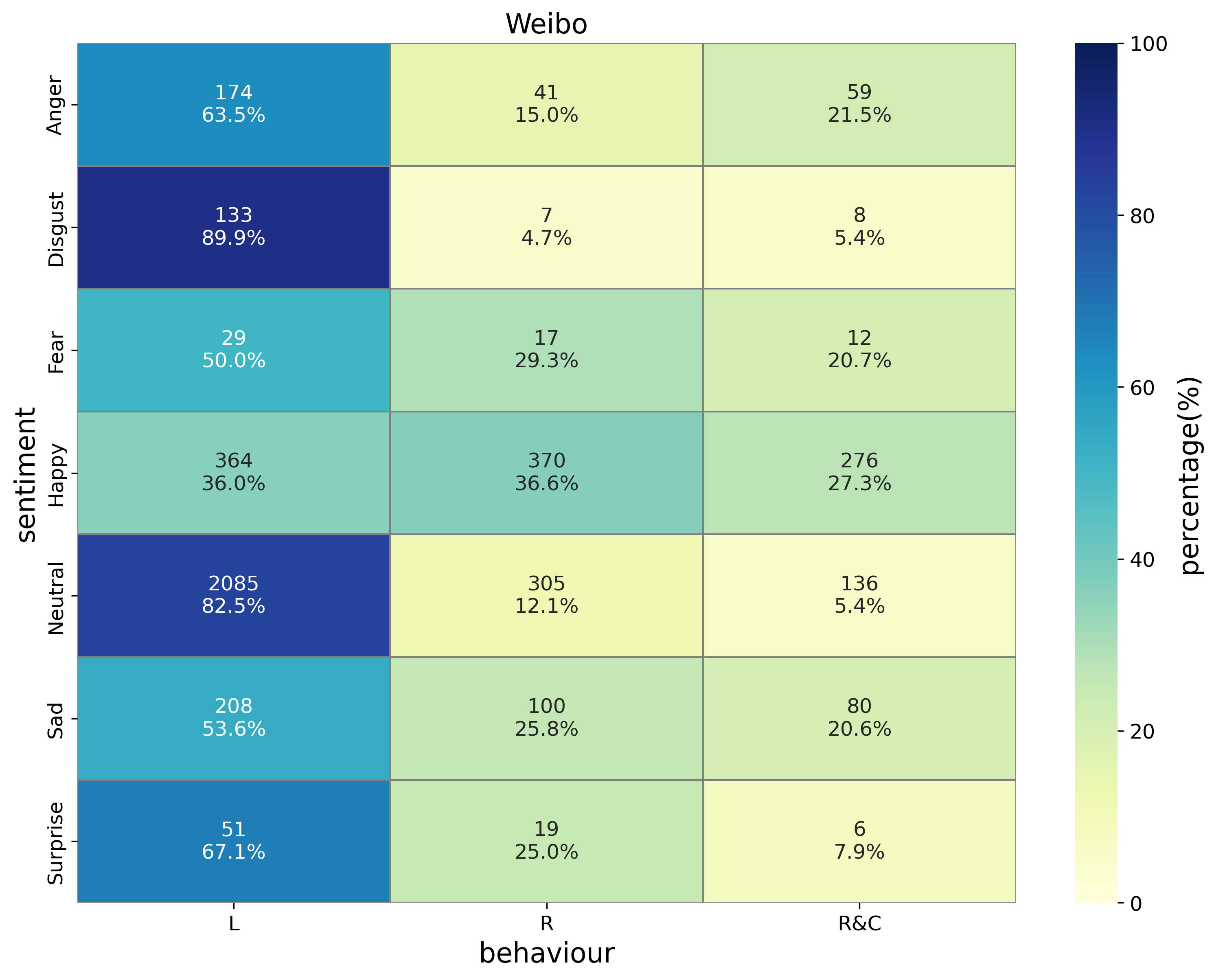}
		\caption{EBCM}
	\end{subfigure}
	\caption{Confusion matrices of emotion-behavior joint prediction using DeepSeek in Weibo.}
	\label{fig:taskd-ds-w}
\end{figure*}

\begin{figure*}[htb]
	\centering
	\begin{subfigure}[b]{0.38\textwidth}
		\centering
		\includegraphics[width=\linewidth,height=5cm]{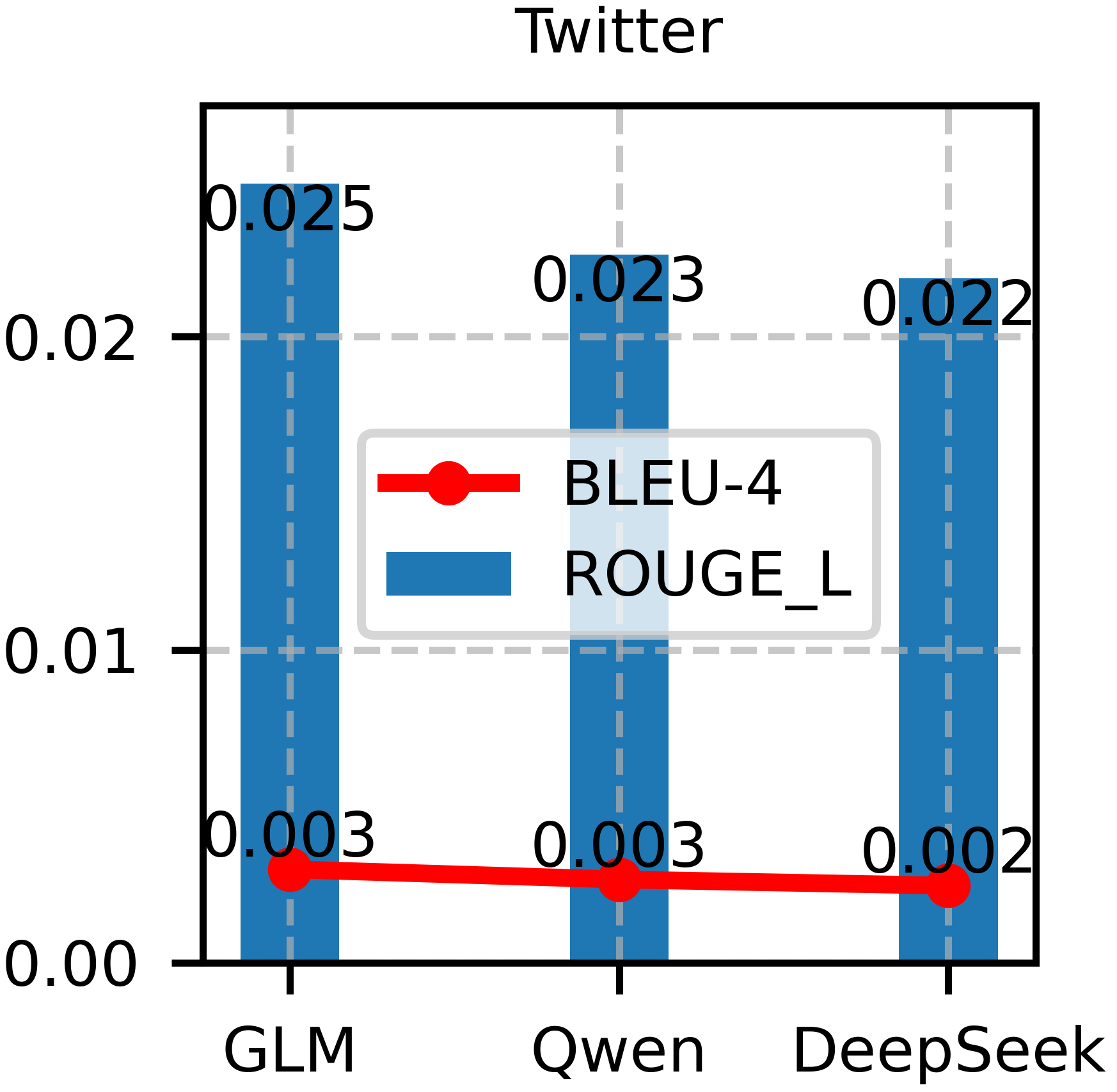}
	\end{subfigure}
	\hfill
	\begin{subfigure}[b]{0.36\textwidth}
		\centering
		\includegraphics[width=\linewidth,height=5cm]{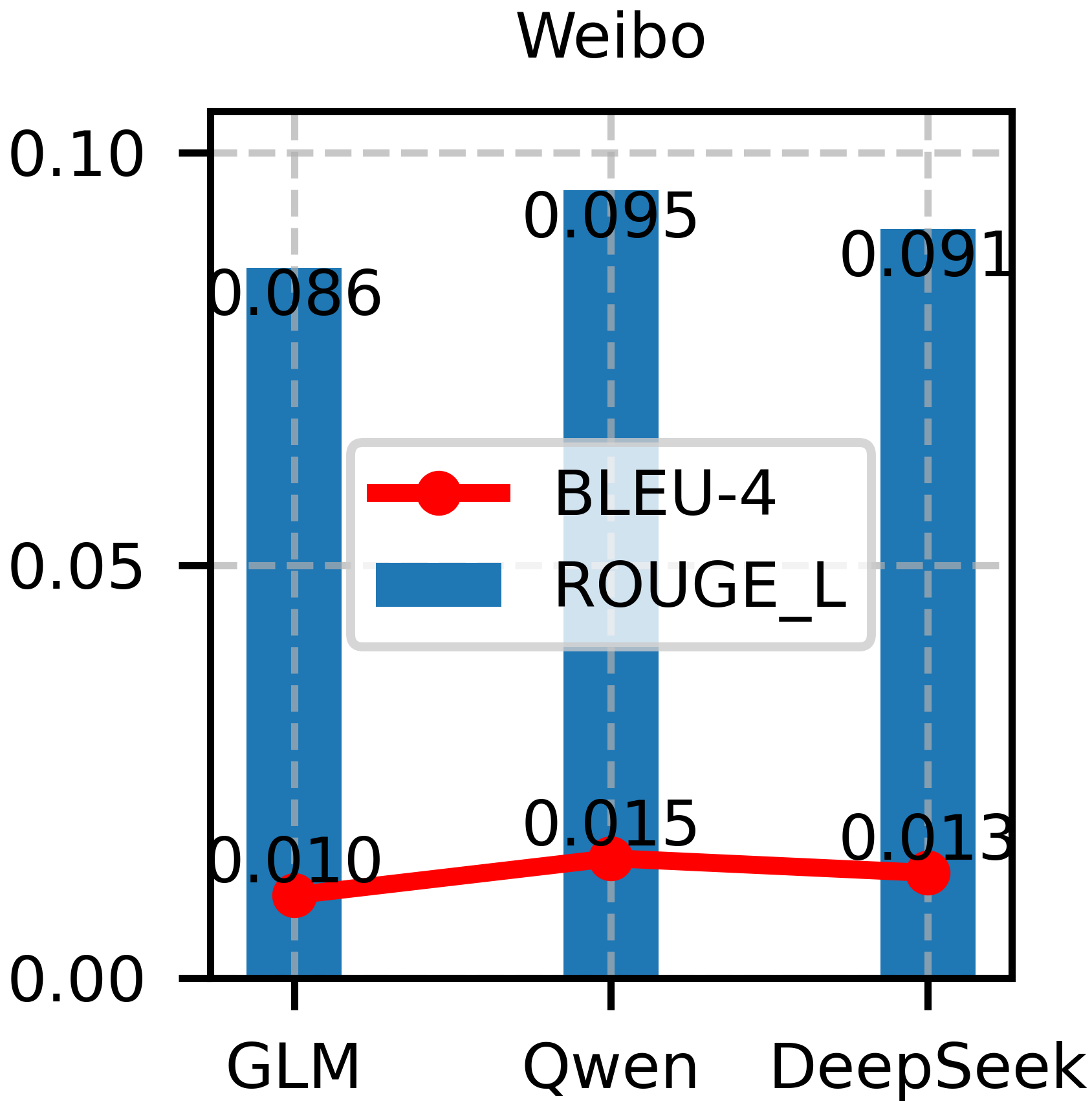}
	\end{subfigure}
	\caption{BLEU and Rouge-L of Task E based on different LLMs.}
	\label{fig:taske-bleurouge}
\end{figure*}

This section provides a comprehensive breakdown of the evaluation results for Tasks A-E. Tables \ref{tab:drtaska_t}-\ref{tab:drtaska_w}, \ref{tab:drtaskb_t}-\ref{tab:drtaskb_w}, and \ref{tab:drtaskc} detail the per-category and overall F1 scores for Task A (emotion prediction), Task B (personalized emotion prediction), and Task C (behavior prediction), respectively. Fig.~\ref{fig:taskC_behavior_hl} further analyzes the performance of Task C by comparing high- and low-activity users. Figs.~\ref{fig:taskd-glm-t}-\ref{fig:taskd-ds-w} present a comparative analysis of different foundation LLMs (GLM, Qwen, DeepSeek) on our core metrics (EECM, BBCM, EBCM). The quality of generated content for Task E is evaluated in Fig.~\ref{fig:taske-bleurouge} using BLEU and Rouge-L scores. 
\begin{table*}[htb]
	\centering
	\footnotesize
	\caption{Detailed results for Task A in Twitter dataset}
	\label{tab:drtaska_t}
		\begin{tabular}{lccccccccc}
			\hline
			\textbf{Models} & \textbf{Happy} & \textbf{Sad} & \textbf{Anger} & \textbf{Surprise} & \textbf{Disgust} & \textbf{Fear} & \textbf{Neutral} & \textbf{M-F1} & \textbf{Acc} \\\hline
			GLM & 0.8661 & 0.7296 & 0.4455 & 0.5648 & 0.3733 & 0.6842 & 0.6137 & 0.5346 & 0.7436 \\
			Qwen & 0.8550  & 0.6832 & 0.5188 & 0.5955 & 0.4556 & 0.7308 & 0.5043 & 0.5429 & 0.7387 \\
			DeepSeek & 0.8687 & 0.7219 & 0.6473 & 0.5250  & 0.2963 & 0.6250  & 0.5833 & 0.5334 & 0.7520 \\\hline
			GLM\_TOC & 0.8411 & 0.6447 & 0.6424 & 0.5567 & 0.3636 & 0.6111 & 0.4170  & 0.5096 & 0.7201 \\
			Qwen\_TOC & 0.8391 & 0.7421 & 0.5420  & 0.5816 & 0.4645 & 0.6279 & 0.3516 & 0.5186 & 0.7161 \\
			DeepSeek\_TOC & 0.8575 & 0.7052 & 0.4416 & 0.5872 & 0.4277 & 0.6750  & 0.4958 & 0.5238 & 0.7334 \\\hline
			GLM\_THOR & 0.8415 & 0.6410  & 0.3318 & 0.4910  & 0.3760  & 0.6301 & 0.4251 & 0.4671 & 0.6953 \\
			Qwen\_THOR & 0.8401 & 0.6826 & 0.4367 & 0.5743 & 0.4362 & 0.7126 & 0.3910  & 0.5092 & 0.7068 \\
			DeepSeek\_THOR   & 0.8430  & 0.6864 & 0.5973 & 0.5229 & 0.3392 & 0.5672 & 0.3962 & 0.4940  & 0.7112 \\\hline
			GLM\_DEBATE      & 0.8381 & 0.5972 & 0.3256 & 0.5054 & 0.3944 & 0.6667 & 0.5393 & 0.4833 & 0.7121 \\
			Qwen\_DEBATE     & 0.8624 & 0.6905 & 0.5936 & 0.5493 & 0.4316 & 0.6667 & 0.5008 & 0.5369 & 0.7383 \\
			DeepSeek\_DEBATE & 0.8501 & 0.6709 & 0.4769 & 0.6163 & 0.4468 & 0.5846 & 0.6288 & 0.5343 & 0.7378 \\\hline
		\end{tabular}%
\end{table*}

\begin{table*}[htb]
	\centering
	\footnotesize
	\caption{Detailed results for Task A in Weibo dataset}
	\label{tab:drtaska_w}
	\begin{tabular}{lccccccccc}
		\hline
		\textbf{Models} & \textbf{Happy} & \textbf{Sad} & \textbf{Anger} & \textbf{Surprise} & \textbf{Disgust} & \textbf{Fear} & \textbf{Neutral} & \textbf{M-F1} & \textbf{Acc} \\\hline
		GLM & 0.7317 & 0.5165 & 0.5670  & 0.4571   & 0.4359  & 0.3729 & 0.6660   & 0.4684 & 0.6437 \\
		Qwen & 0.7930  & 0.6250  & 0.5571 & 0.4252   & 0.4800    & 0.5882 & 0.6953  & 0.5205 & 0.7056 \\
		DeepSeek         & 0.7711 & 0.5976 & 0.5263 & 0.5000      & 0.4828  & 0.4068 & 0.6486  & 0.4916 & 0.6756 \\\hline
		GLM\_TOC         & 0.7528 & 0.5670  & 0.5399 & 0.4545   & 0.2581  & 0.3793 & 0.5922  & 0.4430  & 0.6341 \\
		Qwen\_TOC        & 0.7813 & 0.5647 & 0.5124 & 0.4521   & 0.4565  & 0.5600   & 0.5575  & 0.4856 & 0.6584 \\
		DeepSeek\_TOC    & 0.7735 & 0.6140  & 0.4455 & 0.4828   & 0.4444  & 0.4571 & 0.6279  & 0.4807 & 0.6724 \\\hline
		GLM\_THOR        & 0.7222 & 0.5385 & 0.4481 & 0.3802   & 0.3729  & 0.2807 & 0.5885  & 0.4164 & 0.6086 \\
		Qwen\_THOR       & 0.7383 & 0.6066 & 0.5714 & 0.5068   & 0.3613  & 0.6000   & 0.5130   & 0.4872 & 0.6328 \\
		DeepSeek\_THOR   & 0.7394 & 0.6047 & 0.4508 & 0.4000      & 0.3409  & 0.5000    & 0.5041  & 0.4425 & 0.622  \\\hline
		GLM\_DEBATE      & 0.7604 & 0.4574 & 0.5988 & 0.4375   & 0.3738  & 0.4571 & 0.5604  & 0.4557 & 0.6424 \\
		Qwen\_DEBATE     & 0.7951 & 0.5256 & 0.6842 & 0.5342   & 0.3621  & 0.5429 & 0.5929  & 0.5046 & 0.6711 \\
		DeepSeek\_DEBATE & 0.6916 & 0.6218 & 0.5422 & 0.5072   & 0.4048  & 0.3793 & 0.6493  & 0.4745 & 0.6379 \\\hline
	\end{tabular}%
\end{table*}

\begin{figure*}[htb]
	\centering
	\begin{subfigure}[b]{0.49\textwidth}
		\centering
		\includegraphics[width=\textwidth]{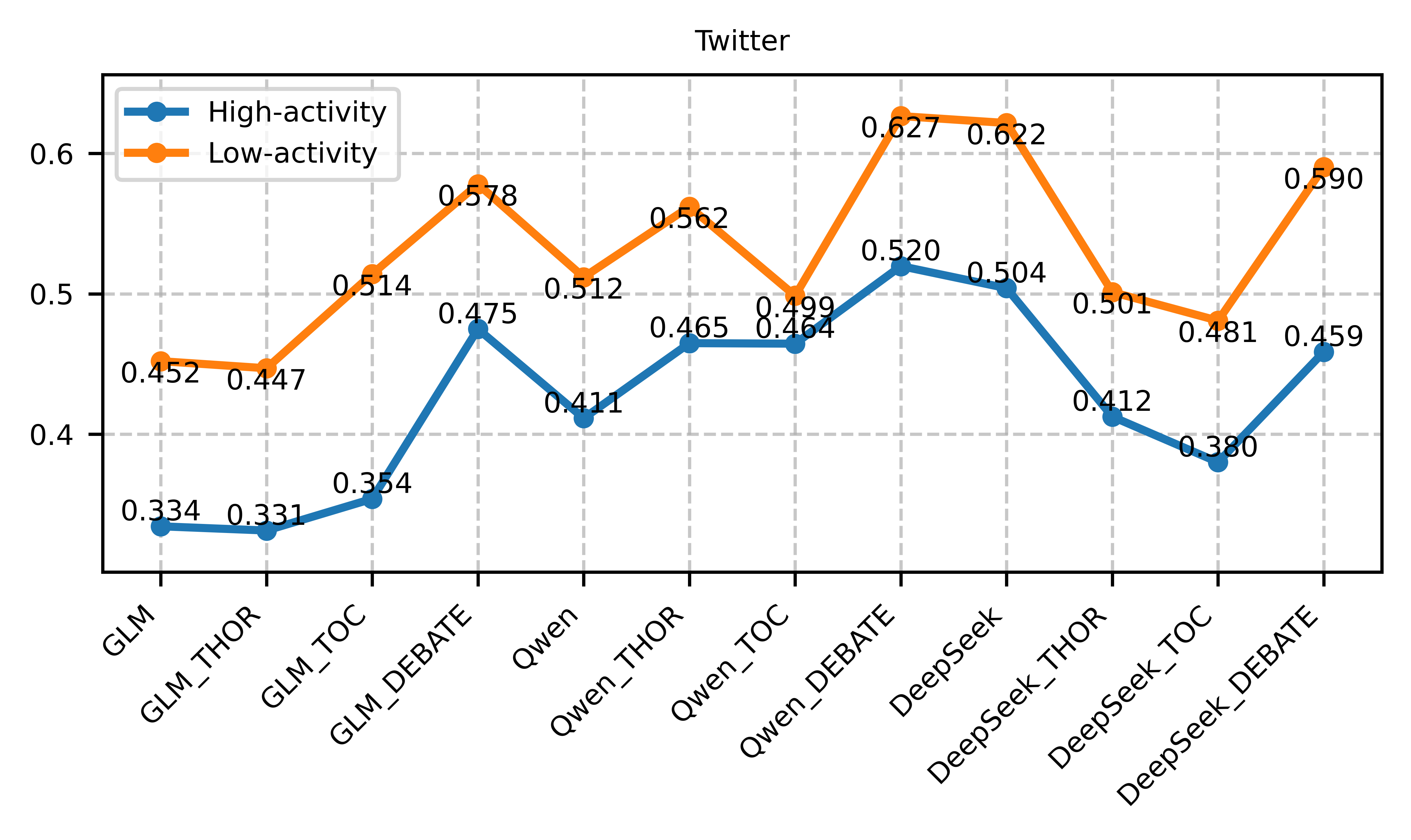}
	\end{subfigure}
	\hfill
	\begin{subfigure}[b]{0.49\textwidth}
		\centering
		\includegraphics[width=\textwidth]{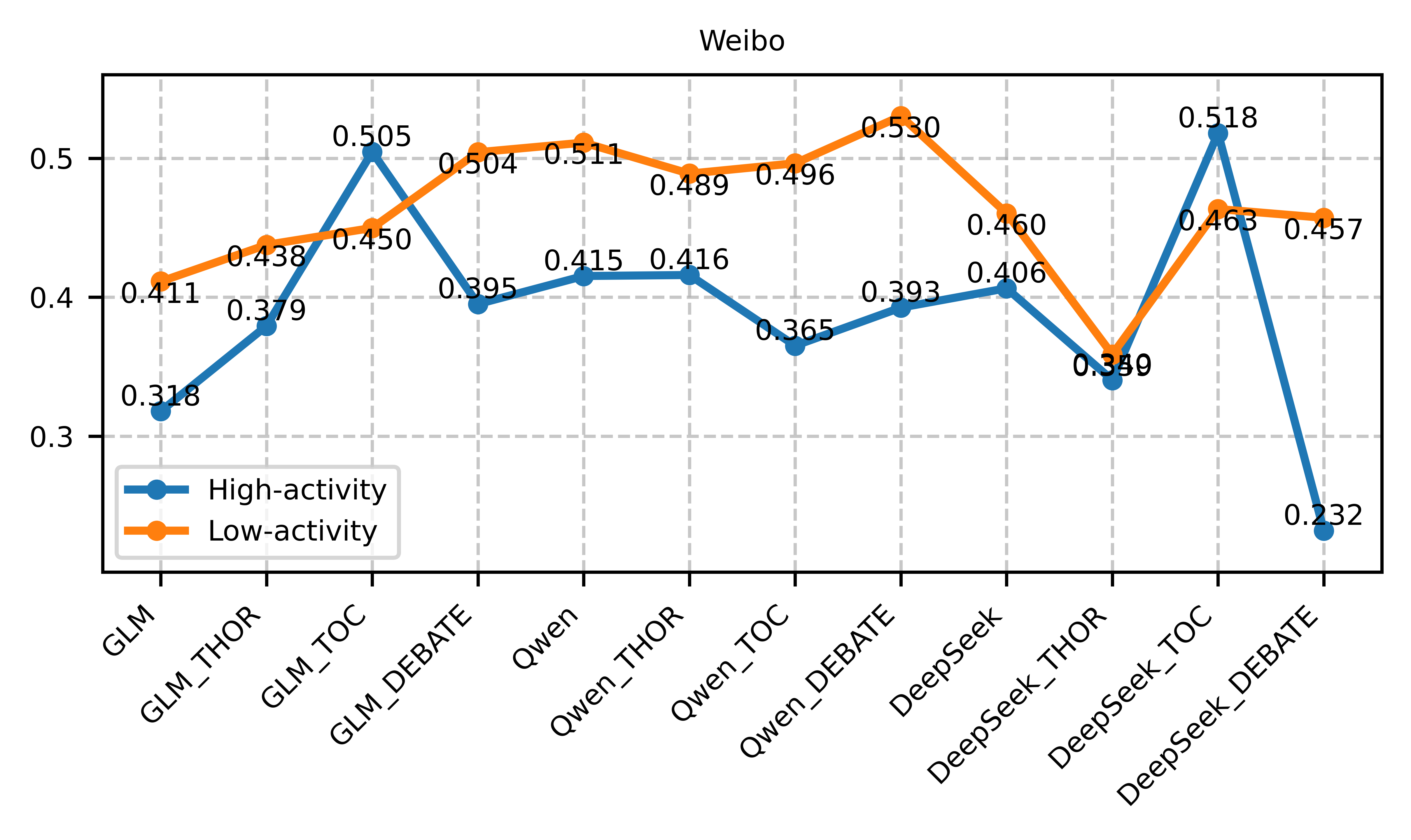}
	\end{subfigure}
	\caption{Comparison of high- and low- activity users for Task B.}
	\label{fig:taskB_sentiment_hl}
\end{figure*}

\begin{table*}[htb]
	\centering
	\footnotesize
	\caption{Detailed results for Task B in Twitter dataset}
	\label{tab:drtaskb_t}
	\begin{tabular}{lccccccccc}
		\hline
		\textbf{Models} & \textbf{Happy} & \textbf{Sad} & \textbf{Anger} & \textbf{Surprise} & \textbf{Disgust} & \textbf{Fear} & \textbf{Neutral} & \textbf{M-F1} & \textbf{Acc} \\\hline
		GLM              & 0.8521 & 0.671  & 0.5045  & 0.3962 & 0.5455   & 0.4186  & 0.5373 & 0.4907    & 0.7152   \\
		Qwen             & 0.8562 & 0.7273 & 0.5493  & 0.5814 & 0.5126   & 0.4564  & 0.6818 & 0.5456    & 0.7436   \\
		DeepSeek         & 0.8618 & 0.6871 & 0.5464  & 0.6714 & 0.5215   & 0.3471  & 0.5455 & 0.5226    & 0.7484   \\\hline
		GLM\_TOC         & 0.8384 & 0.6667 & 0.4023  & 0.6782 & 0.5563   & 0.3918  & 0.5672 & 0.5126    & 0.7219   \\
		Qwen\_TOC        & 0.8327 & 0.6753 & 0.299   & 0.5737 & 0.565    & 0.5342  & 0.5882 & 0.5085    & 0.7046   \\
		DeepSeek\_TOC    & 0.8469 & 0.6795 & 0.4897  & 0.4502 & 0.5643   & 0.3822  & 0.5143 & 0.4909    & 0.7183   \\\hline
		GLM\_THOR        & 0.8394 & 0.6452 & 0.4401  & 0.3366 & 0.4478   & 0.3786  & 0.6479 & 0.4669    & 0.6975   \\
		Qwen\_THOR       & 0.8589 & 0.7176 & 0.5072  & 0.5821 & 0.6286   & 0.5116  & 0.6667 & 0.5591    & 0.7471   \\
		DeepSeek\_THOR   & 0.8363 & 0.7066 & 0.3932  & 0.5891 & 0.4742   & 0.3546  & 0.5625 & 0.4896    & 0.7077   \\\hline
		GLM\_DEBATE      & 0.8485 & 0.6207 & 0.5782  & 0.5159 & 0.5743   & 0.4216  & 0.7013 & 0.5326    & 0.736    \\
		Qwen\_DEBATE     & 0.8657 & 0.6536 & 0.5872  & 0.5965 & 0.551    & 0.4706  & 0.6301 & 0.6221    & 0.7484   \\
		DeepSeek\_DEBATE & 0.8374 & 0.6358 & 0.6129  & 0.5939 & 0.488    & 0.4027  & 0.5397 & 0.5872    & 0.7245   \\\hline
	\end{tabular}%
\end{table*}

\begin{table*}[htb]
	\centering
	\footnotesize
	\caption{Detailed results for Task B in Weibo dataset}
	\label{tab:drtaskb_w}
	\begin{tabular}{lccccccccc}
		\hline
		\textbf{Models} & \textbf{Happy} & \textbf{Sad} & \textbf{Anger} & \textbf{Surprise} & \textbf{Disgust} & \textbf{Fear} & \textbf{Neutral} & \textbf{M-F1} & \textbf{Acc} \\\hline
		GLM              & 0.7111 & 0.5591 & 0.6342  & 0.5161 & 0.4328   & 0.4928  & 0.2308 & 0.4471    & 0.6073   \\
		Qwen             & 0.7643 & 0.6027 & 0.6434  & 0.5455 & 0.3652   & 0.449   & 0.5915 & 0.4952    & 0.6628   \\
		DeepSeek         & 0.7464 & 0.5973 & 0.6062  & 0.5179 & 0.438    & 0.3929  & 0.3214 & 0.4525    & 0.6424   \\\hline
		GLM\_TOC         & 0.7456 & 0.5385 & 0.6059  & 0.4883 & 0.4094   & 0.3279  & 0.339  & 0.4318    & 0.6258   \\
		Qwen\_TOC        & 0.774  & 0.6128 & 0.5718  & 0.5289 & 0.4203   & 0.4096  & 0.5556 & 0.4841    & 0.6533   \\
		DeepSeek\_TOC    & 0.7583 & 0.6161 & 0.6294  & 0.4044 & 0.4615   & 0.4407  & 0.4839 & 0.4743    & 0.6571   \\\hline
		GLM\_THOR        & 0.7342 & 0.5907 & 0.5781  & 0.5484 & 0.4539   & 0.4202  & 0.3273 & 0.4566    & 0.6175   \\
		Qwen\_THOR       & 0.7531 & 0.6518 & 0.5953  & 0.5346 & 0.4341   & 0.3761  & 0.5479 & 0.4866    & 0.6501   \\
		DeepSeek\_THOR   & 0.7248 & 0.5604 & 0.5356  & 0.3542 & 0.4032   & 0.3939  & 0.2642 & 0.4045    & 0.6098   \\\hline
		GLM\_DEBATE      & 0.7375 & 0.5773 & 0.5285  & 0.6367 & 0.5429   & 0.3878  & 0.4478 & 0.4823    & 0.6354   \\
		Qwen\_DEBATE     & 0.7557 & 0.5052 & 0.6019  & 0.6502 & 0.449    & 0.3738  & 0.4516 & 0.4734    & 0.6411   \\
		DeepSeek\_DEBATE & 0.6877 & 0.5414 & 0.6292  & 0.4976 & 0.4286   & 0.3768  & 0.2642 & 0.4282    & 0.6111  \\\hline
	\end{tabular}%
\end{table*}

\begin{table*}[htb]
	\centering
	\footnotesize
	\tabcolsep 5pt
	\caption{Detailed results for Task C}
	\label{tab:drtaskc}
	\begin{tabular}{lccccc|ccccc}
		\hline
		& \multicolumn{5}{c}{\textbf{Twitter}} & \multicolumn{5}{c}{\textbf{Weibo}} \\\cline{2-11}
		\textbf{models}   & \textbf{R\&C}     & \textbf{R} &  \textbf{L}   & \textbf{M-F1}    & \textbf{Acc}   & \textbf{R\&C}     & \textbf{R} &  \textbf{L}   & \textbf{M-F1}    & \textbf{Acc}  \\\hline
		GLM & 0.3949 & 0.4303 & 0.3760 & 0.3003 & 0.4047 & 0.3411 & 0.4884 & 0.6511 & 0.3701 & 0.4896 \\
		Qwen & 0.2801 & 0.5066 & 0.7084 & 0.3738 & 0.5320 & 0.4629 & 0.2788 & 0.3295 & 0.2678 & 0.3167 \\
		DeepSeek & 0.4432 & 0.3474 & 0.6093 & 0.3500 & 0.4745 & 0.248 & 0.4055 & 0.4080 & 0.2654 & 0.3384 \\\hline
		GLM\_TOC & 0.5340 & 0.284 & 0.4269 & 0.3112 & 0.4304 & 0.5129 & 0.3179 & 0.6332 & 0.3660 & 0.4879 \\
		Qwen\_TOC & 0.3923 & 0.4104 & 0.6463 & 0.3622 & 0.4874 & 0.4738 & 0.1468 & 0.5612 & 0.2955 & 0.4227 \\
		DeepSeek\_TOC & 0.4701 & 0.3411 & 0.6661 & 0.3693 & 0.4925 & 0.3881 & 0.3551 & 0.6232 & 0.3416 & 0.4666 \\\hline
		GLM\_THOR & 0.3585 & 0.4436 & 0.2334 & 0.2589 & 0.3764 & 0.4106 & 0.4560 & 0.5449 & 0.3529 & 0.4542 \\
		Qwen\_THOR & 0.1841 & 0.4792 & 0.6303 & 0.3234 & 0.4576 & 0.4079 & 0.4143 & 0.5924 & 0.3536 & 0.4587 \\
		DeepSeek\_THOR & 0.4539 & 0.4118 & 0.6044 & 0.3675 & 0.4819 & 0.2791 & 0.4160 & 0.5857 & 0.3202 & 0.4417 \\\hline
		GLM\_DEBATE & 0.4590 & 0.3087 & 0.2085 & 0.2440 & 0.3511 & 0.4194 & 0.3114 & 0.0481 & 0.1947 & 0.3135 \\
		Qwen\_DEBATE & 0.4532 & 0.3645 & 0.5438 & 0.3404 & 0.4497 & 0.4816 & 0.2407 & 0.6314 & 0.3384 & 0.4689 \\
		DeepSeek\_DEBATE & 0.4678 & 0.3555 & 0.5819 & 0.3513 & 0.4614 & 0.4181 & 0.3104 & 0.6173 & 0.3365 & 0.4464 \\\hline
	\end{tabular}%
\end{table*}

%

\begin{figure*}[htb]
	\centering
	\begin{subfigure}[b]{0.49\linewidth}
		\centering
		\includegraphics[width=\linewidth]{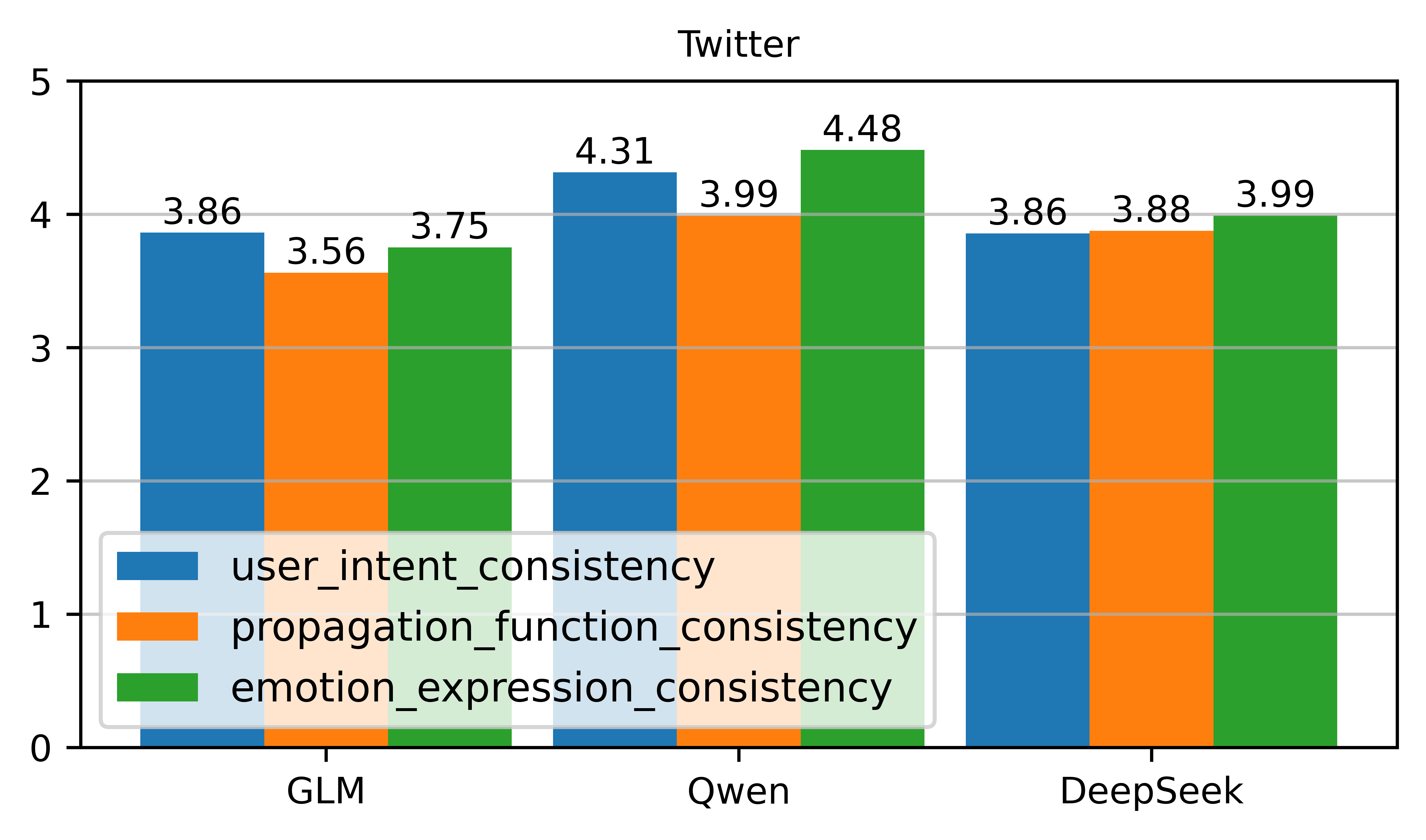}
	\end{subfigure}
	\hfill
	\begin{subfigure}[b]{0.49\linewidth}
		\centering
		\includegraphics[width=\linewidth]{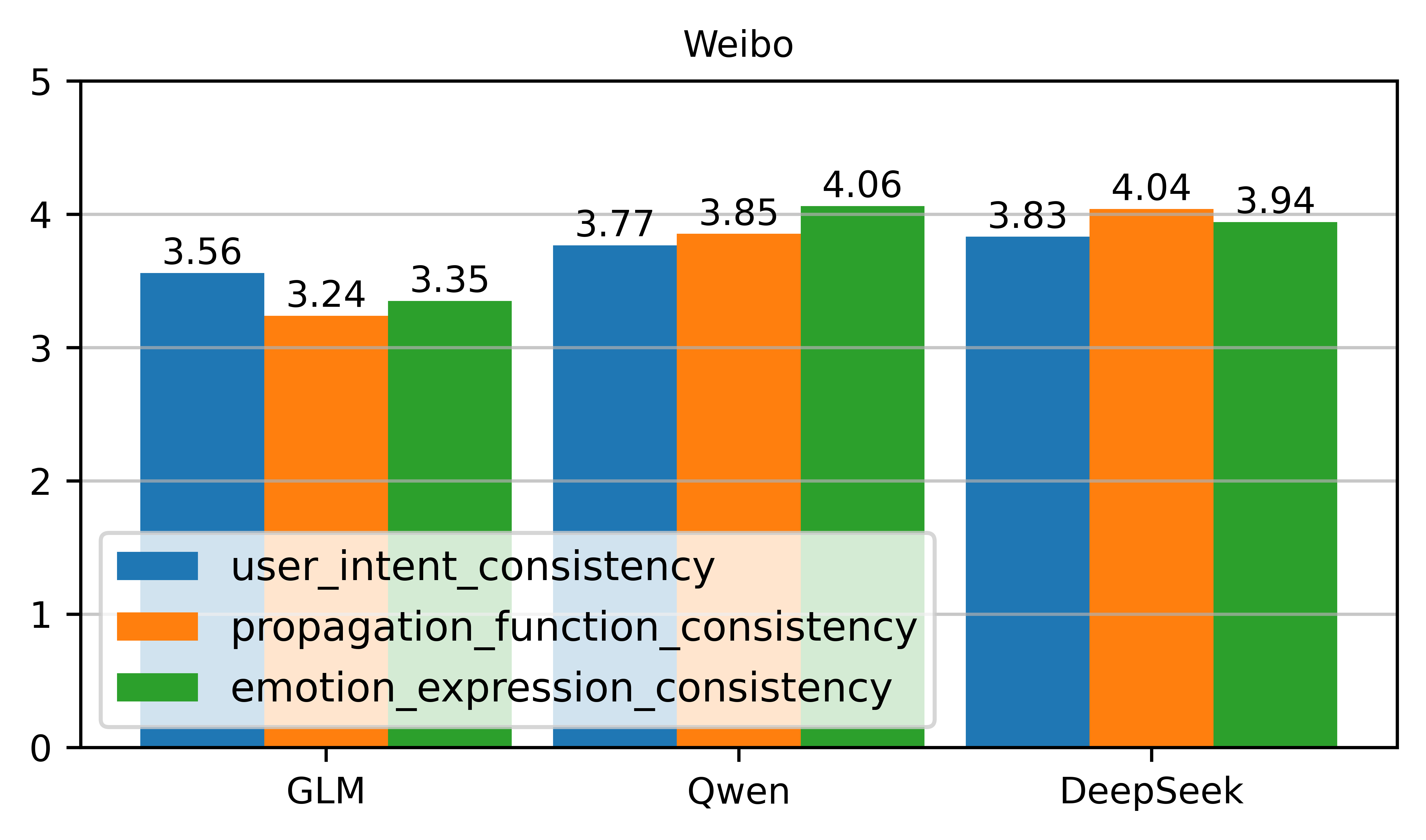}
	\end{subfigure}
	\caption{Comparison of UIC, PFC, and EEC based on different LLMs.}
	\label{fig:taske-upe}
\end{figure*}

\section{The Prompts for LLM-based Evaluations}
\label{appendix:prompts}
\begin{figure*}[htb]
	\centering
	\includegraphics[width=0.8\textwidth]{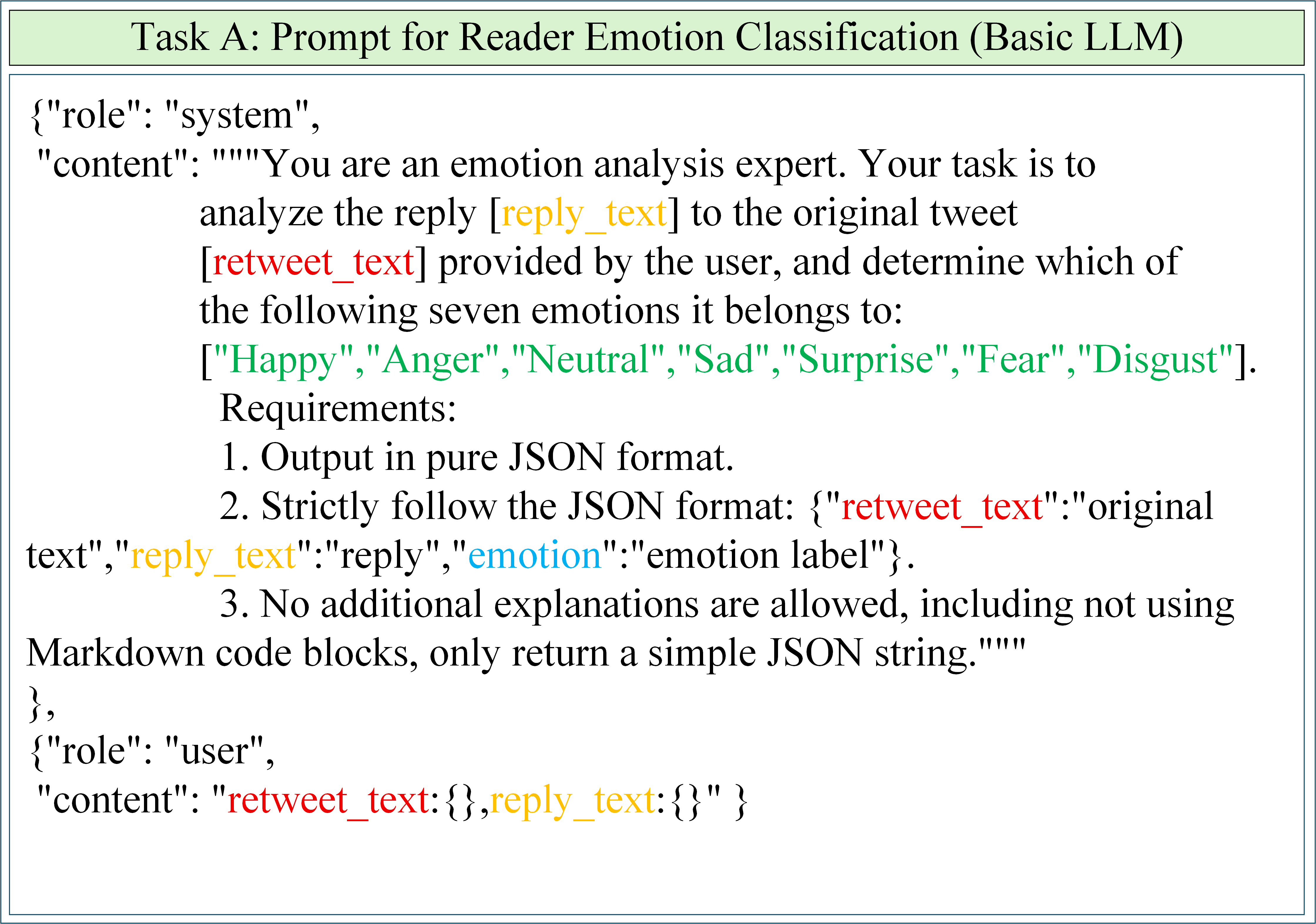}
	\caption{Prompt for Task A (basic LLM).}
	\label{fig:p_taskallm}
\end{figure*}

\begin{figure*}[htb]
	\centering
	\includegraphics[width=0.98\textwidth]{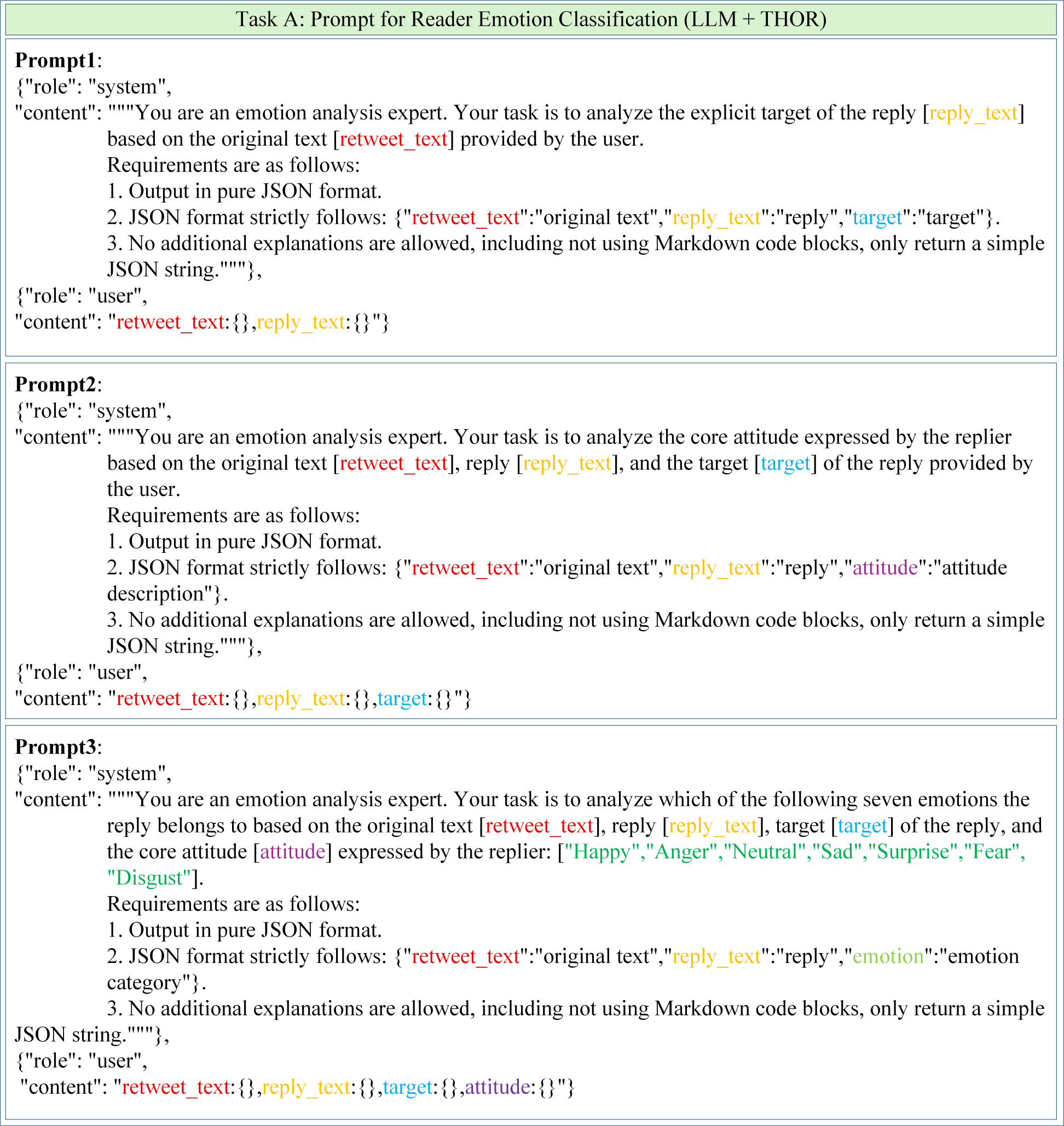}
	\caption{Prompt for Task A (LLM + THOR).}
	\label{fig:p_taska_llm_thor}
\end{figure*}

\begin{figure*}[htb]
	\centering
	\includegraphics[width=0.9\textwidth]{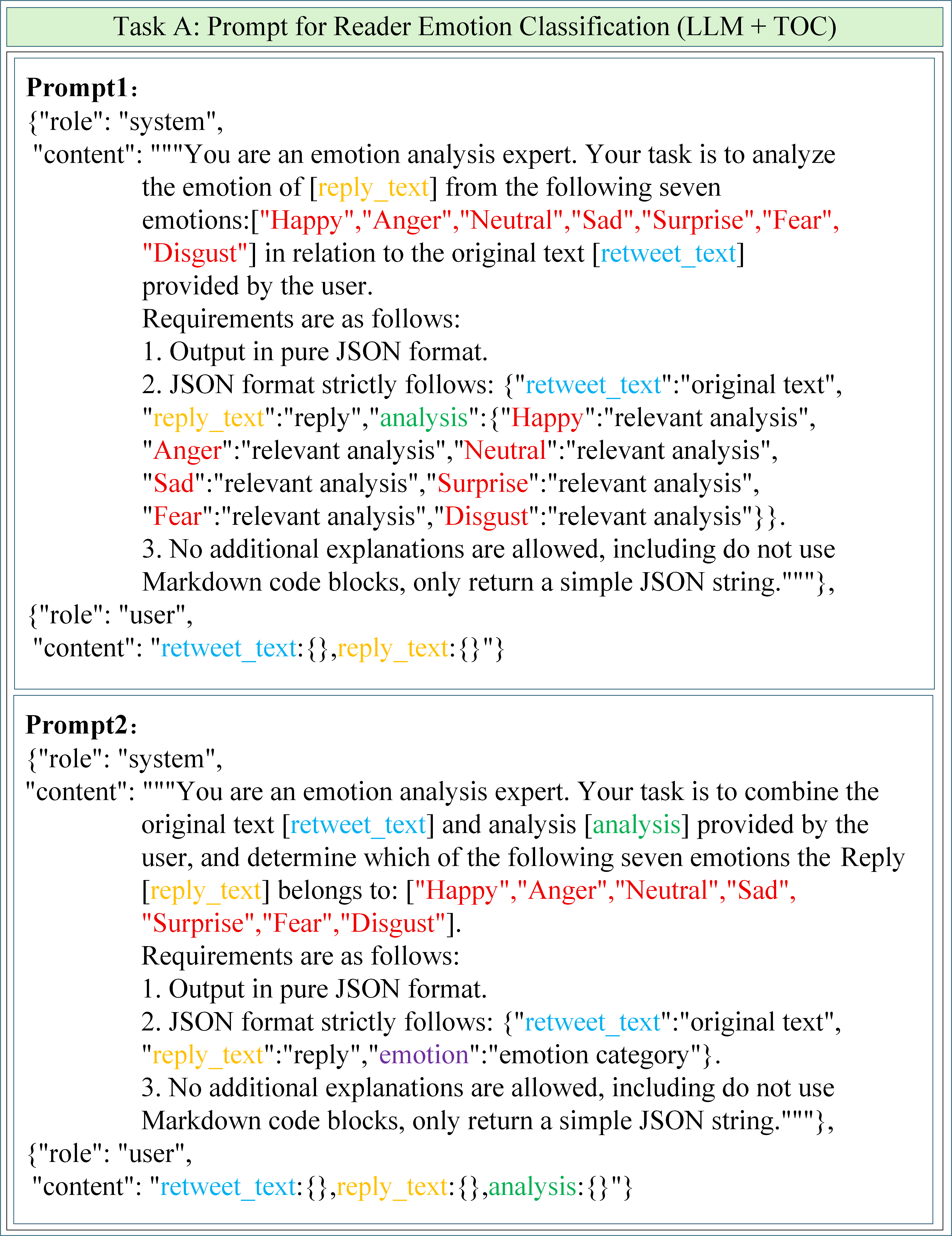}
	\caption{Prompt for Task A (LLM + TOC).}
	\label{fig:p_taska_llm_toc}
\end{figure*}

\begin{figure*}[htb]
	\centering
	\includegraphics[width=0.85\textwidth]{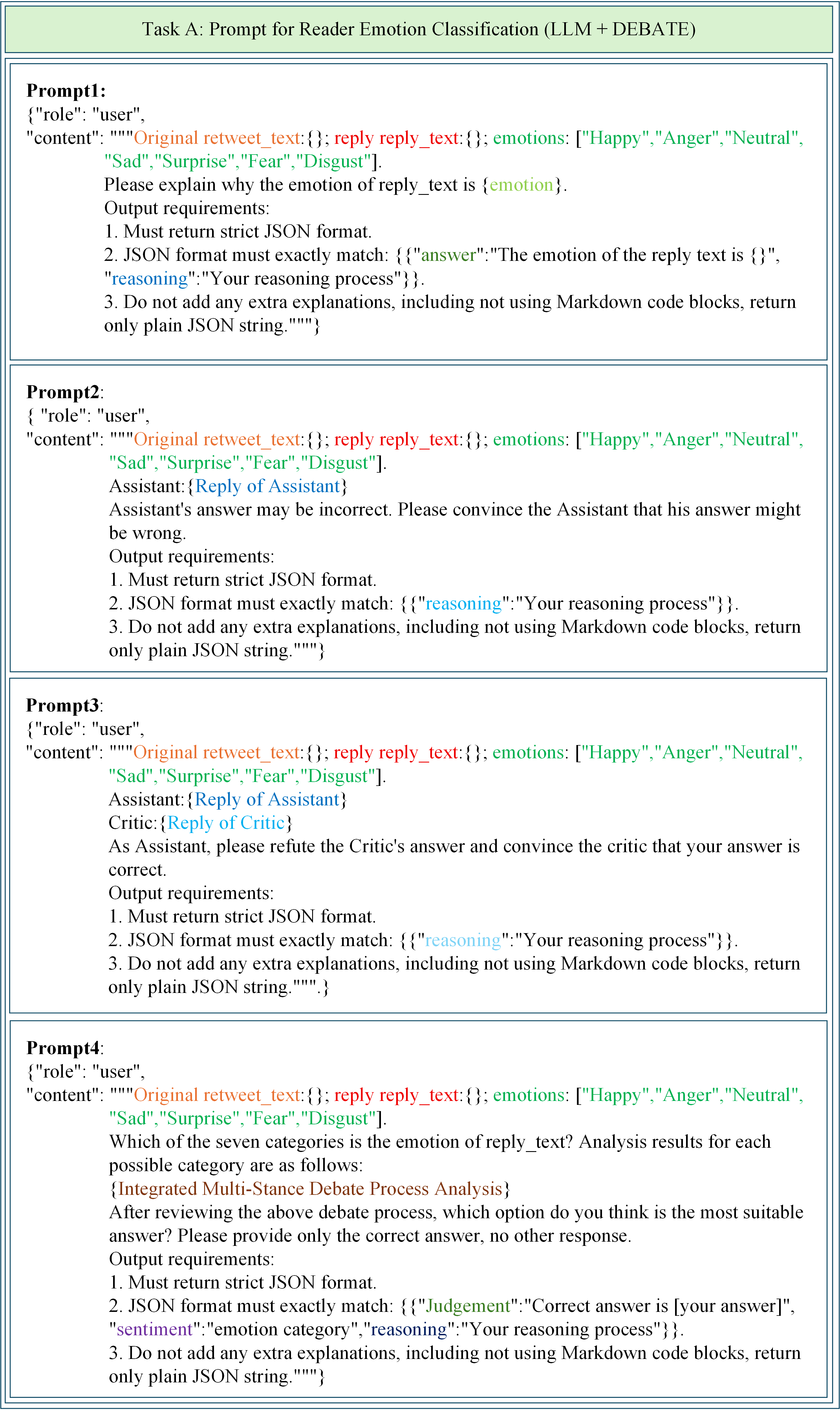}
	\caption{Prompt for Task A (LLM + Debate).}
	\label{fig:p_taska_llm_debate}
\end{figure*}

\begin{figure*}[htb]
	\centering
	\includegraphics[width=0.98\textwidth]{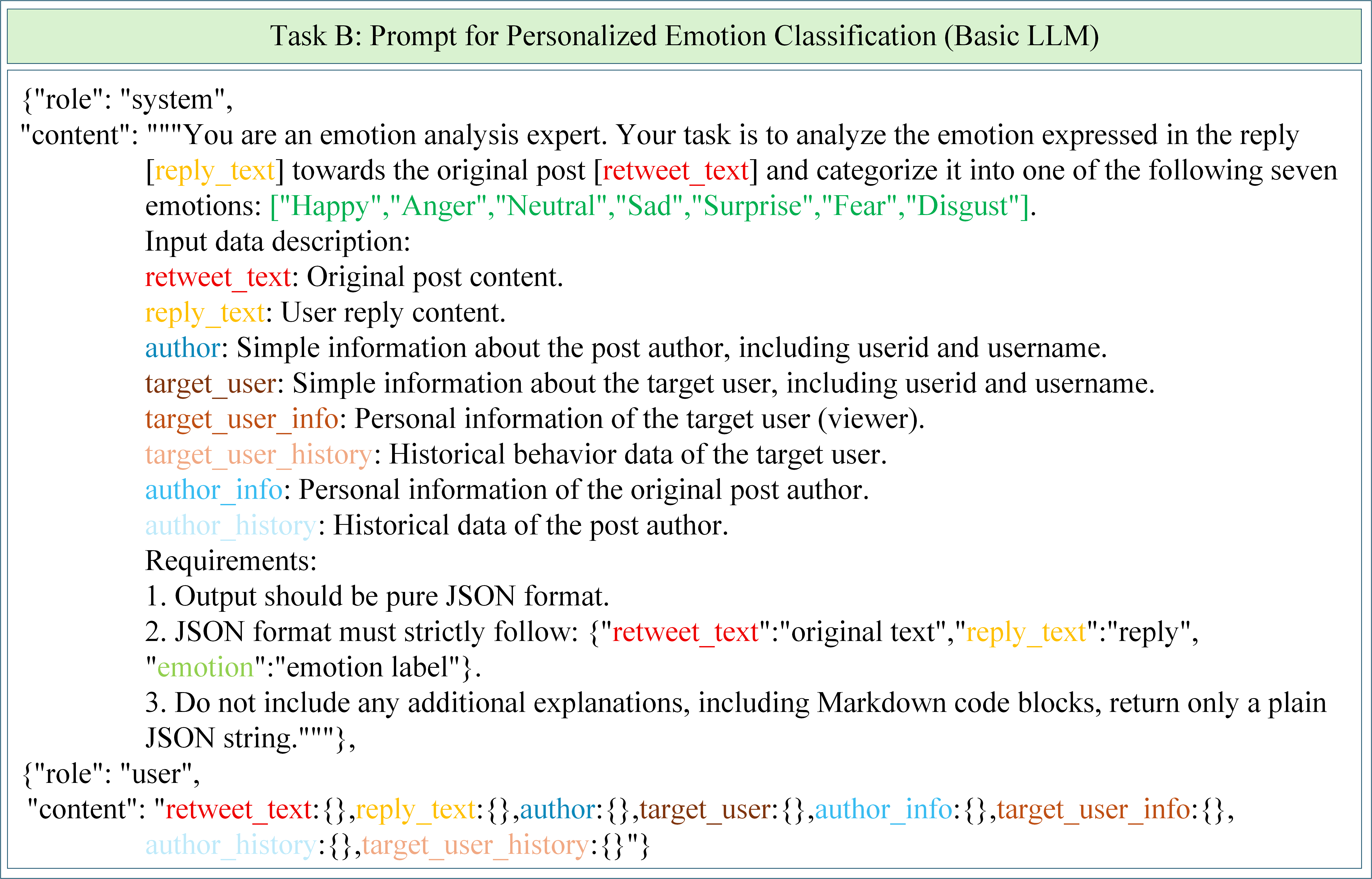}
	\caption{Prompt for Task B (basic LLM).}
	\label{fig:p_taskbllm}
\end{figure*}

\begin{figure*}[htb]
	\centering
	\includegraphics[width=0.85\textwidth]{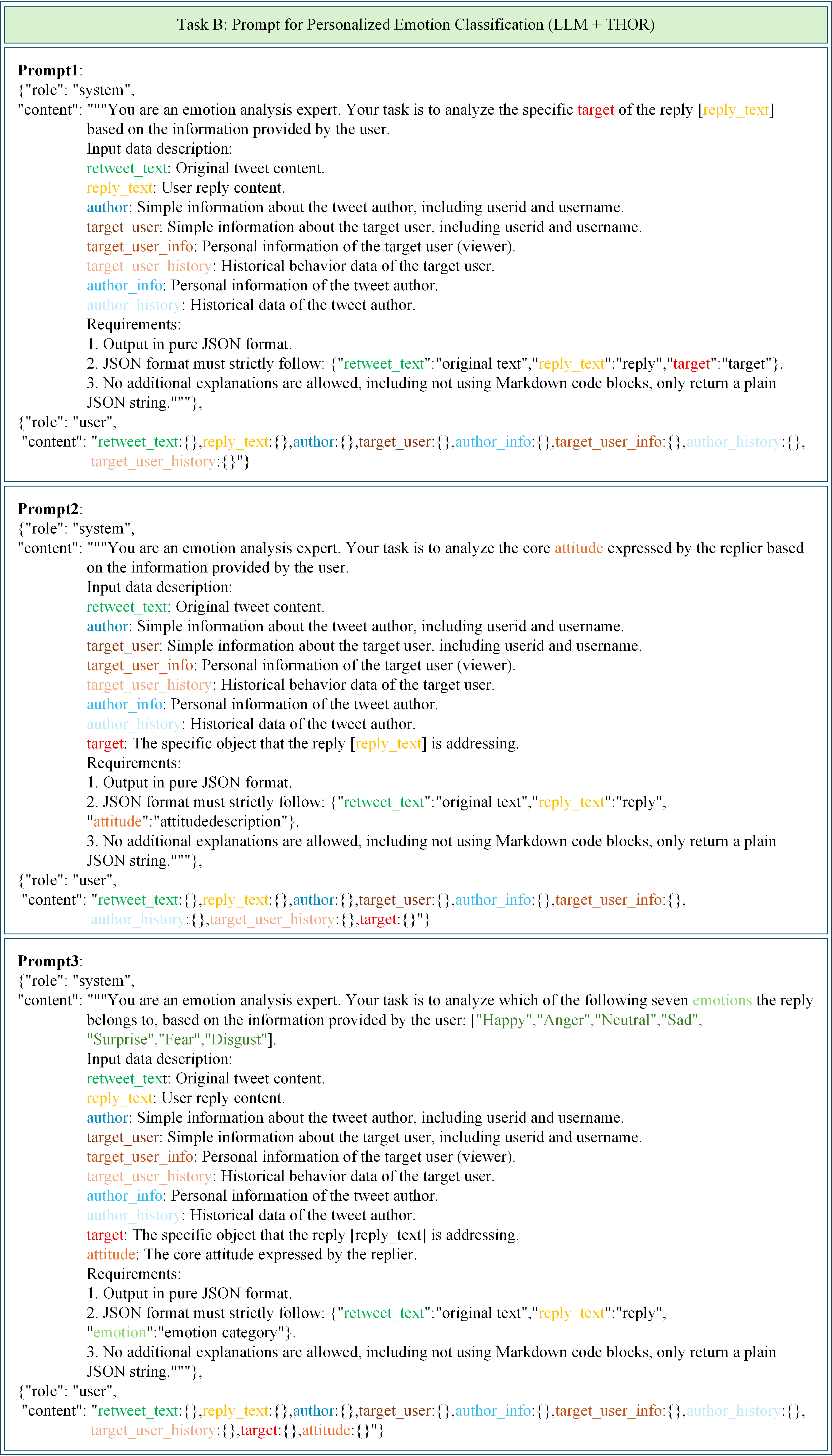}
	\caption{Prompt for Task B (LLM + THOR).}
	\label{fig:p_taskb_llm_thor}
\end{figure*}

\begin{figure*}[htb]
	\centering
	\includegraphics[width=0.98\textwidth]{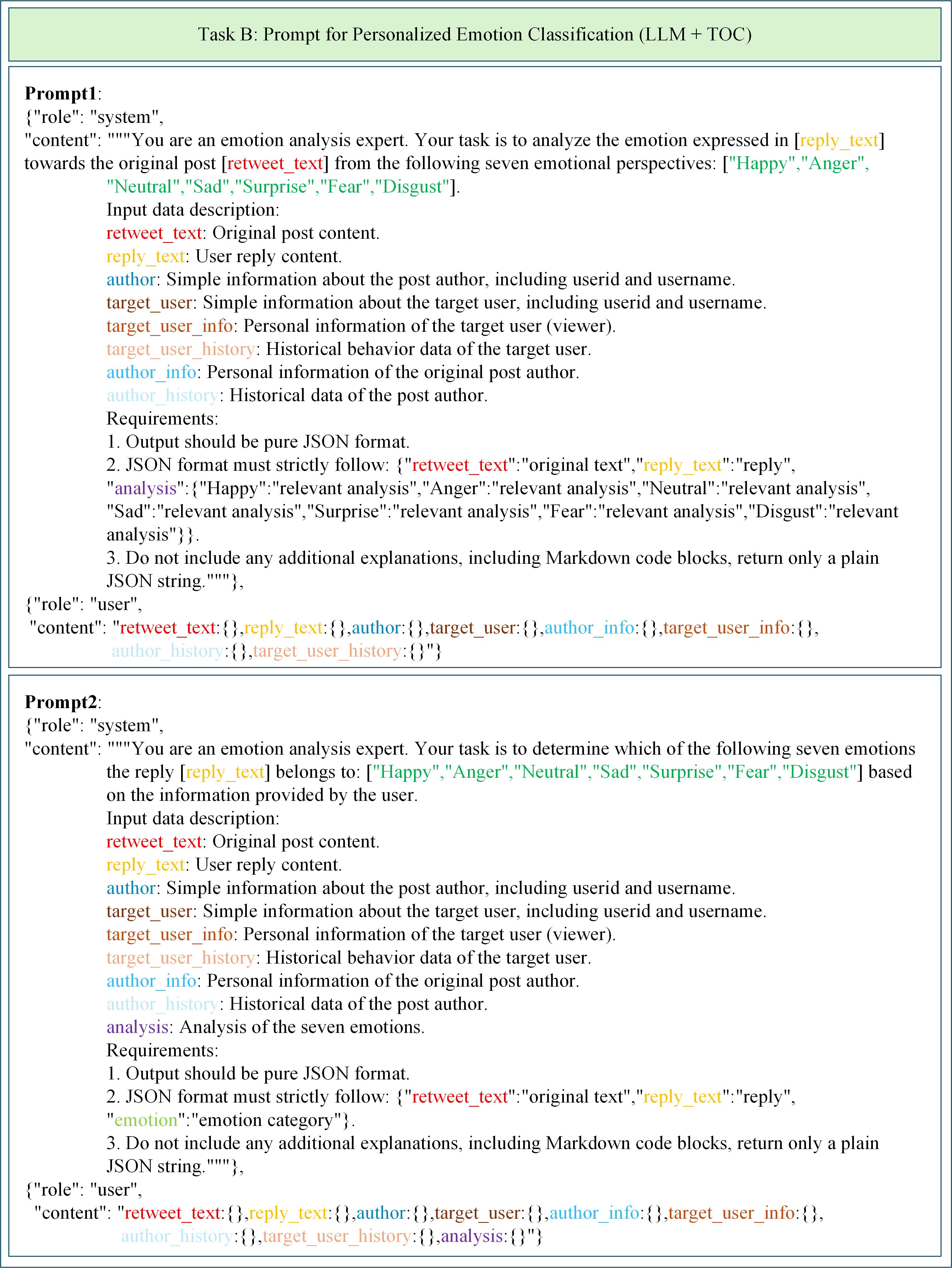}
	\caption{Prompt for Task B (LLM + TOC).}
	\label{fig:p_taskb_llm_toc}
\end{figure*}

\begin{figure*}[htb]
	\centering
	\includegraphics[width=0.9\textwidth]{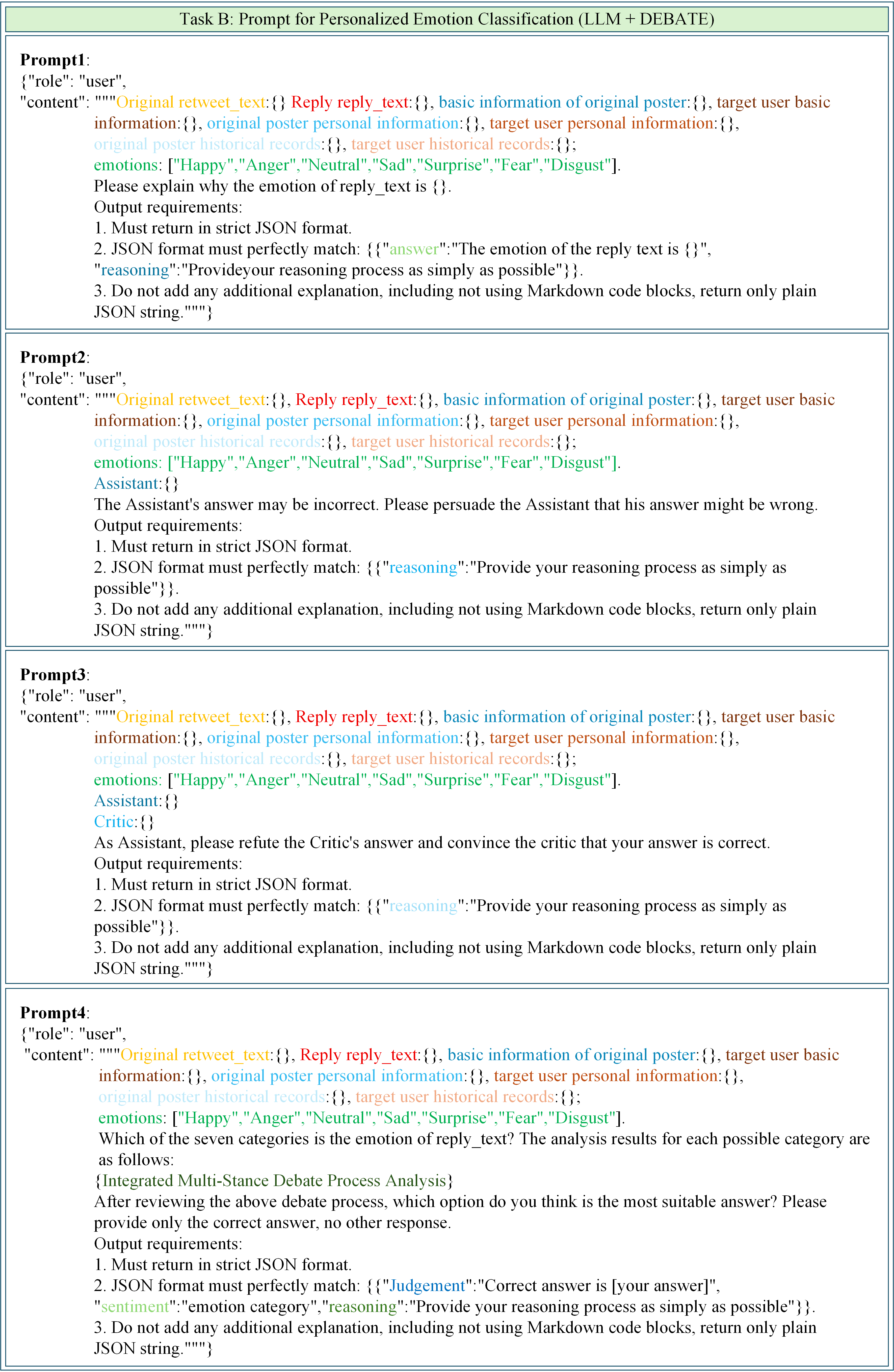}
	\caption{Prompt for Task B (LLM + Debate).}
	\label{fig:p_taskb_llm_debate}
\end{figure*}

\begin{figure*}[htb]
	\centering
	\includegraphics[width=0.98\textwidth]{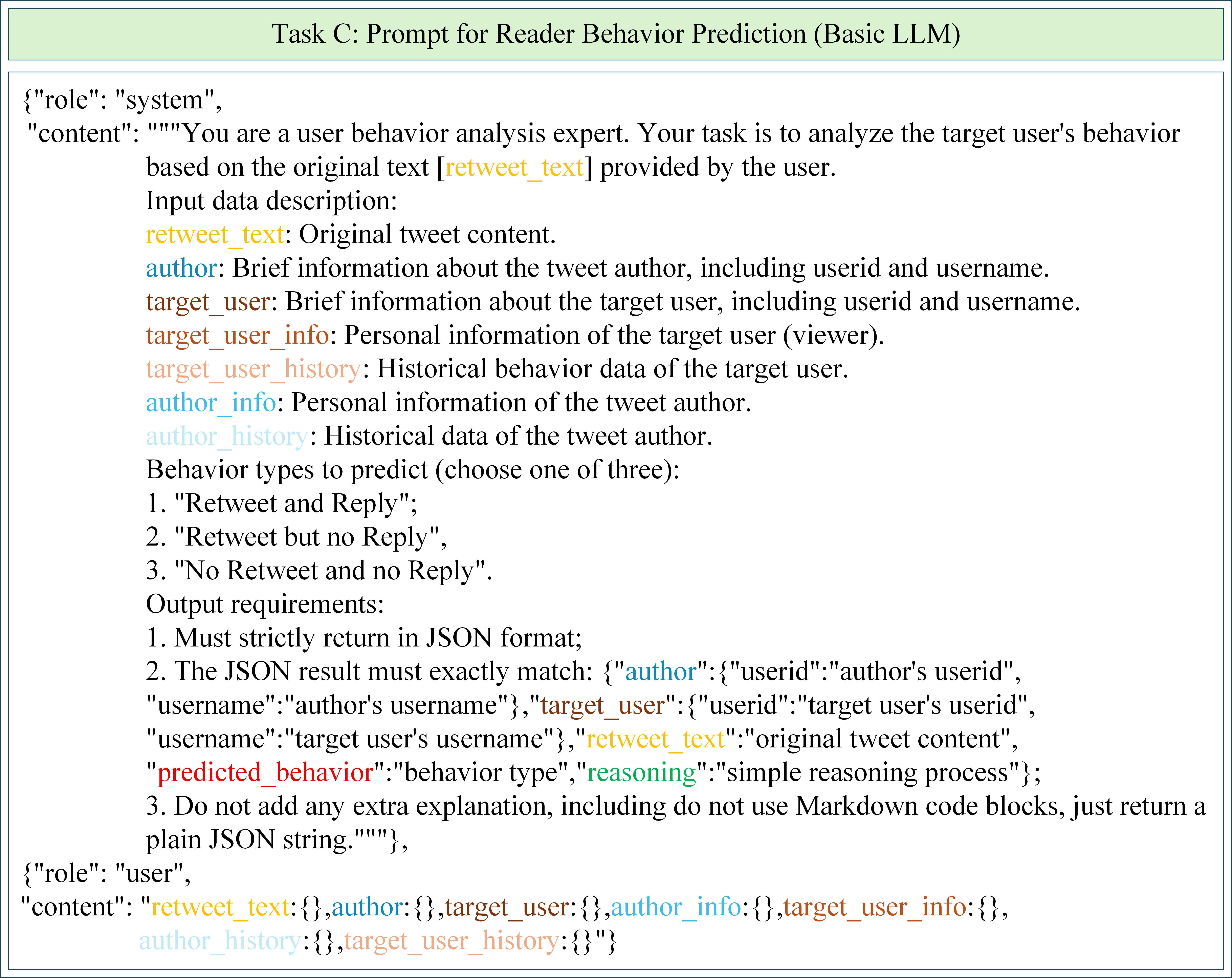}
	\caption{Prompt for Task C (basic LLM).}
	\label{fig:p_taskcllm}
\end{figure*}

\begin{figure*}[htb]
	\centering
	\includegraphics[width=0.9\textwidth, height=\textheight]{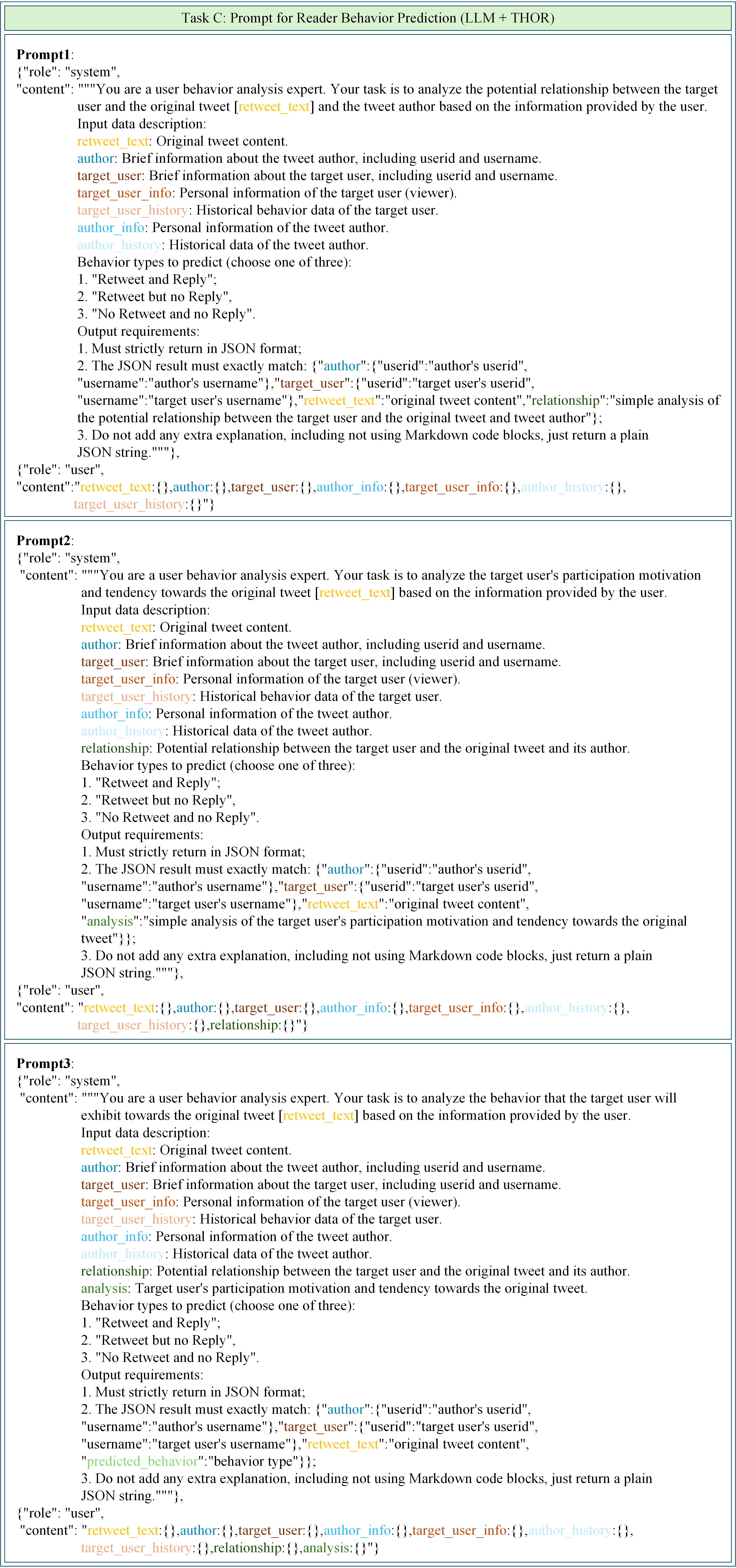}
	\caption{Prompt for Task C (LLM + THOR).}
	\label{fig:p_taskc_llm_thor}
\end{figure*}

\begin{figure*}[htb]
	\centering
	\includegraphics[width=0.98\textwidth]{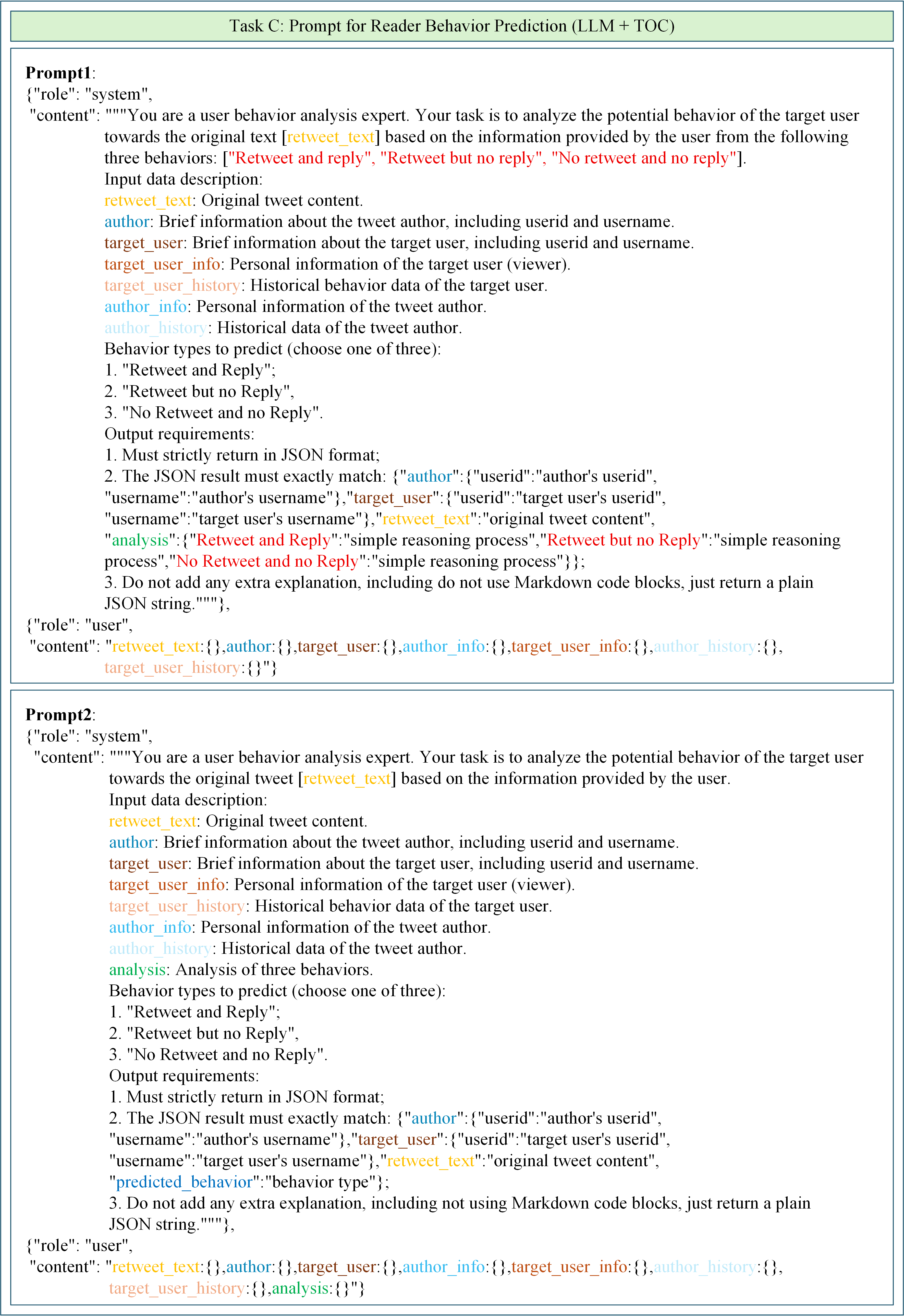}
	\caption{Prompt for Task C (LLM + TOC).}
	\label{fig:p_taskc_llm_toc}
\end{figure*}

\begin{figure*}[htb]
	\centering
	\includegraphics[width=0.98\textwidth]{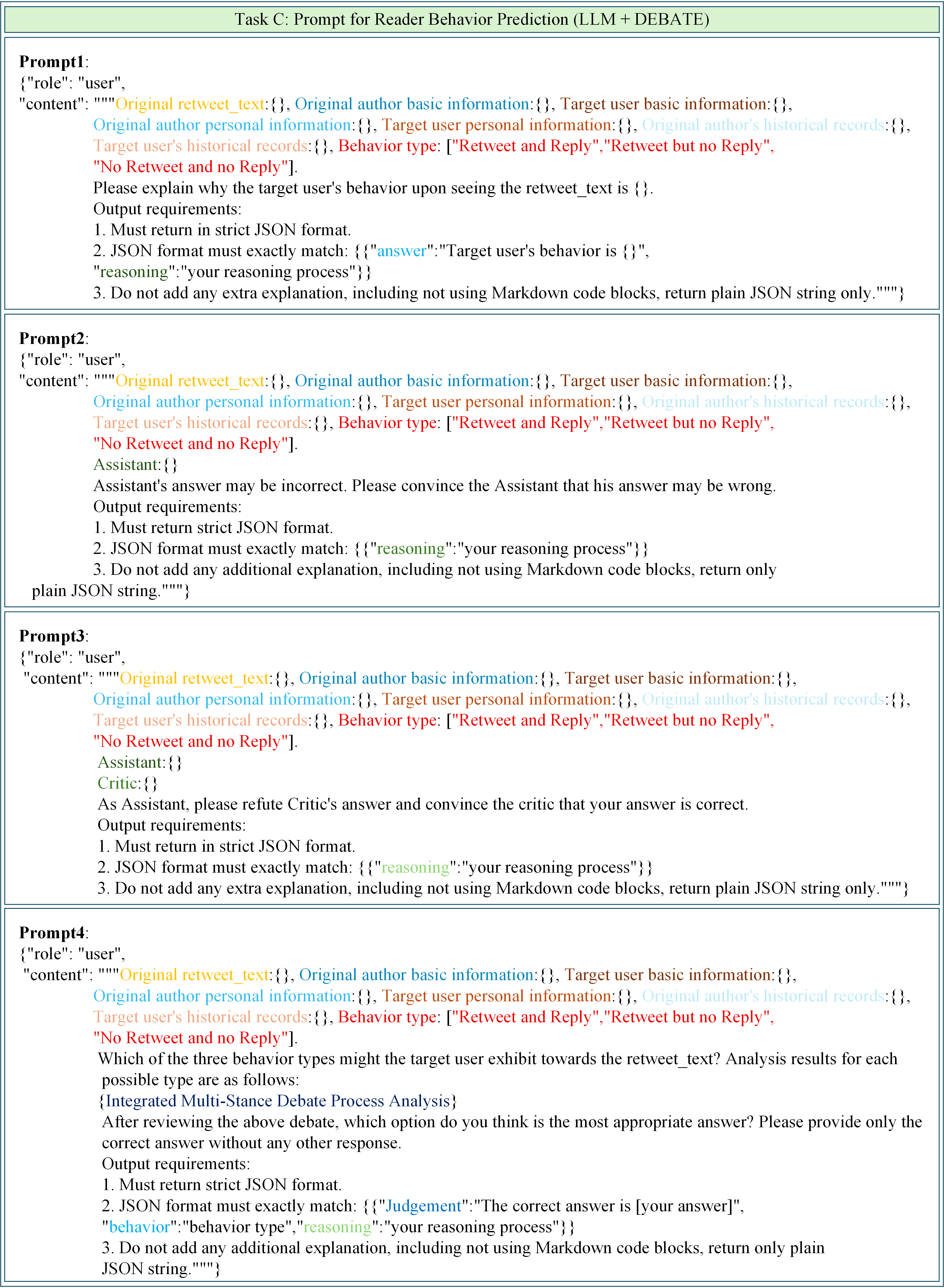}
	\caption{Prompt for Task C (LLM + Debate).}
	\label{fig:p_taskc_llm_debate}
\end{figure*}

\begin{figure*}[htb]
	\centering
	\includegraphics[width=0.9\textwidth, height=\textheight]{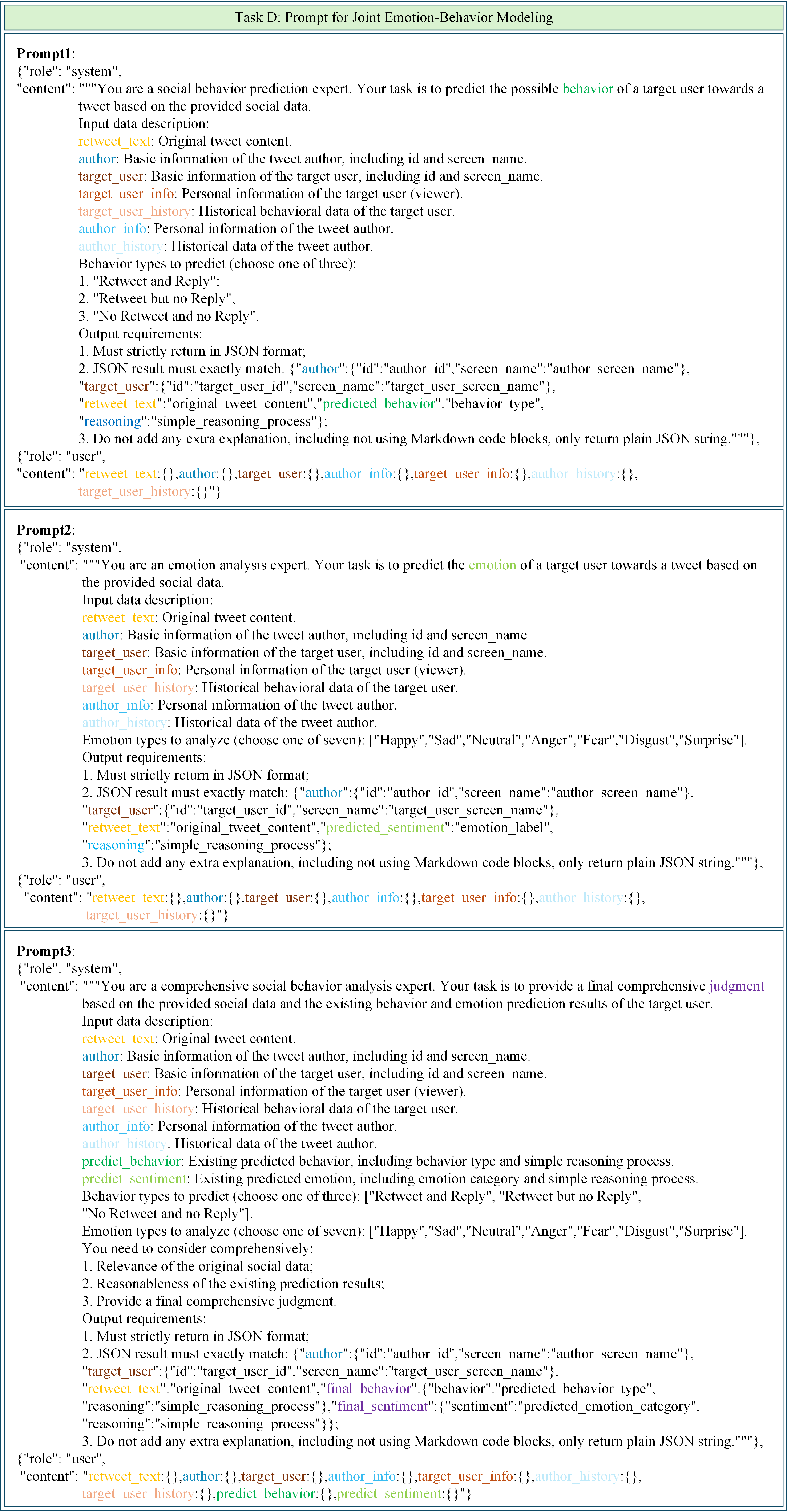}
	\caption{Prompt for Task D.}
	\label{fig:p_taskd}
\end{figure*}

\begin{figure*}[htb]
	\centering
	\includegraphics[width=0.98\textwidth]{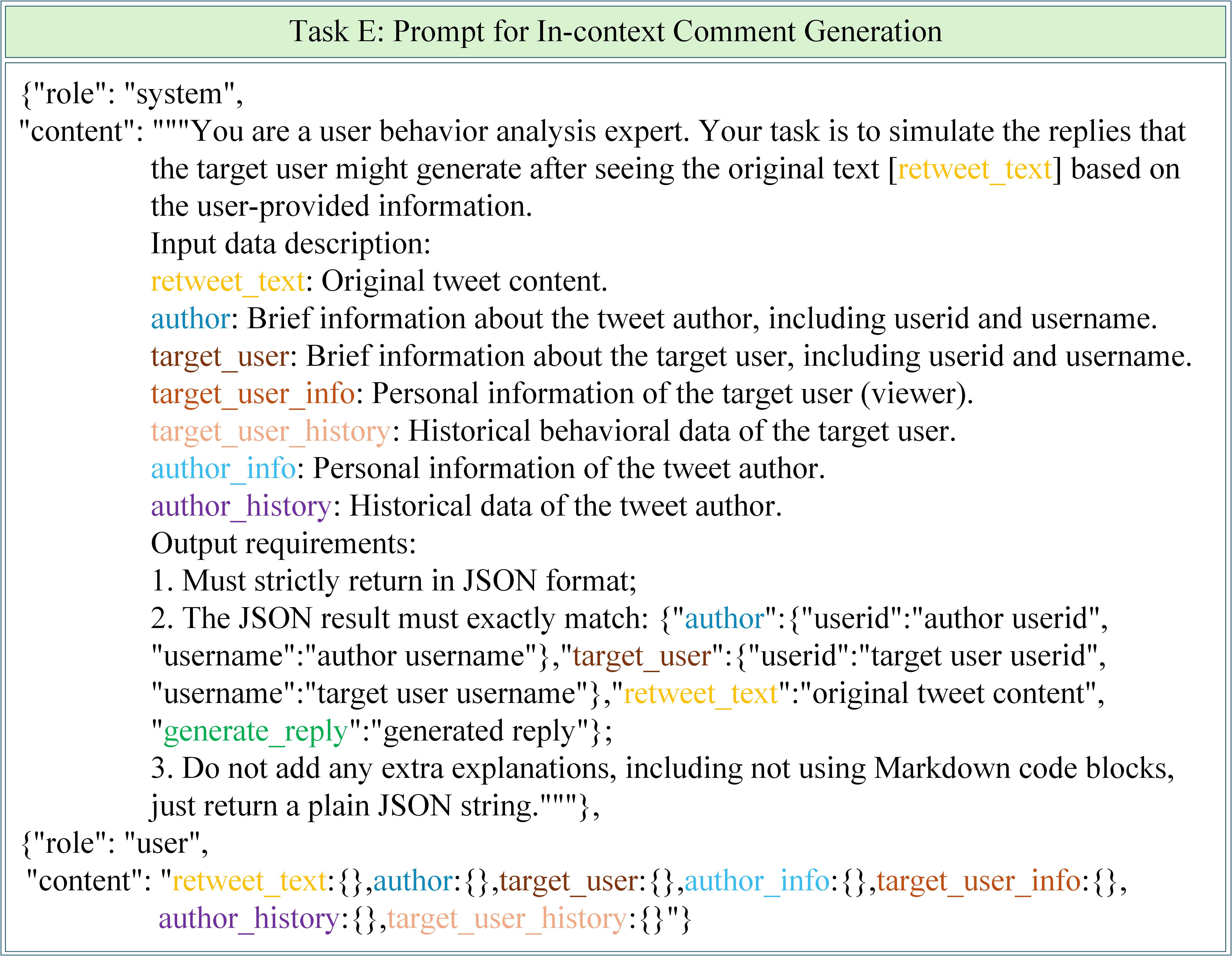}
	\caption{Prompt of in-context comment generation for Task E.}
	\label{fig:p_taske}
\end{figure*}

\begin{figure*}[htb]
	\centering
	\includegraphics[width=0.98\textwidth]{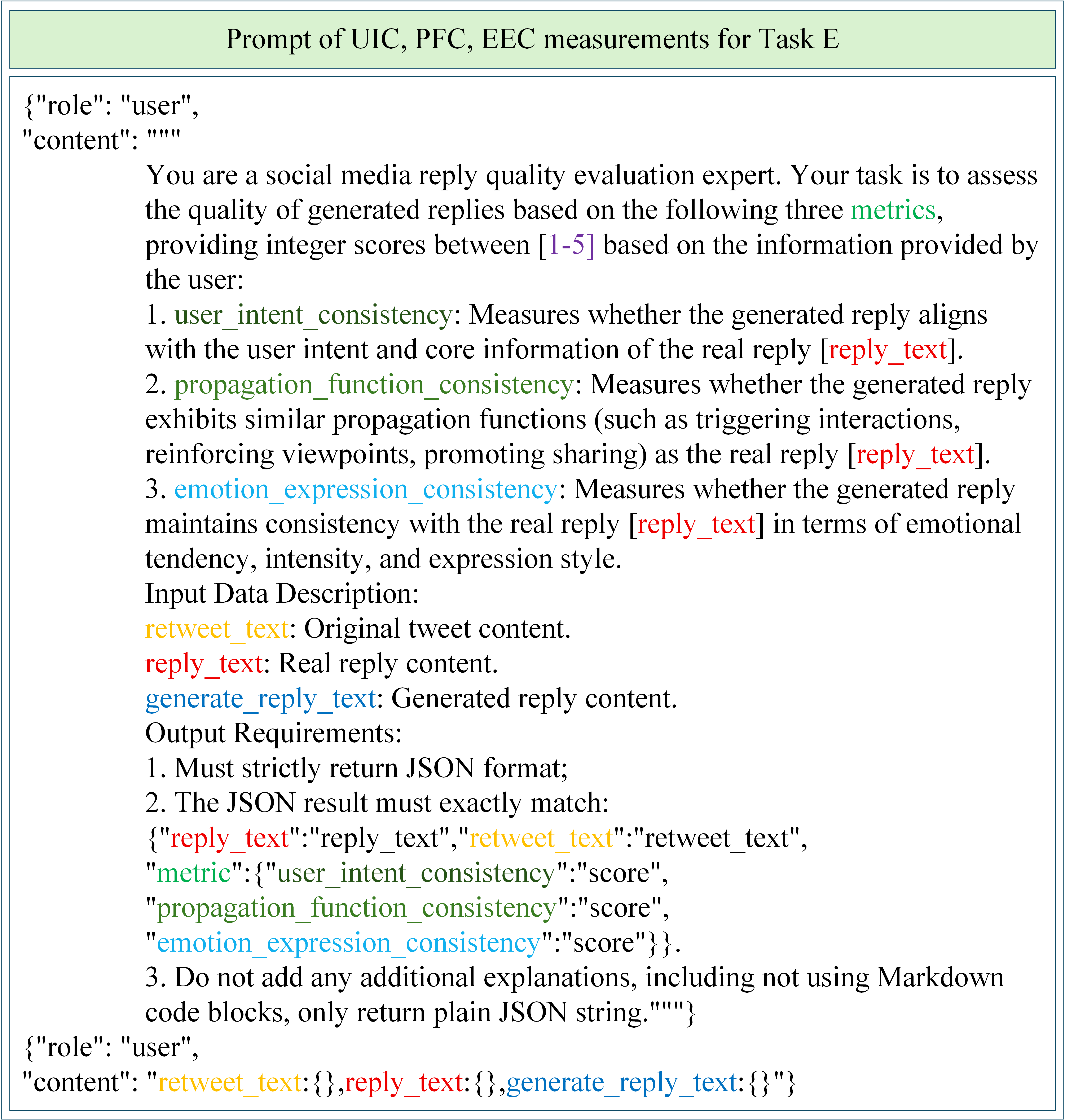}
	\caption{Prompt of UIC, PFC, EEC measurements for Task E.}
	\label{fig:p_taske_upe}
\end{figure*}

The prompts for LLM-based evaluations of each task are illustrated in Figs.~\ref{fig:p_taskallm}-\ref{fig:p_taske}. Fig.~\ref{fig:p_taske_upe} presents the prompt of UIC, PFC, and EEC measurements for in-context comment generation evaluation.

Three LLM-based judgments designed to capture different aspects of generation quality:
	
(1) User Intent Consistency (UIC): Measures whether the generated reply aligns with the user intent and core information of the real reply.
	
(2) Propagation Function Consistency (PFC): Assesses whether the generated reply exhibits similar propagation functions (e.g., triggering interactions, reinforcing viewpoints) as the real reply.
	
(3) Emotion Expression Consistency (EEC): Evaluates whether the generated reply maintains consistency with the real reply in terms of emotional tendency, intensity, and expression style.
	
\end{document}